\documentclass[a4paper,fleqn,usenatbib]{mnras}
\pdfoutput=1
\usepackage{newtxtext,newtxmath}

\usepackage[T1]{fontenc}
\usepackage{ae,aecompl}

\usepackage{graphicx}	
\usepackage{amsmath}	

\usepackage{float}
\usepackage{caption}
\usepackage{subcaption}
\usepackage[skip=0pt]{caption}
\usepackage{physics}
\usepackage{tabularx}
\usepackage{booktabs}
\usepackage{array, multirow}
\usepackage{xparse}

\newcommand{\objid} {\texttt{ObjID}}
\newcommand{\halpha} {${\rm H \alpha}$}
\newcommand{\hbeta} {${\rm H \beta}$}
\newcommand{\mbh}{${M_{\rm BH}}$}
\newcommand{\mdot}{${\dot{M}_{\rm BH}}$}
\newcommand{\mstar}{${M_\ast}$}
\newcommand{\mhalo}{${M_{\rm halo}}$}
\newcommand{\vdisp}{$\sigma_\ast$}
\newcommand{\vmax}{$V_{\rm max}$}
\newcommand{\mbulge}{$M_{\rm bulge}$}
\newcommand{\fgas}{$f_{\rm gas}$}
\newcommand{\mgas}{$M_{\rm H2}$}
\newcommand{\msol}{${\rm M_\odot}$}
\newcommand{\hmol}{${\rm H}_2$}
\newcommand{\msfr}{$M_\ast - {\rm SFR}$}

\title[BH quenching in centrals]{On the quenching of star formation in observed and simulated central galaxies: Evidence for the role of integrated AGN feedback}

\author[J. M. Piotrowska et al.]{
Joanna M. Piotrowska,$^{1,2}$\thanks{E-mail:jmp218@cam.ac.uk}
Asa F. L. Bluck,$^{1,2}$
Roberto Maiolino$^{1,2}$
and Yingjie Peng$^{3}$
\\
$^{1}$Cavendish Laboratory, Astrophysics Group, University of Cambridge, 9 JJ Thomson Avenue, Cambridge, CB3 0HE, UK\\
$^{2}$Kavli Institute for Cosmology, Madingley Road, CB3 0HA, Cambridge, UK\\
$^{3}$Kavli Institute for Astronomy and Astrophysics, Peking University, Yi He Yuan Lu 5, Hai Dian District, Beijing 100871, People's Republic of China\\
}

\date{Accepted XXX. Received YYY; in original form ZZZ}

\pubyear{2021}

\begin{document}
\label{firstpage}
\pagerange{\pageref{firstpage}--\pageref{lastpage}}
\maketitle

\begin{abstract}

In this paper we investigate how massive central galaxies cease their 
star formation by comparing theoretical predictions from cosmological 
simulations: EAGLE, Illustris and IllustrisTNG with observations of 
the local Universe from the Sloan Digital Sky Survey (SDSS). Our 
machine learning (ML) classification reveals supermassive black hole 
mass (\mbh) as the most predictive parameter in determining whether 
a galaxy is star forming or quenched at redshift $z=0$ in all three 
simulations. This predicted consequence of active galactic nucleus 
(AGN) quenching is reflected in the observations, where it is true 
for a range of indirect estimates of \mbh\ via proxies as well as 
its dynamical measurements. Our partial correlation analysis shows 
that other galactic parameters lose their strong association with 
quiescence, once their correlations with \mbh\ are accounted for. 
In simulations we demonstrate that it is the integrated power output 
of the AGN, rather than its instantaneous activity, which causes 
galaxies to quench. Finally, we analyse the change in molecular gas 
content of galaxies from star forming to passive populations. We 
find that both gas fractions (\fgas) and star formation efficiencies 
(SFEs) decrease upon transition to quiescence in the 
observations but SFE is more predictive than \fgas\ in the ML 
passive/star-forming classification. These trends in the SDSS are 
most closely recovered in IllustrisTNG and are in direct contrast 
with the predictions made by Illustris. We conclude that a~viable 
AGN feedback prescription can be achieved by a~combination of 
preventative feedback and turbulence injection which together quench 
star formation in central galaxies.
\end{abstract}

\begin{keywords}
galaxies: evolution -- galaxies: nuclei -- galaxies: star formation
\end{keywords}




\section{Introduction}

The launch of large surveys like the Sloan Digital Sky Survey (SDSS) 
\citep{York00}, revealed that local galaxies reside either in the 
`blue cloud' or the `red sequence' in the colour-magnitude diagram 
\citep[e.g.][]{Strateva01, Baldry04, Wyder07} and that this division 
persists to as far back as redshift $z \sim 3$ 
\cite[e.g.][]{Giallongo05, Brammer09, Willmer06}. The distribution 
in galactic colour, however, reflects a more fundamental bimodal 
distribution in the specific star formation rate (sSFR, star formation 
rate per unit stellar mass) \citep[e.g.][]{Kauffmann03, Santini09}, 
owing to significantly different optical signatures between young 
and old stellar populations \citep[e.g.][]{Bruzual03, Maraston11}. 
When observed across cosmic time, the mass density in the red, 
passive population increases in size, while it remains roughly 
constant in the star forming cloud \citep[e.g.][]{Bell04, Faber07}. 
This observation is generally interpreted as a change in the object 
membership between the two populations as a consequence of star 
formation slowing down. As a result, quenching is at large defined
as the process of galaxy transition between the blue cloud and the 
red sequence over time. For the purpose of our research, however, 
quenching refers specifically to the change in location in the 
stellar mass (\mstar) -- star formation rate (SFR) plane during 
which galaxies fall off the star forming Main Sequence (MS, 
\citealt{Noeske07}; \citealt{Brinchmann04}) to join the `passive' 
population with low sSFRs.

Quenching has been shown to correlate well with galaxy stellar mass
\citep[e.g.][]{Baldry06, Peng10, Liu19, Bluck19}, halo mass 
\citep[e.g.][]{Woo13, Wang18}, morphology (e.g. \citealt{Cameron09a};
\citealt{Cameron09b}; \citealt{Bell12}; \citealt{Bluck14}; 
\citealt{Omand14}), 
the angular momentum of inflowing gas (i.e. angular momentum
quenching, \citealt{Peng20}, \citealt{Renzini20}), stellar kinematics \citep{Wang20},
central velocity dispersion (e.g. 
\citealt{Wake12}; \citealt{Teimoorinia16}; \citealt{Bluck16}; 
\citealt{Bluck19}; \citealt{Bluck20a}; \citealt{Bluck20b})
and, more recently, with dynamically measured supermassive black 
hole mass \citep[e.g.][]{Terrazas16, Terrazas17, Martin-Navarro18}. 
In the case of satellite galaxies, an additional connection between 
environment and quiescence was revealed, where the star forming 
state of an object depends on local overdensity and its location 
within the parent dark matter halo \citep[e.g.][]{Bosch08, Peng12, 
Woo13, Bluck16, Liu19}. In this study we choose to focus on central 
galaxies only, in order to investigate the intrinsic quenching 
mechanisms, unaffected by the galactic environment.
 
X-ray observations of galaxy clusters brought a crucial 
understanding of the macrophysics of quenching, showing that the 
bulk of baryonic matter associated with massive elliptical galaxies 
resides in their surrounding haloes in a form of hot, ionised gas 
\citep[e.g.][]{Paolillo02, Fukazawa06, Humphrey11}. The main cooling 
channel for this plasma is via the free-free interaction and 
predicts much higher cold gas mass accretion rates onto the galaxies 
than found in the observations \citep[see][for a review]{McNamara07}. 
Therefore, it is apparent that one fundamental operational mechanism 
of quenching is to provide enough heat to offset cooling and keep 
galactic halos hot. This way, a substantial reservoir of gas is 
prevented from collapsing onto a galaxy and not allowed to deliver 
fuel for active star formation.

The microphysics of how the temperature of the haloes is kept high, 
however, remains an open question. One proposed solution is the 
shock-heating of gas as it falls through deep halo potential wells 
as suggested by \cite{Dekel06} and observationally supported by 
e.g. \cite{Woo13}, \cite{Tal14} and \cite{Wang18}. Alternatively, 
accretion onto supermassive black holes in active galactic nuclei
(AGN) is a strong candidate for a heating source, since it can 
generate enough energy to offset high cooling rates in galaxy 
clusters \citep[e.g.][]{Silk98, Binney04, Scannapieco04}. In fact, 
modern cosmological simulations can only successfully shut down star 
formation in massive galaxies using some form of AGN feedback, as 
other processes like supernova explosions are not able to prevent 
galaxies from constantly forming stars at redshift $z=0$ 
\citep[e.g][]{Bower06, McCarthy11, Dubois16, Weinberger17}. 

The ongoing AGN debate revolves around the exact energy deposition
mechanism within the galactic halo -- whether it is achieved through 
violent outflows of gas in the high accretion `quasar mode' 
\citep[e.g.][]{Hopkins08, Hopkins10, Villar-Martin11, Maiolino12} or 
rather through slow and steady energy injection in the low accretion 
`preventative' feedback mode \citep[e.g.][]{McNamara00, Birzan04, 
Fabian12, Hlavacek-Larrondo12, Hlavacek-Larrondo15}. From an 
observational perspective, there exists evidence in favour of both 
heating avenues, however the exact mechanism of coupling the AGN power 
to the halo gas is not well determined yet 
\citep[see][for a~review]{Werner19}. 

More recently, a breakthrough in our understanding of quenching came 
along with the first integrated and spatially resolved surveys of
directly measured molecular gas content in galaxies \citep{Bolatto17, 
Saintonge17, Saintonge18, Tacconi18, Sorai19, Aravena19, Lin20}.
By translating CO line luminosities to \hmol\ masses, multiple authors 
found a significant variation in gas fractions 
($f_{\rm gas}=M_{\rm H2}/M_\ast$) across the \msfr\ plane with 
galaxies below the MS showing significantly lower \fgas\ than their 
MS counterparts (e.g. \citealt{Genzel15}; \citealt{Lin17}; 
\citealt{Saintonge17}; \citealt{Belli21}; \citealt{Ellison21};
\citealt{Dou21a}; \citealt{Dou21b}). Even 
more interestingly, a significant fraction of these studies 
additionally found decreasing star formation efficiencies
($\rm{SFE= SFR/}M_{\rm gas}$, the inverse of gas depletion time 
$t_{\rm dep}$), away from the Main Sequence with the magnitude
of the decrease in SFE exceeding that of \fgas\ 
\citep[e.g.][]{Tacconi18, Lin20, Colombo20, Ellison20, Dou21a, Dou21b}.
In \cite{Piotrowska20}, we used the reddening of optical SDSS spectra 
to estimate the in-fibre molecular gas masses for $\sim 60\, 000$ 
galaxies and confirmed the above trends in a significantly larger 
sample of local galaxies. This combined observational evidence 
suggests a new macrophysical property of a successful quenching 
paradigm -- galaxies do not cease their star formation by solely 
depleting their gas reservoirs. Instead, it is the combined effect of 
the lack of star forming fuel and its reduced potential for 
gravitational collapse, which work in tandem to drive massive local
galaxies towards quiescence.

Much like in the case of heating in the galactic haloes, the 
microphysics of \fgas\ reduction offers room for debate. On the one 
hand, low gas fractions can stem from molecular gas consumption via 
star formation, in the absence of fresh gas supply from the constantly
heated surrounding halo. However, one can easily imagine a scenario 
in which cold gas is also removed from the galaxy, owing to momentum
kicks from supernova explosions \citep[e.g.][]{Kay02, Marri03} or 
quasar activity \citep[e.g.][]{Maiolino12, Harrison14, Zakamska14}
or both \citep[e.g.][]{Fluetsch19}. As shown in a body of theoretical
work \citep[e.g.][]{Springel03, Hopkins11, Pontzen17, Henriques19}
supernova feedback is only likely to be relevant in low-mass galaxies, 
where gravitational potential wells are shallow enough to prevent gas
re-accretion via galactic fountains. The picture is less clear about 
AGN -- although quasar outflows have been found to carry significant 
masses of gas in both the ionised \citep[e.g.][]{Carniani15, Rupke17} 
and molecular phases \citep[e.g.][]{Feruglio10, Veilleux13, 
Fluetsch19, Fluetsch20}, their outflow speeds rarely exceed the 
escape velocity required to leave massive galactic hosts.

The microphysics of SFE reduction received little attention in the 
literature, in contrast to \fgas. One interesting hypothesis suggests 
a~direct influence of galactic morphology, in which gas is prevented 
from collapse via stabilising torques from the central bulge 
\citep{Martig09}. On the opposite end of physical scales, 
interstellar turbulence and magnetic fields have been suggested as 
regulatory mechanisms controlling local SFE 
\citep[e.g.][]{Krumholz05, Federrath12}. The origin of such turbulent behaviour 
in gas remains unclear, however recent observations have shown
that the injection of energy in the ISM from weak radio jets can play 
an important role in this context \citep{Venturi21}. Finally, AGN 
activity could dramatically increase the cooling times in the 
circumgalactic medium (CGM) gas by increasing its entropy via 
kinetic-mode feedback \citep{Zinger20}.

Throughout the quenching debate a general picture emerges in which
there is a unanimous observational support for the macrophysics
of quenching, like the necessity for galactic haloes to remain hot
or for the star formation efficiency to drop within the galaxies.
In contrast, the microphysics of what mechanisms give rise to these
trends escapes our direct observation. The nature of complex 
processes like gas accretion, jet launching or powerful explosions
cannot be inferred from their electromagnetic signature without 
appropriate theoretical modelling.

This is exactly where cosmological simulations prove invaluable.
Because the simulated universe is built with clearly defined 
treatment of unresolved baryonic physics, fluid dynamics and gravity, 
one can make a connection between small-scale physical processes 
and their predicted observable consequences accessible to our 
instruments. With the knowledge of implemented prescriptions and 
their limitations we can make detailed testable predictions to validate 
or challenge model assumptions when we compare these predictions 
with the observable Universe. Hence, if there exists a~close 
correspondence between the simulations and the observations, we can 
use the former to explain a possible physical origin of the trends 
we see in the observable Universe. As of today, 
\mbox{(magneto-) hydrodynamical} 
simulations are becoming increasingly successful at 
reproducing the observed Universe on scales from $\sim$kpc to 
$\sim$Gpc, despite their necessarily simplified treatment of small 
scale `subgrid' physics (e.g. star formation, AGN feedback, cooling).
It is now standard for them to reproduce a wide range of 
observable properties of galaxy populations (e.g. \citealt{Furlong15}; 
\citealt{Tayford15}; \citealt{Crain17}; \citealt{Vogelsberger14a}; 
\citealt{Sparre15}; \citealt{Snyder15}; \citealt{Genel18}; 
\citealt{Nelson18}; \citealt{Pillepich18a}) and even allow for 
meaningful statistical comparisons with individual objects
(e.g. \citealt{Taylor16}; \citealt{Zhu18}; \citealt{Pawlowski20}).

In this work we investigate the intrinsic physical mechanisms 
responsible for quiescence in massive central galaxies as observed 
at redshift $z=0$. In order to do that, we extract testable predictions 
about the observable signatures of quenching from three of the most 
successful cosmological hydrodynamical simulations run to date: 
EAGLE \citep{Schaye15, Crain15}, 
Illustris \citep{Vogelsberger14a, Vogelsberger14b, Genel14, Sijacki15} 
and IllustrisTNG \citep{Marinacci18, Naiman18, Nelson18, Pillepich18a, Springel18}. 
Since all three suites primarily 
use supermassive black holes to suppress star formation in their 
massive centrals, these predictions inform us about the observational 
consequences of different implementations of AGN feedback quenching. 
We then rigorously test the predicted trends against local galaxies 
observed with the SDSS, to better interpret the observations with the 
aid of theoretical models. Due to the complex nature of quenching 
we choose to implement a~machine learning algorithm, in order to 
explore the non-linear relationships among multiple variables 
simultaneously. Having established that black hole mass (\mbh) is 
the most important quenching parameter in both simulations and
observations, we then explore the gas properties of observed and 
simulated galaxies as a~function of the critical \mbh\ parameter. 
By using empirical calibrations to estimate both black hole and 
molecular gas masses in the SDSS galaxies we perform our analysis on 
a~large sample of $\sim~200\, 000$ galaxies, exceeding the sample 
size of direct measurements by several orders of magnitude.
			 
This article is structured as follows: in Sec.~\ref{sec:DATA} we 
present our data along with sample selection criteria. 
Sec.~\ref{sec:METHOD} describes our use of empirical calibrations 
to estimate the in-fibre molecular gas masses and black hole masses 
in the SDSS. It also provides a detailed description of the random
forest classifier method. In Sec.~\ref{sec:RESULTS} we present our 
results, followed by a discussion in Sec.~\ref{sec:discussion} and 
a brief summary in Sec.~\ref{sec:SUMMARY}. 
In order to ensure reproducibility all of our analysis is
available at 
\url{https://hub.docker.com/u/jpiotrowska}.

\section{DATA}
\label{sec:DATA}

In this study, we conduct a~consistent analysis of the connection 
between galactic parameters and quenching across observations 
and simulations. In the case of the observed Universe, we utilize 
the Sloan Digital Sky Survey (SDSS) photometric and spectroscopic 
data products derived by multiple groups over the years. In the 
simulated rendition, we explore three independent universe realisations 
as obtained within the EAGLE, Illustris and IllustrisTNG suites. 
 
\subsection{SDSS}
We choose to analyse the SDSS DR7 \citep{Abazajian09} in this study 
because it is the largest sample of local ($z \sim 0$) galaxies 
observed in both photometry and spectroscopy. Consequently, the 
survey gives us an opportunity to statistically explore a host of 
physical properties in relation to star formation and quenching. More 
specifically, we analyse a sample of 230 636 central galaxies, which 
meet our sample selection criteria described in 
Sec.~\ref{sec:sample-selection}. 

\subsubsection*{Morphological parameters}

We first extract morphological parameters from the \cite{Simard11} 
catalogue of bulge+disk photometric decompositions (including galaxy 
S\'ersic index, galaxy ellipticity and bulge semi-major effective 
radius from Tables~3~\&~1 in \cite{Simard11}). We then add 
information about stellar mass estimates for the bulge and disk 
components from the \cite{Mendel14} catalogue, matching entries 
between the catalogues on the \objid\ identifier. In particular, we 
extract the S\'ersic mass and maximum observable redshift for a~given 
object ($z_{\rm max}$) from Table 3 in \cite{Mendel14} as well as 
the bulge mass, disk mass and $\Delta_{\rm B+D}$ (difference between
the total mass and the sum of bulge and disk masses in units of 
standard error) parameters from Table 4 in the same publication.
The $z_{\rm max}$ parameter is required for calculating volume 
weights $V_{\rm max}$ described in Sec.~\ref{sec:qf}, while bulge 
mass (\mbulge) measurement is necessary to estimate \mbh\ in two 
different calibrations listed in Sec.~\ref{sec:bh-mass-estimate}. 

In order to match the morphological information with other published 
catalogues we retrieve \texttt{SpecObjID}, \objid, \texttt{redshift}, 
\texttt{ra} and \texttt{dec} entries for all 793 272 spectroscopically 
observed galaxies (identified with \texttt{objType=0}) from the 
SpecObj table in the Catalog Archive Server Jobs 
System\footnote{\url{https://skyserver.sdss.org/casjobs/}} 
(\texttt{CasJobs}) online workbench. We then join these entries on 
\objid\ with our combined morphological table, obtaining a total of 
651 567 objects.

\subsubsection*{Stellar masses, SFRs and emission line fluxes}

We extract stellar mass and star formation rate estimates for SDSS DR7
from the MPA-JHU release of spectrum 
measurements\footnote{\url{https://wwwmpa.mpa-garching.mpg.de/SDSS/DR7/}}.
Because each SDSS optical fibre has an aperture of 3" in diameter, 
the spectroscopic measurement is usually limited to only the central 
region of a galaxy. Hence, for each object in the sample the MPA-JHU
release provides an estimate of both the in-fibre quantities 
(labelled with subscript `fib' throughout this paper) and their 
total values for the whole galaxy.

Both the total and in-fibre 
stellar masses were estimated from spectral energy distribution (SED) 
fitting to \textit{ugriz} photometry, following the philosophy of 
\cite{Kauffmann03} and \cite{Salim07}. As described in detail in 
\cite{Brinchmann04}, the in-fibre SFRs for galaxies classified as 
star-forming were estimated from dust-corrected emission line fluxes 
(dominated by \halpha\ emission). In AGN-dominated and low 
signal-to-noise ratio (S/N) spectra the in-fibre SFRs were estimated 
from the strength of the $4000$\AA\ break (D4000). \cite{Salim07}
then estimated total galaxy SFRs by using a relationship between the 
in-fibre SFR values and the in-fibre colours to add SFR outside of 
the fibre based on galaxy colours from the SDSS photometry.

We also use the MPA-JHU data release to extract fluxes for the 
\halpha, \hbeta, $[{\rm N II}] 6584$ and $[{\rm O III}] 5007$ 
emission lines along with their true uncertainty estimates derived 
from duplicate observations. In this release, the fluxes were 
calculated following continuum subtraction with the unpublished 
Bruzual \& Charlot (2008) stellar population synthesis spectra.

We then match the MPA-JHU table on ra, dec and redshift with the 
morphological data, obtaining a~total of 582 804 galaxies.

\subsubsection*{Halo masses and central classification}

We make use of publicly available group 
catalogues\footnote{\url{https://gax.sjtu.edu.cn/data/Group.html}} 
constructed with the identification technique outlined in \cite{Yang07},
to extract halo masses and galaxy classification into centrals and 
satellites. More precisely, we focus on the \textit{modelC} catalogue 
with halo masses extended below $10^{12} {\rm M_\odot}$ using an 
empirical fit to the $M_\ast - M_{\rm halo}$ relation in Eq.~(20) of 
\cite{Yang09}. In the abundance matching technique X. Yang and 
collaborators assumed WMAP5 cosmology and used model absolute 
magnitudes in calculations of the galactic stellar masses.The authors 
kindly provided us with a complete data set in a private email 
exchange, giving permission to include it as part of
our analysis script available online. 

After matching the group catalogue against our already compiled 
data on ra and dec, we arrive at a~total of 512 675 galaxies, 
which comprise our parent sample.

\subsubsection*{Velocity dispersions}

In order to estimate supermassive black hole masses ($M_{\rm BH}$) 
we require the knowledge of stellar velocity dispersions (\vdisp) in 
our objects. To this end we extract \vdisp\ measurements 
from the NYU Value-Added Galaxy Catalog\footnote{\url{http://sdss.physics.nyu.edu/vagc/}} 
published by \cite{Blanton05} along with the median signal-to-noise 
ratios (S/N) in the spectra to impose quality cuts before estimating 
$M_{\rm BH}$. Since all of the previously matched galaxies in the 
group catalogue have a NYUVAGC release identifier, there is no 
matching step required.

\subsection{Cosmological simulations}
\label{sec:cosmosims}

In this study we compare the observable consequences of quenching 
expected from three cosmological simulation suites: 
EAGLE, Illustris and IllustrisTNG (hereafter TNG)
with our local Universe as seen through the SDSS. 
To this end, we 
choose the same simulation volume of $\sim{\rm(100\ cMpc)^{3}}$ and runs 
which include full physics treatment at the highest resolution 
available for this box size. We choose to focus 
on cosmological simulations only, in order to explore theoretical 
predictions arising from the most complex treatment of physics in 
large statistical samples of galaxies. In this context, a~theoretical 
prediction constitutes any outcome of the simulation which was
not calibrated for against the observable Universe. Among all 
cosmological simulations completed to date we choose to focus on 
EAGLE, Illustris and TNG data because of their public 
availability, thorough documentation and support provided by each of 
the collaborations. In the remainder of this section we briefly describe each of
the simulation suites and discuss their implementation of 
subgrid physics most relevant to our study - the AGN feedback 
model.

One challenge common to all cosmological simulations 
is their inability to directly follow the evolution of black holes 
and their accretion disks due to resolution limits.
For this reason, EAGLE, Illustris and TNG implement black hole 
particles - collisionless sink particles which contain subgrid black 
holes and are allowed to accrete gas 
from their surroundings. The mass of a subgrid black hole 
$(M_{\rm BH})$ usually differs from that of the whole particle 
$(M_{\rm dyn})$ and these two are applied in different 
calculations throughout the simulation. All black hole-specific 
processes make use of $M_{\rm BH}$, while gravitational 
interactions between the particle and the rest of the simulation
involve $M_{\rm dyn}$ instead.

Much like the evolution of black holes, their potential
origin from processes like e.g. the collapse of metal-free
massive stars cannot be traced directly either. 
Hence, all three simulations `seed' black hole particles
by placing them in unoccupied haloes above a chosen
mass threshold on-the-fly within the runs. Seeding 
parameters like the mass of a black hole seed or halo mass 
threshold differ among the subgrid models, hence EAGLE,
Illustris and TNG have different lower limits on \mbh\
present in simulated galaxies.

\subsubsection*{EAGLE}
The EAGLE\footnote{\url{http://icc.dur.ac.uk/Eagle/}} 
(Evolution and Assembly of GaLaxies and 
their Environments) project is a set of cosmological hydrodynamical 
simulations performed with the \textsc{GADGET-3} tree-SPH 
(smoothed particle hydrodynamics) code \citep{Springel05}.
The simulations assume a~$\Lambda{\rm CDM}$ universe
with \mbox{$\Omega_m=0.307$,} $\Omega_\Lambda=0.693$,
\mbox{$\Omega_b=0.04825$}, $\sigma_8 =0.8288$, $n_s=0.9611$
and \mbox{$h=0.6777$} as estimated by 
\cite{Planck14}.
An interested reader can find all subgrid model description
and calibration for EAGLE in \cite{Schaye15} and \cite{Crain15},
while for the details of subhalo catalogue compilation we refer them
to \cite{McAlpine16}. 

Black holes in EAGLE are seeded with
$M_{\rm BH} = 10^5\, {\rm M_\odot}h^{-1}$ in all unoccupied 
haloes once these reach $M_{\rm halo} >10^{10}\, {\rm M_\odot}h^{-1}$.
Once seeded, they then grow through Bondi–Hoyle-Lyttleton accretion
\citep{Hoyle39, Bondi44, Bondi52} extended to account for 
the angular momentum of
gas around a black hole according to the prescription 
by \cite{Rosas15}. Accretion rates in the suite are not allowed
to reach arbitrarily large values and are constrained by 
the Eddington limit.

EAGLE simulations implement a single form of AGN feedback
\citep{Booth09}, which best corresponds to powerful quasar winds 
launched in consequence of cold mode accretion. In this 
prescription, each black hole carries a feedback energy `reservoir'
$E_{\rm BH}$ which after each time step $\Delta t$ is increased by
$\epsilon_{\rm f} \epsilon_{\rm r} \dot{M}_{\rm BH} c^2\Delta t$,
where $\epsilon_{\rm r} =0.1$ is the radiative efficiency of an
accretion disk \citep{Shakura73},  
$\dot{M}_{\rm BH}$ is the accretion rate within it, 
$\epsilon_{\rm f}=0.15$ is the fraction of radiated
energy which couples into the ISM and $c$ is the speed of light.
Once $E_{\rm BH}$ is large enough to increase the temperature 
of at least one gas particle adjacent to the black hole by 
$\Delta T_{\rm AGN}=10^{8.5}\ {\rm K}$, the black hole particle
stochastically heats each of its neighbours by increasing their
temperature by $\Delta T_{\rm AGN}$. In this way, the simulation 
captures the unresolved process in which a hot accretion disk 
emits significant amounts of  radiation, a small fraction of which 
couples thermally into its immediate surroundings.

From the EAGLE RefL0100N1504 run we extract halo, stellar and 
black hole masses as well as the black hole accretion and star 
formation rates for non-spurious ($M_\ast > 0$) subhalos 
in the redshift $z=0$ snapshot, 
obtaining 325~496 galaxies as the parent sample in this suite. 

\subsubsection*{Illustris}
The Illustris\footnote{\url{https://www.illustris-project.org/}} 
cosmological simulations were performed with a~moving-mesh 
code \textsc{AREPO} \citep{Springel10} assuming a~$\Lambda {\rm CDM}$
universe with cosmological parameters as estimated by WMAP~9
\cite{Hinshaw13}: $\Omega_m=0.2726$, $\Omega_\Lambda=0.7274$,
$\Omega_b=0.0456$, $\sigma_8 =0.809$, $n_s=0.963$ 
and $h=0.704$. The project is introduced in 
\cite{Genel14} and \cite{Vogelsberger14a} with an overview of galactic 
populations in \cite{Vogelsberger14b} and details of 
black hole evolution and treatment in \cite{Sijacki07} and 
\cite{Sijacki15}.

Illustris simulations place black hole particles in haloes 
for which $M_{\rm halo} > 5 \times 10^{10}\, {\rm M_\odot} h^{-1}$
with a seed black hole mass of $M_{\rm BH} = 10^5\, {\rm M_\odot} h^{-1}$.
Black holes then grow smoothly through accretion according to the 
Bondi-Hoyle-Lyttleton formula with an added boost factor, which
accounts for the influence of unresolved ISM structure. Similarly to
EAGLE, \mdot\ is bound by an upper limit equal to the 
Eddington rate ($\dot{M}_{\rm Edd}$). Black hole accretion rate is 
directly linked to the AGN feedback model, which consists of three separate 
prescriptions referred to as `quasar', `radio' and `radiative' modes 
\citep{Sijacki15}.

The `quasar' mode operates at high accretion rates for which
\mbox{$\chi = \flatfrac{\dot{M}_{\rm BH}}{\dot{M}_{\rm BH}} > 0.05$}.
In this prescription, a fraction \mbox{$\epsilon_{\rm f}=0.05$} of the bolometric
luminosity of the disk ($L_{\rm disk}$) is thermally coupled to 
the surrounding gas in an isotropic fashion,
such that the rate of thermal energy injection is given by:
\begin{equation}
\dot{E}_{\rm inj} = \epsilon_{\rm f} L_{\rm disk} = \epsilon_{\rm f} \epsilon_{\rm r} \dot{M}_{\rm BH} c^2\, ,
\end{equation}
where $\epsilon_{\rm r}=0.1$ is the radiative 
efficiency of a thin disk \citep{Shakura73}. 
This way, the simulation effectively models 
energy-driven outflows caused by the AGN, provided that radiative 
losses are negligible.

At low accretion rates for which $\chi < 0.05$, AGN 
feedback in Illustris switches into the `radio' mode. In this model,
once a black hole increases its mass by  
\mbox{$\delta_{\rm BH}=\delta M_{\rm BH}/M_{\rm BH} = 1.15$},
an AGN-driven bubble is created within the circumgalactic medium (CGM) 
and placed at random within twice the bubble radius $R_{\rm bubble}$ 
away from the galactic centre. Each bubble is assigned an energy 
$E_{\rm bubble}$ linked to $\delta M_{\rm BH}$ via
\begin{equation}
E_{\rm bubble} = \epsilon_{\rm m} \epsilon_{\rm r} c^2 \delta M_{\rm BH}\, ,
\end{equation}
where $\epsilon_{\rm m}=0.35$ is the efficiency of mechanical 
heating by the bubbles. Bubble radius is then calculated from 
solutions for the radio cocoon expansion in a spherically 
symmetric case \citep{Heinz98}, where 
\mbox{$R_{\rm bubble} \propto \qty(\flatfrac{E_{\rm bubble}}{\rho_{\rm ICM}})^{1/5}$}
and $\rho_{\rm CGM}$ is the density of the CGM. 

In this fashion, Illustris models the influence of AGN jets 
which are not resolved in the simulation and are thought to inflate
hot bubbles in the CGM around massive galaxies \cite[e.g.][]{McNamara00, Fabian12}. 
The implemented
model yields larger bubbles for more powerful jets and accounts for 
the influence of ambient density on the bubble size. This feedback 
process is, in principle, also self-regulatory. Inflating a bubble increases 
the temperature of the CGM to offset cooling and temporarily cut-off 
the precipitation of new gas for future accretion and star formation. 
The black hole then slows down its growth and the next
bubble injection is pushed further in time, preventing 
the black holes from injecting too many bubbles into the CGM.
Although this implementation of radio mode AGN feedback proved
very successful at suppressing star formation in massive galaxies,
it has done so at the cost of excessive gas removal from galactic
haloes. In consequence, the gas fractions of groups of galaxies 
and clusters in Illustris are significantly lower than in the observed
Universe, while their central galaxies grow too large in mass
\citep{Genel14}.

The final mode of AGN feedback - the radiative one
- operates at all accretion rates and acts to modify the net cooling
rate of gas in the presence of an ionising radiation field 
associated with nearby black holes. This mode has the least 
influence on the ISM out of all three and is at its most effective
in the `quasar' mode at accretion rates close to the 
Eddington limit.

In our study we focus on the \mbox{Illustris-1} 
\citep{Nelson15} run, extracting black hole, stellar and halo masses
as well as star formation and black hole accretion rates for
subhalos with non-zero $M_\ast$ from the redshift $z=0$ group 
catalogues, which yields a~parent sample size of 157~241 objects.
Additionally, we supplement galaxies with 
the following entries in the (HI+\hmol) 
content catalogue: radial profiles in the subhalo stellar mass, \hmol\ mass and star
formation rate as well as the total molecular gas mass in a~given 
object. In order to obtain the HI and \hmol\ abundances 
in the simulations, \cite{Diemer18} use HI/\hmol\ transition models 
- numerical prescriptions for calculating molecular hydrogen 
fraction in a given gas cell based on locally averaged properties 
such as gas state variables, its metallicity and the estimate of 
the local UV background. Because the published catalogues use 
four different HI/\hmol\ transition models to obtain the \hmol\ 
masses, we choose to present our results with only one of them
- the \cite{Gnedin14} model - in the main text. We then show   
in Appendix~\ref{sec:appendix:h2-models} how our
results are consistent across all models provided in the catalogue.

\subsubsection*{IllustrisTNG}
The IllustrisTNG\footnote{\url{https://www.tng-project.org/}} 
(The Next Generation) cosmological simulations were 
run with an updated version of \textsc{AREPO} extended to 
solve the equations of ideal magnetohydrodynamics. 
The suite also differs from Illustris in its treatment of subgrid 
physics, among which the changes in AGN feedback prescription 
are most relevant for our study. IllustrisTNG (hereafter TNG) is presented 
in a series of five simultaneous papers: \cite{Marinacci18}, 
\cite{Naiman18}, \cite{Nelson18}, \cite{Springel18} and \cite{Pillepich18b}.
Details of the galaxy formation model are described in \cite{Pillepich18a},
while the prescription for black hole feedback is introduced in
\cite{Weinberger17}. 

Black holes in TNG are seeded with 
$M_{\rm BH} = 8 \times 10^5\, {\rm M_\odot}h^{-1}$ in all unoccupied 
haloes once these reach $5 \times 10^{10}\, {\rm M_\odot}h^{-1}$. 
Black holes then grow in mass through pure Bondi-Hoyle-Lyttleton accretion, 
with an upper limit set by the Eddington rate. Similarly to Illustris,   
the AGN feedback affects the galaxy in three different modes.
A high-accretion state corresponds to a `quasar'-like mode, 
while the low-accretion, `kinetic' mode aims to capture
the currently unobservable kinetic winds launched from the
AGN at low accretion rates.
At both accretion states gas cells in the vicinity of a black hole particle also
experience different cooling rates due to the presence of a
radiation field from the AGN. The threshold below which a black hole
is accreting in a low \textit{state scales with black hole mass}:
\begin{equation}
\chi = {\rm min} \qty[0.002 \qty(\frac{M_{\rm BH}}{10^8 {\rm M_\odot}})^2 , \ 0.1  ] \, ,
\end{equation}
and hence the switch between the two main AGN feedback modes 
occurs around 
\mbox{$\log(M_{\rm BH}/{\rm M_\odot})\sim8$} \citep{Weinberger17}. 

The high-accretion mode in TNG follows the `quasar' mode 
prescription in Illustris with the product of efficiencies increased to
$\epsilon_{\rm f} \epsilon_{\rm r} = 0.02$ from 0.005. In contrast,
the Illustris low-accretion mode is replaced by a new, kinetic mode
feedback, which is no longer implemented at a distance from the AGN.
Instead, at low accretion rates the AGN in TNG interact with gas cells
in the same local neighbourhood within which the thermal, quasar 
feedback is injected. At each time 
step, black holes in their low-accretion state accumulate kinetic 
feedback energy at a rate proportional to the mass accretion rate 
of the gas:
\begin{equation}
\dot{E}_{\rm kin} = \epsilon_{\rm kin} \dot{M}_{\rm BH} c^2\, ,
\end{equation}
where
\begin{equation}
\epsilon_{\rm kin}= {\rm min}\qty[\frac{\rho}{0.05 \rho_{\rm SF\, thresh}},\, 0.2]\, ,
\end{equation} 
$\rho$ is the gas density around the black hole particle and 
$ \rho_{\rm SF\, thresh}$ is the threshold
density for star formation. Once the available feedback energy 
reaches a threshold for its release (determined by the dark matter 
velocity dispersion and gas mass within the 
feedback region), the AGN injects this feedback energy in a form 
of a momentum kick in a randomly chosen direction.
This solution 
avoids potential numerical artefacts associated with more complex 
momentum injection patterns at the given resolution in the simulations.
Although the random choice of direction does not strictly conserve 
momentum at a given time step, when averaged over time the total 
momentum is conserved, while yielding the desired energy injection 
into the ISM. The deposited feedback energy is then carried by the 
gas further away from the AGN, increasing the entropy of the immediate
ISM as well as the CGM once it percolates outside of the galactic host.

In this work we make use of the TNG100-1 run \citep{Nelson19}
and extract the same set of parameters as we did in Illustris for 
101~798 subhalos with non-zero $M_\ast$ from the redshift $z=0$ group 
catalogues. These constitute our parent TNG sample.

\subsection{Sample selection}
\label{sec:sample-selection}

Across all simulations and observations we select galaxies with 
$10^9\, {\rm M_\odot} < M_\ast < 10^{12}\, {\rm M_\odot}$ 
and $M_{\rm halo} > 10^{11} {\rm M_\odot}$. In our main analysis we focus on central 
galaxies only, since we expect their transition to quiescence not to depend on surrounding
environment \mbox{\citep[e.g.][]{Peng10, Peng12, Bluck16, Bluck20b}}. This allows us to investigate the
intrinsic quenching mechanisms, among which AGN (black hole) feedback is one possible
candidate. This criterion yields a~sample of 7~116 objects in EAGLE, 14~133 in Illustris,
11~623 in TNG and 389~371 galaxies in the SDSS. 
The stark contrast between the sample size in simulations 
and observations is a consequence of a simulation box size of 
$\sim ({\rm 100\ cMpc})^3$, which is smaller than the volume of the 
Universe probed by the SDSS. The differences between simulations 
themselves are due to the differences in galaxy stellar mass functions
produced by each run (e.g. see \citealt{Lim17} for a comparison between 
EAGLE and Illustris and \citealt{Pillepich18b} between Illustris and TNG).

\setlength{\abovedisplayskip}{5pt}
\setlength{\belowdisplayskip}{5pt}

In order to estimate black hole masses in the SDSS, we require a~clean measurement of central
velocity dispersion, unaffected by the contribution from galactic rotation in edge-on objects.
Because this problem affects disk-dominated galaxies, we apply an inclination cut solely to
objects with bulge-to-total mass ratio $(B/T) \leq 0.5$, removing those with axial ratio $b/a < 0.9$. 
The axial ratio is calculated from ellipticity via $b/a = 1 - e$ and the $(B/T)$ parameter from 
$(B/T) = M_{\rm B}/(M_{\rm B}+M_{\rm D})$, where $M_{\rm B}$ and $M_{\rm D}$ are the
bulge and disk masses respectively. At this stage all galaxies with $\Delta_{\rm B+D}$ 
parameter greater than unity are also discarded from the sample because their estimates of bulge
and disk masses are unreliable. Following the cut on galaxy inclination we also calculate 
a~correction factor, which enters our quenched fraction and correlation analysis in the form of 
weight $w_{\rm inc}$:
\begin{equation}
w_{\rm inc} = \begin{cases}
					\frac{1}{N}\sum_{i=1}^{N}\frac{n_{{\rm tot}, i}}{n_{{\rm cut}, i}}  &  \text{for}\ (B/T) \leq 0.5 \\
					1 & \text{for}\ (B/T) > 0.5
					\end{cases} ,
\label{eq:inc}
\end{equation}
where $N=10$ is the number of $(B/T)$ bins we split our data into between $(B/T)=0$ 
and $(B/T) \leq 0.5$, $n_{{\rm tot}, i}$ is the total number of objects in a~given $(B/T)$ bin $i$
and $n_{{\rm cut}, i}$ is the number of objects left in the $(B/T)$ bin following the $b/a$ cut. 

Effectively, there are two weights assigned within our SDSS sample -- 1.00 or 10.33, 
depending on whether a galaxy is disk- or bulge-dominated. By applying an inclination
cut we trade a decrease in sample completeness for an increase in \mbh\ estimation accuracy.
We find that this choice does not impact our results, demonstrating this explicitly in our
Random Forest analysis in Appendix~\ref{sec:discussion:rf-corr}. We further impose stellar velocity 
dispersion quality cuts, requiring median S/N in the spectrum of 3.5 and 
$\sigma_\ast < 450\, {\rm kms^{-1}}$. We test the impact of our selection criteria by 
increasing the S/N threshold and finding the results stable against more conservative cuts. 
This selection yields a~total of 230~636 central galaxies in the SDSS spanning a~redshift 
range of $0.02 < z < 0.2$.

In the final part of our analysis we estimate \hmol\ gas masses in 
the SDSS galaxies, using dust extinction of optical spectra as proxy 
for the line-of-sight molecular gas content, as described in 
Sec~\ref{sec:gas-mass-estimate}. For this purpose we require a~good 
quality measurement of the \halpha\ and \hbeta\ Balmer lines,
imposing a~S/N ratio cut of 6 and 2 on each of the emission line 
fluxes respectively. We further remove all AGN candidates as 
identified by the NII-BPT \citep{BPT81} diagram, since their emission
line ratios dominated by the AGN cannot be used to estimate \mgas\ 
from dust extinction. To this end we  require a~S/N cut of 3 on the 
two remaining emission lines - $[{\rm N II}] 6584$ and 
$[{\rm O III}] 5007$. The final sample with molecular gas mass 
estimates comprises 35 964 objects in total.

Throughout this work we classify the observed and simulated galaxies into two 
categories: `star forming' and `passive' (quiescent), using a simple criterion applied
to their sSFR (${\rm sSFR=SFR/}M_\ast$). All figures and results presented in this 
article classify objects with log(sSFR/${\rm yr^{-1}}$)  $< -11$ as passive, however we explore
a range of log(sSFR/${\rm yr^{-1}}$)  between $-10.5$ and $-11.5$ for the selection criterion
and find that our conclusions are robust against the choice of sSFR value within
that range.

\section{Method}
\label{sec:METHOD}

In this section we describe our methods used to infer quantities not directly observable with the 
SDSS fibre spectroscopy - black hole and molecular gas masses. We also provide a~brief overview
of the Random Forest classification technique which enables us to determine the relative importance
of different galactic parameters for predicting the star formation state of local central galaxies.

\begin{figure}
\includegraphics[width=\columnwidth]{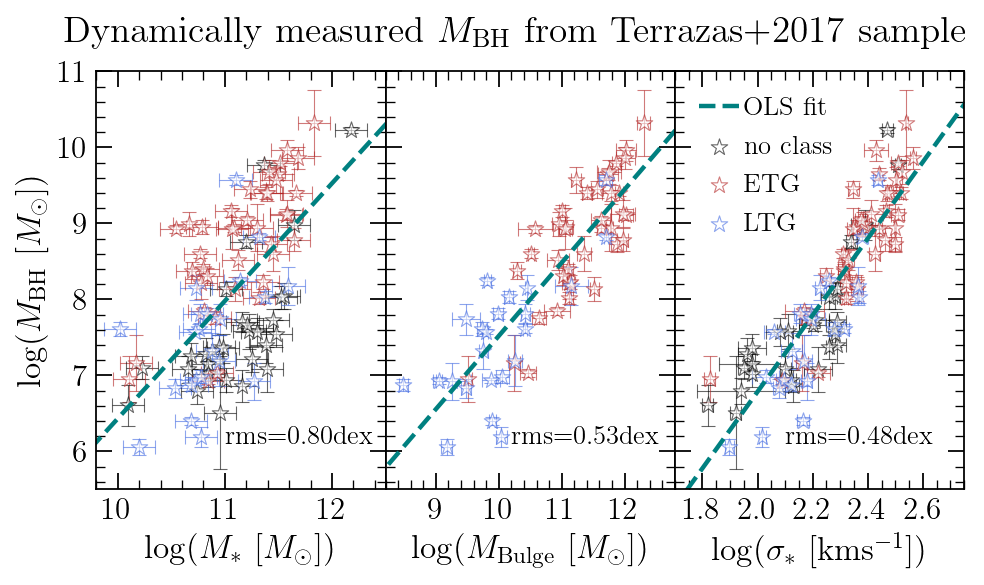} 
\caption{Comparison of scatter in the \mstar--\mbh, \mbh--\mbulge\ and 
		 \mbox{\mbh--\vdisp} relation in a sample of dynamically measured 
		 black hole masses taken from \citet{Terrazas17}. 
		 It is clear that the relationship between \mbh\ and \vdisp\ is the strongest of 
		 the three, supporting the choice of \vdisp\ to estimate black hole masses in our 
		 SDSS sample. Please note that the sample of galaxies without ETG/LTG
		 (early-type/late-type galaxy)
		 classification do not have an associated bulge mass measurement.}
\label{fig:mbh-sigma-mstar}
\end{figure}

\subsection{$M_{\rm BH}$ estimation in the SDSS}
\label{sec:bh-mass-estimate}

\setlength{\belowdisplayskip}{4pt}
\setlength{\abovedisplayskip}{4pt}

A direct measurement of a~supermassive black hole mass requires high resolution 
spectroscopic observations in order to accurately model the orbital dynamics of gas 
and stars within its gravitational sphere of influence. Given the complexity of such 
measurements and objects selection criteria, the current literature can only provide 
around a~hundred dynamically measured \mbh\ in massive central galaxies observed 
in the local Universe \citep[e.g.][]{Terrazas16, Terrazas17}. 
Despite their small sample sizes, these studies deliver important inferences about
quenching processes, e.g. indicating a connection between quiescence and high black
hole mass. When comparing observations with theoretical models, however, 
large samples are often superior since they allow one to explore the behaviour of 
statistically average galaxy populations. 

Fortunately for this work, the direct measurement studies uncovered an intimate 
connection between central black holes and their host galaxies 
\citep[see][for a~review]{KormendyHo13}, finding strong correlations 
between \mbh\ and \mstar \citep{Reines15}, $M_{\rm bulge}$ \citep{HaringRix04} 
or \vdisp\ \citep{Hopkins07, MM13, Saglia16}. In Fig.~\ref{fig:mbh-sigma-mstar}
we show a~comparison among these three scaling relations for central galaxies in 
the \citet{Terrazas17} sample, which indicates that the connection between
\mbh\ and \mstar\ is the weakest of them all. An~ordinary least squares (OLS) 
fit to logarithmic values 
shows a~RMS scatter of 0.80~dex for the \mbh--\mstar\ relation, while in the 
\mbh--$M_{\rm bulge}$ and \mbh--\vdisp\ relations the scatter is significantly 
smaller, reaching 0.53~dex and and 0.48~dex respectively.

In lieu of dynamical measurements of black hole masses in our sample, 
we test a variety of calibrations published in the literature to estimate \mbh\ in the SDSS galaxies.
We focus primarily on \vdisp, due to its tightest correlation with the black hole mass,
illustrated in Fig.~\ref{fig:mbh-sigma-mstar}. However, we also use a prescription for estimating
\mbh\ from $M_{\rm bulge}$ in Eq.~\ref{eq:HR04}, which allows us to explore a larger
number of objects since this parameter does not require a cut on galaxy inclination
or a requirement on high S/N of the galactic spectra . 
We also include separate calibrations for different galaxy morphologies: 
early-type (ETG) / late-type (LTG) galaxies in Eq.~\ref{eq:MM13} and pseudo-/classical bulges 
in Eq.~\ref{eq:S16-ps}. These distinct morphologies are associated with different modes of black 
hole growth \citep[e.g.][]{Kormendy11, KormendyHo13} and have been found to exhibit
different scaling relations between galactic properties and \mbh\ 
\citep[e.g.][]{Mathur12, MM13, Saglia16}. 

In our study we label galaxies with $(B/T) > 0.5$ as ETG and 
$(B/T) \leq 0.5$ as LTG. This criterion was not used directly by 
\cite{MM13}, who classified galaxy morphology through visual 
inspection, however \cite{Simard11} show that a selection in 
bulge-to-total light ratio of above 0.7 is a good proxy for visual
classification of ETGs. This bulge-to-total light ratio corresponds to 
$(B/T) > 0.5$ in mass, as demonstrated by \cite{Bluck19}. 
Additionally, we also classify bulges
as classical when the galaxy's S\'ersic index is greater than 2 and 
as a pseudobulge otherwise. All parametrisations we use in our 
study are listed in Eq.~\ref{eq:S16} through \ref{eq:S16-ps} below, 
along with the intrinsic scatter associated with
each relation, denoted as~$\epsilon$:
\begin{equation}
	\begin{split}
		\log{M_{\rm BH}} = 5.246 \times \log{\sigma_{\rm c}} -&3.77  \\ 
		&\text{\citep{Saglia16}},\ \epsilon=0.417
	\end{split}
\label{eq:S16}
\end{equation}
\begin{equation}
	\begin{split}
		\log{M_{\rm BH}} = 0.54\times\log{(M_{\rm bulge}/10^{11})} + 2.18\times\log{(\sigma_{\rm c}/200)}& + 8.24  \\
		\text{ \citep{Hopkins07}},\  \epsilon=0.22&
	\end{split}
	\label{eq:H07}
\end{equation}
\begin{equation}
	\begin{split}
		\log{M_{\rm BH}} =  1.12 \times \log{(M_{\rm bulge}/10^{11})} + 8.20 \qquad \quad& \\
		\text{ \citep{HaringRix04}},\ &\epsilon= 0.30
	\end{split}
	\label{eq:HR04}
\end{equation}
\begin{equation}
\begin{split}
\log{M_{\rm BH}} = \begin{cases}
					5.20 \times \log{(\sigma_{\rm c}/200)} + 8.39  \quad \text{ETG} \quad \epsilon=0.34 \\
					5.06 \times \log{(\sigma_{\rm c}/200)} + 8.07 \quad \text{LTG} \quad \epsilon=0.46 
					\end{cases} \\
					\text{\citep{MM13}}
\end{split}
\label{eq:MM13}
\end{equation}
\begin{equation}
\begin{split}
\log{M_{\rm BH}} = 
	\begin{cases}
					4.546 \times \log{(\sigma_{\rm c}/200)} - 2.030 \\
					\qquad \qquad \qquad \qquad \text{classical bulge},\ \epsilon=0.348 \\
					2.129 \times \log{(\sigma_{\rm c}/200)} + 2.526 \\
					\qquad \qquad \qquad \qquad \text{pseudo bulge},\ \epsilon=0.455 
					\end{cases} \\
					\text{\citep{Saglia16}}
\end{split}
\label{eq:S16-ps}
\end{equation}
where \mbulge\ is given in units of ${\rm M_\odot}$ and 
$\sigma_c$ denotes the central velocity dispersion in units of ${\rm kms^{-1}}$.
The central velocity dispersion \citep{Jorgensen95} is calculated from \vdisp\ via:
\begin{equation}
\sigma_c = \sigma_\ast \qty(\frac{R_{\rm bu}/8}{R_{\rm fib}})^{-0.04},
\end{equation}
where $R_{\rm bu}$ and $R_{\rm fib}$ denote the bulge and the SDSS 
fibre radii respectively.

In this work we make use of all calibrations in the SDSS, 
explicitly showing how our results in Sec.~\ref{sec:RESULTS} are 
robust against the exact parametrisation of the observed galaxy 
scaling relations. 

\subsection{Gas mass estimation in the SDSS}
\label{sec:gas-mass-estimate}

We follow the~molecular gas mas estimation method outlined in 
\cite{Piotrowska20}, using dust reddening of the optical spectra to 
infer the mass of molecular hydrogen (\mgas) in a~given SDSS
fibre. More specifically, we use an empirical relation between 
hydrogen number density and colour excess ($E(B-V)$) observed
for different lines of sight within the Milky Way \citep{Gudennavar12},
to estimate the amount of gas associated with the optical extinction
observed for each galaxy in our sample. Under the assumption of 
a~uniform foreground dust screen and a~linear metallicity dependence 
in the gas-to-dust ratio the average gas surface density within the 
fibre, $\Sigma_{\rm gas}$, is given by:
\setlength{\abovedisplayskip}{5pt}
\setlength{\belowdisplayskip}{5pt}
\begin{equation}
\Sigma_{\rm gas} = \qty[(31.6 \pm 1.0)A_V + 1.0]
\qty(\flatfrac{Z_\odot}{Z})\ 
{\rm M_\odot {pc^{-2}}}\, ,
\label{eq:sigmaH}	
\end{equation}
where $Z$ is the gas phase metallicity, $Z_\odot$ is its solar value
and \mbox{$A_V = E(B-V)/R_V$} is the dust attenuation in the optical
\mbox{$V$-band} with \mbox{$R_V \sim 3.1$} for the Milky Way. 
In this work we use the O3N2 calibration extended to non purely 
star-forming regions by \cite{Kumari19} to calculate gas phase
metallicity for all galaxies which meet our emission line quality
selection criteria.

In order to estimate
$A_V$ we assume a Galactic extinction curve parametrised by 
\cite{Cardelli89} and compare the observed ratio of \halpha\
and \hbeta\ emission line fluxes 
($\flatfrac{F_{\rm H\alpha}}{F_{\rm H\beta}}$) to its theoretical
intrinsic value of 2.86 as determined by Case B calculations in 
\cite{Hummer87}. More specifically, we calculate $A_V$ using the 
following formula: 
\begin{equation}
A_\lambda = \frac{K_{\lambda}}{K_{\rm H\beta} - K_{\rm H\alpha}} 
2.5 \log_{10}\Bigg(\frac{{F_{\rm H\alpha}}/{F_{\rm H\beta}}}{2.86}\Bigg)\, ,
\label{eq:AV}
\end{equation}
where the $K_{\lambda}=A_{\lambda}/A_V$ coefficients are taken 
from the \cite{Cardelli89} extinction law 
($K_{\rm H\alpha}=0.817, K_{\rm H\beta}=1.164$), however any 
extinction or attenuation law of choice can be applied in this method.

As we explicitly show in \cite{Piotrowska20} this method 
accurately traces hydrogen in its molecular phase and when we compare the 
molecular gas masses obtained with this method against \mgas\ 
inferred from the CO emission in the COLDGASS sample 
\citep{Saintonge17}, we find an RMS of 0.4~dex and a Spearman 
correlation strength of 0.84. For a~further discussion of the details 
and robustness of this method we refer the interested
reader to \cite{Piotrowska20}. 

The use of proxies to estimate black hole and molecular gas masses 
in the SDSS allows us to extend the sample size by a factor of 
a~thousand in comparison with the more direct measurement studies. 
This substantial increase in sample size is critical for 
a~statistically robust comparison between the theoretical predictions 
from the simulations and the observed Universe.

\subsection{Random Forest analysis}
\label{sec:rf-method}

Given that quenching is a~complex process, one expects the connection between large-scale
galactic properties and the star-forming state of galaxies to be non-trivial. 
Hence, an appropriate analysis should not be restricted to a~search for power-law relationships
and determination of scatter, but rather embrace the non-linear character of the problem
along with the whole range of potential parameters involved. 
A~natural approach meeting these criteria is machine learning, which allows us to explore
and identify relationships among our data without making a~priori assumptions about said
relationships. 

In particular, we opt for a~Random Forest (RF) classifier to ask which of
the following variables: \mbh, \mhalo\, \mstar\ or \mdot\ (in simulations only) 
has the most influence on determining whether a~galaxy is star forming or quiescent.
We choose these particular quantities because each of them is associated with
a separate quenching mechanism under consideration. 
\mstar\ serves as a proxy for the strength of supernova feedback, \mhalo\ is directly
linked to CGM gas heating via virial shocks, \mbh\ traces the integrated energy input
into the gas via AGN feedback and \mdot\ describes the instantaneous influence of 
AGN on its surroundings, measuring the feedback energy output at a given snapshot.

\subsubsection*{Random Forest classification overview}
A Random Forest classifier is a simple machine learning 
algorithm which assigns discrete labels (`classes') to elements
within a dataset, using a set of inputs (`features') for each 
element. It is a form of supervised learning, which means
that each element in the dataset used for algorithm training 
(i.e. the algorithm learning process) has a~class label assigned
to it. To put this in the context of our dataset, each element 
is a single galaxy labelled as `star-forming' or `quenched', based
on its specific star formation rate with the default classification
threshold chosen as $\log({\rm sSFR/ yr^{-1}}) = -11$ . 
The input features for each object 
are \mhalo, \mstar\ and \mbh\ and the RF learns how to best assign
the `quenched' and `star-forming' labels to galaxies 
given all these three masses.

One of the main advantages of a Random Forest is its 
straightforward architecture. This property of the RF allows
one to follow the learning logic throughout the algorithm
and removes the risk of a `black box' approach commonly associated
with more elaborate machine learning techniques. Most importantly
for our research, however, this straightforward architecture
enables an explicit calculation of the \textit{relative importance}
of each input feature for determining the final classification,
referred to as the \textit{feature importance}.
Putting this in a relevant context, once the RF learns to 
classify galaxies into star-forming and quenched, we can then
calculate how important \mhalo, \mstar\ and \mbh\ were 
for separating objects into these two categories. We can also
directly check how the classification was performed and even
visualise the exact structure of the decision-making within 
the algorithm.

A machine learning Forest, by analogy with a physical one,  
is a collection of multiple decision trees. A single tree consists of
a~series of splits on a dataset, each of which results in two subsets.
These subsets in RF nomenclature are known as `nodes' and hence 
the tree begins with a `root' node (the whole input dataset) and 
grows through subsequent splits on `parent' nodes into `daughter' 
nodes. These splits are performed by choosing a boundary value in 
one of the input features and each boundary value \& feature 
combination is optimised to yield daughter nodes more 
homogeneous than the parent one, i.e. aiming to deliver 
subsets consisting of only a single class.

There are several different ways in which trees determine 
the optimal splits. In our RF architecture we choose to minimise 
Gini index $G(n)$, which measures how often drawing an 
object from a daughter node at random would result in 
a~misclassification. The Gini index is defined as:
\begin{equation}
G(n)  = 1 - \sum^{J}_{i=1} p_i(n)^2\, ,
\label{eq:gini}
\end{equation}
where $p_i(n)$ is the probability of selecting an element with 
class $i$ when taking a random draw from elements occupying
a node $n$. The total number of available classes is given by $J$
and in our RF $J=2$. The Gini index evaluates to 0 for entirely
pure daughter nodes and takes a maximum value of 
$0.5$ in the case of a random draw of each of the two classes in 
a daughter node being equally probable. The trees then grow in 
depth by choosing feature \& boundary value combinations 
which minimise $G(n)$ (maximise the reduction of impurity)
at every split. This exponential growth 
continues until a tree reaches a~user-defined maximum depth
or until all elements in the dataset occupy their own individual 
`leaves' (terminal nodes not subject to further splitting).
Once the tree is complete, each element is assigned a classification
probability based on the distribution of classes within a given
leaf.

The true predictive power of a Forest comes from its 
Randomness. A collection of weakly 
correlated outcomes of individual trees creates an ensemble 
prediction which outperforms that of any given tree. Differences
among the trees render the combined result less sensitive to outliers 
and peculiarities of the parent dataset, allowing the algorithm to 
find general patterns present within the data. This property underlies 
the main application of RFs - predicting classification of previously
unseen objects given the complex, general relationships between
features and classes learned from a training dataset.

What makes the Forest Random is a random 
sampling of the data which comprise root nodes of individual 
decision trees. The number of trees in a forest as well as the 
sampling method are determined by a user and
we choose to draw a 100 bootstrap samples with replacement 
to construct 100 trees. Additionally, one can further increase 
the level of randomness within individual trees themselves 
by forcing the algorithm to only consider a random subset of features
to split on at each node. Empirical tests suggest that restricting 
the number of features to as few as two at each node makes the 
algorithm more robust with respect to noise in classification 
problems \citep[e.g.][]{Breiman01}. Hence, it is a common practice 
to randomly choose $\sqrt{p}=2$ features to split on at every node 
in RF classifiers, where $p$ is the number of input features 
\citep[e.g.][]{Hastie09}. 
In contrast, allowing the algorithm to choose from all features
at every split is most effective for isolating causality between
the input features and classification labels, as shown explicitly
by \cite{Bluck21}. We perform our analysis both using all input 
features at every node and selecting a random sample of $\sqrt{p}$ 
features to find that our results are consistent between the two 
forest designs. However, in the interest of brevity we only present 
results using all input features in Sec.~\ref{sec:rf} and show the other 
forest architecture outcome in Appendix~\ref{sec:discussion:rf-robust}.

\subsubsection*{Training and validation samples}

\begin{table} 
\caption{Summary of the forest parameters and sample properties in our RF analysis.
		The $\langle {\rm AUC} \rangle$ is the 
		median AUC (Area Under the Curve - a measurement of algorithm performance;
		equal to 1.0 for a perfect classifier and to 0.5 for a random guess)
		of the test set, while $\gamma$ denotes the \texttt{min\_samples\_leaf}
		hyperparameter optimised to achieve the highest AUC score in a~given sample.}
\begin{tabularx}{\columnwidth}{cccXX}
\toprule
Data set & Input sample size & \# passive objects & \centering $\langle {\rm AUC} \rangle$ & \centering $\gamma$ \tabularnewline
\midrule
SDSS & 148 954 & 153 550 & \centering 0.85 & \centering 145 \tabularnewline
SDSS HR04 & 321 522 & 174 425 & \centering 0.86 & \centering 117 \tabularnewline
SDSS gas & 13 104 &  6552 & 	\centering 0.98 & \centering 40 \tabularnewline
\addlinespace
EAGLE & 1786 & 893 & \centering 0.81 & \centering 148 \tabularnewline
Illustris & 1224 & 612 & \centering 0.98 & \centering 5 \tabularnewline
TNG & 2778 & 1389 & \centering 0.98 & \centering 15 \tabularnewline
\bottomrule
\end{tabularx}
\label{tab:auc-params}
\end{table}

The main task required of an~RF algorithm is finding an optimal general relationship between
the class label and input features for each element in a dataset, which can then be used
to make predictions about previously unseen data. Hence, the RF is first trained using
a~\textit{training set} drawn from the data and its performance is then evaluated 
by quantifying the accuracy of predicted classifications on a~\textit{test set}. 

In our study, both the training and test sets are drawn from samples of galaxies selected
according to the criteria listed in Sec.~\ref{sec:sample-selection} for each simulated and 
observed local universe. In order to ensure that the algorithm is not driven by either the
star-forming (SF) or passive (PA) population which may dominate a~given data set, 
we always consider an equal number of SF and PA galaxies 
in input samples for the RF. We hence create 
a~`balanced sample' consisting of 50\% PA and 50\% SF galaxies
by randomly choosing a~subset of the larger population,
such that the numbers of passive and star-forming galaxies are equal. 
Such a sample results in a better performance of the classifier by maximising 
the number of splits with a meaningful decrease in impurity. This approach 
also leads to a more intuitive interpretation of the results, especially when 
we focus on the relative importance of input features for a given classification, 
rather than the predictive power of the algorithm.

The SDSS sample is dominated by quiescent galaxies and hence
in order to construct a balanced input for the RF we include all SF galaxies and 
subsample the PA population. In simulations the situation is reversed
because their galaxy populations are dominated by star-forming
objects. Table~\ref{tab:auc-params} summarises 
the process of sample balancing by comparing the final input RF sample size
against the number of passive galaxies in a given data set. In the simulations the input 
sample size is twice the number of passive objects in the whole sample, showing that
the algorithm is trained on the whole quiescent population and on a random subset
of the star-forming one. In contrast, the SDSS galaxies are mostly passive and hence
the star-forming population is used in full, while the quiescent objects are randomly
subsampled. The `SDSS HR04' data set refers to RF runs with \mbh\ estimated from \mbulge\
according to Eq.~\ref{eq:HR04} and does not require an inclination cut on disk-dominated
galaxies, which preferentially removes star-forming objects from the RF samples using
\vdisp\--derived \mbh. The `SDSS gas' data set describes the input for the 
star-forming/passive classification using \fgas\ and SFE as input features. In this case 
the bulk of passive SDSS galaxies is removed via S/N cuts on emission line fluxes and hence
the input sample size is dictated by the leftover quiescent objects.

Once the sample consists of an equal number of star-forming and passive
galaxies, we proceed to randomly split it into a~training and a~test
set with an equal number of objects in each. This way we maximise the RF training
opportunity and ensure its exposure to a~large test set of previously unseen data. 
Finally, in order to account for the randomness in our balanced sample selection, we repeat the 
RF experiment 500 times for each data set, subsampling the larger of populations at random
with each repetition. Each of the experiments results in 
different feature importances which together yield a set of 
importance distributions (one for each feature among \mbh, \mhalo\ and 
\mstar). We then take our final result for a given feature to be the 
median value of its importance distribution and show errorbars
spanning the 5th and 95th percentiles of said distribution.

\subsubsection*{Feature importances and algorithm optimisation}

The main purpose of our RF analysis is the \textit{identification 
of the most important quenching parameter}, i.e. estimating the feature 
importances. This can be done in various
ways \citep[see Chapter 6 in][for a~summary]{Loupe14} and in our case is done 
by calculating the Mean Decrease in Impurity, otherwise known
as the Gini Importance. This metric calculates each feature 
importance as the sum over the number of splits within the algorithm 
which include the feature, weighted by the number of elements the
feature splits. More precisely, the relative importance $I_R(k)$
of feature $k$ is calculated via:
\begin{equation}
I_R(k) = \frac{I_k}{\sum_j I_j} = 
\frac{\sum_{n_k} N(n_k) \Delta G(n_k)}{\sum_n N(n) \Delta G(n)}\, ,
\end{equation}
where 
\begin{equation}
\Delta G(\rm{parent}) = G({\rm parent}) - G({\rm daughter})
\end{equation} 
is a change in the Gini index between a daughter and a parent node.
The $n_k$ sum is performed over all splits the feature $k$ contributes 
to while the $n$ sum corresponds to all splits within the forest. The 
change in Gini index is also weighted by the number of elements in 
a~given parent node $N$, which results in a higher importance 
calculated for features which have an impact on a~larger number
of elements in the dataset.

Another popular method for estimating feature importance
relies on randomising the input feature values and evaluating the
impact this action has on the final classification outcomes.
This method, called the permutation importance,
breaks the connection between features and classification labels 
established through training and hence a~drop in the algorithm 
classification accuracy indicates how much a given label depends on 
a~given feature \citep{Breiman01}. We check that our main conclusions 
are consistent with the two importance calculation techniques
and present results based on Gini importance in Sec.~\ref{sec:rf}.
For a detailed discussion on the Gini importance and its broader
applications we refer the interested reader to \cite{Bluck21}.

A final point to consider in the Random Forest architecture
is the \textit{risk of over-fitting}, which occurs when the algorithm
perfectly learns the relationship between features and class labels
in a given training set and then spectacularly fails upon exposure
to previously unseen data. In such a scenario the performance of
an RF would be ranked very high for the training set and then very
low for a test set in our experiment. One can avoid over-fitting by
controlling the number of elements (galaxies) which comprise the
leaves of individual trees. By allowing the trees to grow without
constraints, each leaf node ends up occupied by a single galaxy
only and hence the algorithm perfectly learns all the patterns 
within a training data set and achieves a high score in
a training set. In contrast, when one imposes a minimum
occupancy for a leaf node, the algorithm can only learn the general
patterns within the dataset recovered both in the training and
testing sets, which come at a price of a lower classification 
accuracy within the training set. Hence, controlling the minimum 
number of elements which comprise a leaf node in a tree allows 
one to avoid over-fitting in the Random Forest algorithm.

In order to optimize the performance of our RF classifier we use 
the Receiver Operating Characteristics (ROC) graphs \citep{Fawcett06}, 
which show the True Positives Rate (TPR) as a~function of False 
Positives Rate (FPR). The output of our RF is a~set of class probabilities 
(i.e. how likely a~given galaxy is to be classified as quenched), hence the 
ROC curve is computed by varying probability thresholds for passive 
classification and calculating their corresponding TPR \& FPR. We further 
reduce the curves to a~single parameter by taking the area under the 
curve (AUC) in the ROC space. The AUC parameter ranges between 
${\rm AUC}=0.5$ for a classifier equivalent to a random class assignment 
and $\rm{AUC}=1.0$ for a perfect classification algorithm. In this work we 
optimise the Random Forest architecture to maximise $\rm AUC_{train}$ 
under the criterion that $\rm AUC_{train}-AUC_{test} \leq 0.01$ in
order to avoid the risk of over-fitting in the algorithm. 
In this way we maximise the performance of our Random Forest on 
both the training and test data sets, making sure the algorithm 
is generalisable and performs well on previously unseen data 
\citep[see][for examples of this approach]{Bluck20a, Bluck20b}. 

We implement the RF algorithm using the \texttt{RandomForestClassifier} class
in the \texttt{sklearn.ensemble} module of the \texttt{scikit-learn} \citep{scikit-learn} 
open source machine learning package for 
python\footnote{\url{https://scikit-learn.org}}. 
As mentioned before, we set the \texttt{max\_features} value to either
"sqrt" or "None", \texttt{n\_estimators}~$=200$ and optimise the \texttt{min\_samples\_leaf}
parameter to maximise AUC in all data sets individually, while retaining default values 
for all other parameters. The final \texttt{min\_samples\_leaf} values along with 
AUC scores are presented in Table~\ref{tab:auc-params}.

\begin{figure*}
\includegraphics[width=\textwidth]{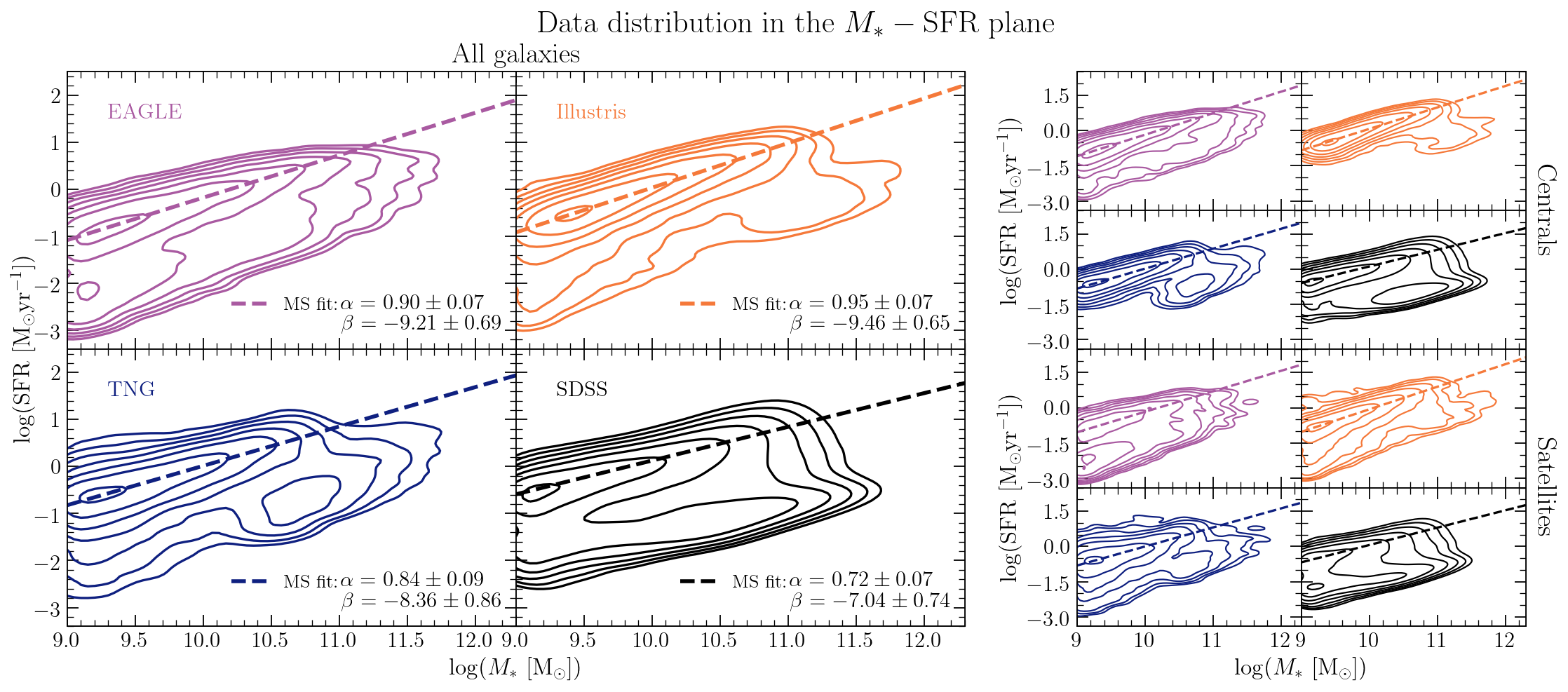} 
    \caption{The distribution of galaxies in the \msfr\ plane
    		 for EAGLE (purple), Illustris (orange), TNG (blue) and SDSS (black).
    		 Dashed lines show ordinary least squares fits to the star-forming 
    		 Main Sequence defined as galaxies with log(sSFR/${\rm yr^{-1}}$)  $> -10.5$. 
    		 Left panel: all galaxies in the sample, 
    		 panels on the right: galaxies split into centrals (top) and
    		 satellites (bottom). MS fits in the whole population agree within 
    		 uncertainties in the simulations and their slopes are consistently
    		 steeper than in
    		 the SDSS. It is also apparent that for both Illustris and TNG passive
    		 centrals are primarily high-mass objects, while EAGLE seems to have 
    		 quenched centrals spanning all mass ranges. Please note: low sSFR 
    		 values in the simulations are redistributed by drawing 
    		 from the SDSS passive sequence distribution to aid visual comparison
    		 (see Fig.~\ref{fig:appendix-ssfr} in appendix~\ref{ssfr-function}). 
    		 The SDSS contours were corrected to account for the Malmquist bias. }
    \label{fig:ms}
\end{figure*}

\section{Results}
\label{sec:RESULTS}

We begin our analysis with a~qualitative comparison of observed and simulated
local galaxies by considering the data distributions in the 
\msfr\ plane, presented as density contours in Fig.~\ref{fig:ms}. In order
to fairly compare the observed and simulated universes, we weight the distribution of
SDSS objects in the plane by the inverse of the maximum comoving volume they can be observed
within, given their intrinsic brightness and the magnitude limit of the survey. 
This way we account for the preferential loss of low-luminosity objects
with increasing redshift of observation, known as the Malmquist bias \citep{Malmquist1922}.
In simulations, we make sure to show all galaxies in a~finite range of the logarithmic SFR
axis by relocating objects with extremely low specific star formation rates (sSFR). 
To this end, for every galaxy with log(sSFR/${\rm yr^{-1}}$)  $< -12$
we draw an sSFR value from the observed distribution in the SDSS passive sequence 
(see appendix~\ref{ssfr-function} for details). This manoeuvre is only intended
for presentation purposes, in order for us to make a~fair visual comparison of galaxy
populations between simulations and the SDSS. These redistributed values can
then be treated as upper limits on SFR, much like the SFR estimates in the SDSS passive
sequence derived from the strength of the D4000 break, instead of ${\rm H}\alpha$ fluxes.
In this way we avoid neglecting low-SFR objects in simulations or, alternatively, 
extending the $\log(\rm{SFR})$ range to arbitrarily large negative values. 

In Fig.~\ref{fig:ms} we present density contours of galaxy distributions in the \msfr\
plane, which show very good qualitative agreement across all data sets for all galaxies
(left panel), centrals (top right panel) and satellites (bottom right panel). 
All simulation suites have well-pronounced star-forming 
Main Sequences (MS) and less abundant passive populations, much like the SDSS.
However, the jagged contours indicate a~smaller coverage of the parameter space
in the simulations compared to the observations. Illustris and TNG also show a~hint of 
transition between the dominance of star formation at low \mstar\ and quiescence 
at high \mstar. This feature is even more pronounced in the subset of central galaxies, 
where quenched low-\mstar\ objects are absent in these suites. The 
raw SDSS measurements exhibit a~similar behaviour in centrals, 
however once the volume correction is accounted
for, the population of quiescent central galaxies spans the whole range in stellar mass, 
peaking at around $\log(M_\ast/{\rm M_\odot})=10.5$. Satellite galaxies can be
both quenched and star-forming at all \mstar\ in all data sets. 
The distribution of satellites in EAGLE follows that of the SDSS, 
peaking at low stellar mass objects. In Illustris and TNG
no such behaviour is observed and their distribution is dominated by star-forming objects
at all \mstar.

\setlength{\tabcolsep}{4.5pt}
\begin{table} 
\caption{Summary of OLS fit parameters to the star-forming MS.} 
\begin{tabularx}{\columnwidth}{lccccc}
\toprule
 & & SDSS & EAGLE & Illustris & TNG \\  
\midrule
\parbox[t]{3mm}{\multirow{2}{*}{all}} & $\alpha$ & 
	$0.72 \pm 0.07$ & $0.90 \pm 0.07$ & $0.95 \pm 0.07$ & $0.84 \pm 0.09$ \\
& $\beta$ & $-7.04 \pm 0.74$ & $-9.21 \pm 0.69$ & $-9.46 \pm 0.65$ & $-8.36 \pm 0.86$ \\
\addlinespace \addlinespace
\multicolumn{2}{c}{\parbox[t]{3mm}{\multirow{2}{*}{all with scatter}}}
& \multicolumn{1}{r}{$\alpha$} & $0.89 \pm 0.09$ & $0.92 \pm 0.09$ & $0.82 \pm 0.10$ \\
&& \multicolumn{1}{r}{$\beta$} & $-9.03 \pm 0.93$ & $-9.16 \pm 0.94$ & $-8.18 \pm 1.05$ \\
\addlinespace \addlinespace
\parbox[t]{3mm}{\multirow{2}{*}{cen}} & $\alpha$ & 
	$0.71 \pm 0.07$ &  $0.92 \pm 0.07$ &  $0.91 \pm 0.06$ &  $0.85 \pm 0.08$ \\
& $\beta$ & $-6.98 \pm 0.71$ & $-9.33 \pm 0.67$  & $-9.06 \pm 0.57$ &  $-8.46 \pm 0.75$\\
\addlinespace \addlinespace
\parbox[t]{3mm}{\multirow{2}{*}{sat}} & $\alpha$ &
	$0.74 \pm 0.08$ & $0.87 \pm 0.11$ & $0.96 \pm 0.08$ & $0.81 \pm 0.13$ \\
& $\beta$ &  $-7.31 \pm 0.81$ & $-8.89 \pm 1.10$ & $-9.66 \pm 0.81$ & $-8.10 \pm 1.24$ \\
\bottomrule
\end{tabularx}
\label{tab:ms-fits}
\end{table}

\begin{figure*}
\includegraphics[width=\textwidth]{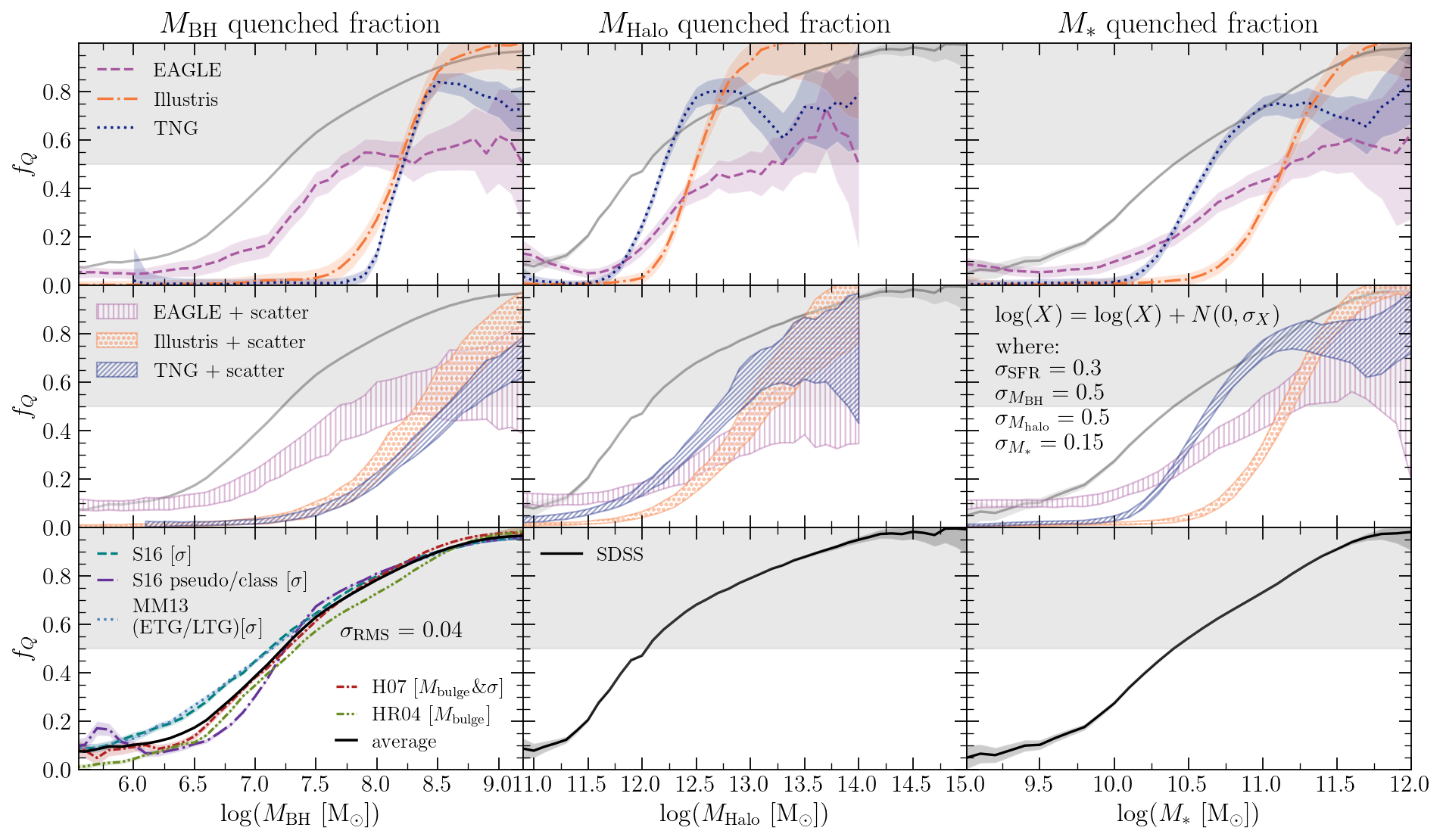} 
    \caption{Comparison of quenched fractions ($f_Q$) as function of $M_{\rm BH}$,
    				 $M_{\rm halo}$ and $M_\ast$
    				 in simulations (top row), in simulations with observation-like scatter 
    				 added (middle row) and observations (bottom row). In the SDSS $M_{\rm BH}$ 
    				 quenched fraction calculation five different $M_{\rm BH}$ calibrations are included, listed 
    				 in the legend. The simulation panels also include SDSS curves (solid gray lines) 
    				 to aid direct comparison. 
    				 For all quantities, the simulations show steeper trends than 
    				 SDSS, with the exception of EAGLE, which shows shallower slopes and 
    				 levels off at $f_{Q} \sim 0.6$. Adding scatter to each of the quantities listed 
    				 in the figure flattens the simulated curves, making
    				 the transition significantly less abrupt for Illustris and TNG. Nevertheless,
    				 the simulations still under-predict quenching for all parameters 
    				 (as quantified in Table~\ref{tab:fq05}).
    				The different $M_{\rm BH}$ calibration curves demonstrate that the conclusions 
    				drawn from SDSS do not strongly depend on the choice of $M_{\rm BH}$ indirect inference
    				method. }
    	\label{fig:fq}
\end{figure*}

Table~\ref{tab:ms-fits} summarises MS fit parameters corresponding to dashed lines 
in Fig.~\ref{fig:ms}. We perform an~OLS fit to
$\log({\rm SFR}) = \alpha \log(M_\ast) + \beta$ for Main Sequence objects selected 
with a~log(sSFR/${\rm yr^{-1}}$)  $> -10.5$ criterion. 
We divide MS galaxies into 0.1 dex bins in $\log(M_\ast)$ for 
${9 < \log(M_\ast/{\rm M_\odot}) < 11}$ 
and fit to the median $\log({\rm SFR})$ values in each bin.
In order to estimate uncertainties on $\alpha$ and $\beta$ we use the median absolute 
deviation (MAD) of $\log({\rm SFR})$ in the bins, treating them as heteroskedastic, 
independent errors on median $\log({\rm SFR})$.

Table~\ref{tab:ms-fits} shows that MS fits across all simulations agree with each 
other within the calculated uncertainties, while the SDSS shows a~significantly shallower
slope. When we focus on the central-satellite split we notice that in EAGLE and TNG 
the satellite MS is flatter, while for Illustris and SDSS it is steeper, however slopes in the 
two populations agree within their uncertainties for all data sets. Finally, we check how
the Main Sequence fits to the whole population change in the simulations 
when we add observation-like scatter to the raw data. 
For each simulated galaxy we add a random draw from a Gaussian 
to both $\log({\rm SFR})$ and $\log(M_\ast)$ to imitate the addition of
measurement uncertainty present in the observations. The widths of scatter distributions are 
$\sigma_{M_\ast}=0.15$ and $\sigma_{\rm SFR}=0.3$ respectively to reflect the median
uncertainties in the SDSS. The resulting MS slopes are flatter in all three simulations,
however not flat enough to match the SDSS. They also agree with the raw MS fits
within their estimated errors.

Our simplistic method to fitting the MS allows us to compare SDSS and simulations
without additional selection choices on emission line ratios and S/N values, which are not
available in the simulation suites. This approach is similar to the objective MS definition in
\cite{Renzini15} and for the SDSS yields parameters consistent with their study.  
The broad qualitative agreement between simulations and observations 
in the \msfr\ distributions of centrals and satellites is promising.
We also note that the quantitative discrepancies between the 
observed and simulated Main Sequences can be significantly reduced 
when different techniques are applied to estimating SFR in the observations
\citep[e.g. as demonstrated by][for TNG50 at redshift $z=1$]{Nelson21}. 
Hence, encouraged by this initial comparison we investigate 
similarities and differences between the simulated and observed data sets 
in more detail in the subsequent sections.

\subsection{Behaviour of quenched fractions and sSFR as a~function of \mbh, \mhalo\ and \mstar}
\label{sec:qf}

\begin{table} 
\caption{Parameter values at which quenched fractions reach $f_Q=0.5$. All values have
				an uncertainty of $\pm 0.2 \,{\rm dex}$. Values in brackets show how adding 
				observation-like scatter to simulations influences the point at which the galaxy
				population is mostly quiescent.}
\begin{tabularx}{\columnwidth}{XXXXX}
\toprule
 &  \centering  SDSS &  \centering EAGLE &  \centering Illustris &  \centering TNG \tabularnewline  
\midrule
 \centering  $\log(M_{\rm BH})$ & \centering   7.2 &  \centering 7.8 (8.1) 
 		& \centering  8.2 (8.4)  & \centering 8.2 (8.6)  \tabularnewline
 \centering  $\log(M_{\rm Halo})$ &  \centering 12.0  &  \centering  13.2 (13.6)
 		&   \centering 12.5 (13.0) &   \centering 12.2 (12.8)  \tabularnewline
 \centering  $\log(M_\ast)$ &  \centering  10.4  &  \centering 11.2 (11.1) 
 		&   \centering 11.1 (11.1) &  \centering  10.6 (10.6) \tabularnewline
\bottomrule
\end{tabularx}
\label{tab:fq05}
\end{table}

\begin{figure*}
\includegraphics[width=\textwidth]{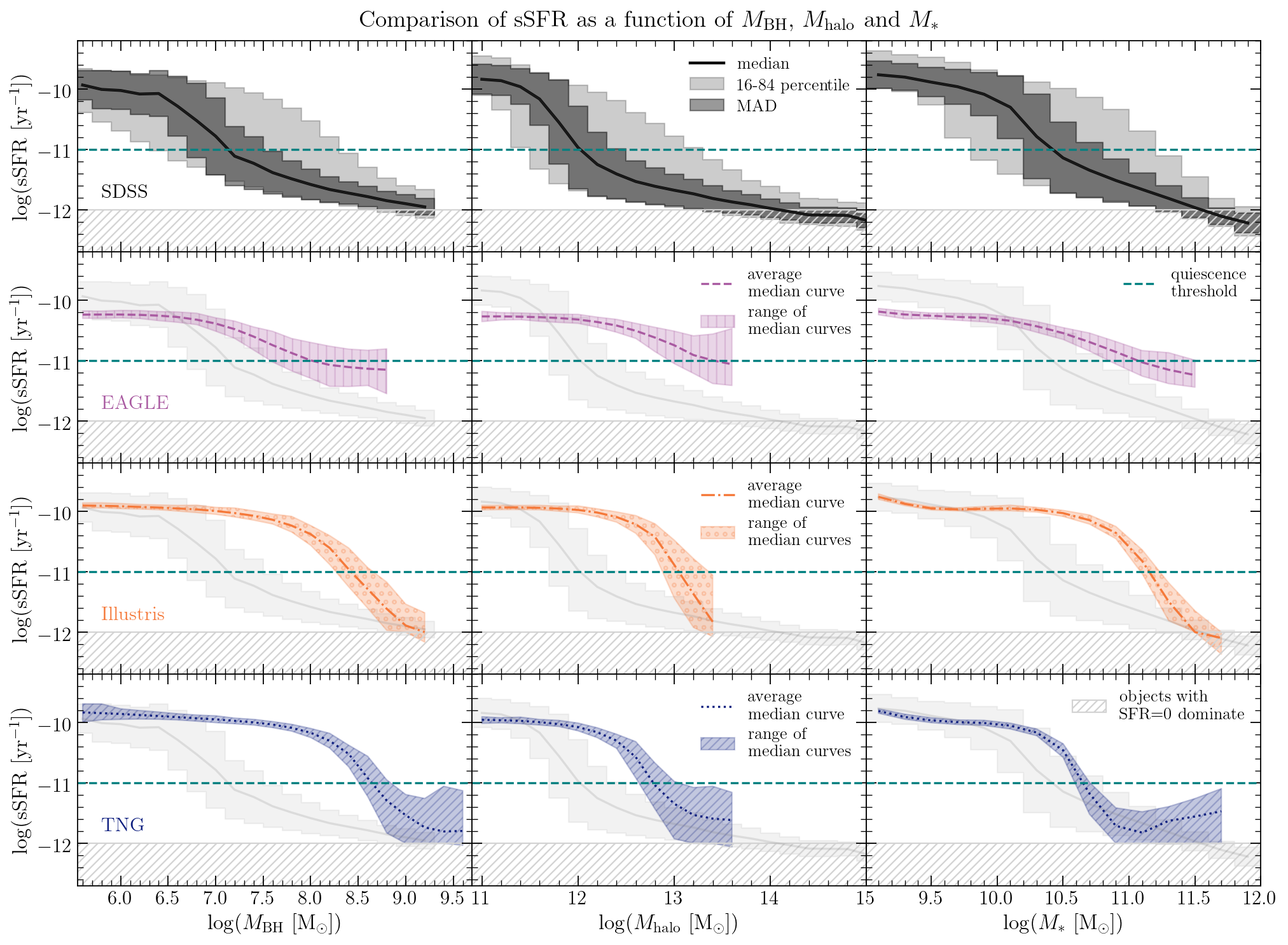} 
    \caption{Comparison of sSFR as a~function of \mbh\ (left column), \mhalo\
    				(middle column) and \mstar (right column) in the SDSS and 
    				in the simulations. In the SDSS the calculation includes 
    				$1/V_{\rm max}$ and inclination corrections and the shaded regions
    				correspond to the median absolute deviation (MAD) and
    				the 16-84 percentile range of sSFR distribution in a~given bin. 
    				The hatched coloured 
    				regions in the simulations show the range of median curves
    				resulting from 500 realisations of adding random scatter to the data
    				to mimic measurement uncertainty.
					The minimum number of galaxies in a~given 0.2 dex bin is 20.
    				In EAGLE all curves show similar slopes to the SDSS, however 
    				they hardly reach log(sSFR/${\rm yr^{-1}}$) $=-11$, highlighting the
    				lack of quenched objects in the suite. Illustris and TNG show
    				steeper slopes than the observations, quenching galaxies 
    				at higher values of all parameters than the SDSS.}
    	\label{fig:ssfr-line}
\end{figure*}

We first look at the transition to quiescence by comparing quenched fractions $f_Q(X)$ 
calculated as a function of \mbox{$X=M_{\rm BH}$, \mhalo\ and \mstar.} We define the quenched 
fraction as :
\begin{equation}
f_Q = \frac{n_{{\rm Q},i}}{n_{{\rm Q},i} + n_{{\rm SF},i}}
\end{equation}
where $i$ labels a~bin in quantity $X$, and $n_{{\rm Q},i}$,  $n_{{\rm SF},i}$ 
are the number of quenched and star-forming objects in~a given bin respectively. 
Each bin is 0.4~dex wide, while bin centres are separated by 0.1~dex
in order to obtain smoother curves. In simulations $n_i$ is simply the 
number of galaxies in a~given population, while in the SDSS we correct for 
the inclination cut and the Malmquist bias by substituting $n_i$ with:
\begin{equation}
n_i = \sum_j \omega_j, \qquad \text{where}\ \omega_j=\frac{w_{{\rm inc},j}}{V_{{\rm max},\, j}}.
\label{eq:sdss-weight}
\end{equation}
where \vmax\ is the maximum comoving volume a~galaxy can be observed in,
$j$ enumerates objects of a~given population in a~bin and $w_{\rm inc}$ 
is a~correction factor associated with inclination selection, defined in Eq.~\ref{eq:inc}. 
We verify that
$w_{\rm inc}$ achieves its intended effect by comparing the survey
\mstar\ quenched fractions with and without inclination selection 
criteria and corrections, finding very good
agreement between the two resulting curves.

Fig.~\ref{fig:fq} presents quenched fractions as a~function of \mbh, \mhalo\ 
and \mstar\ for all simulations together in the top panel
and the SDSS in the bottom panel. It is apparent that quenched fractions show
an increasing trend with all quantities both in the observed and simulated 
universes. Illustris and TNG show more rapid transitions between star forming
and quenched populations in all parameters, demonstrated by
visibly steeper curves than those seen in the SDSS. The steep 
$f_Q(M_\ast)$ curves in these two suites are consistent with recent
studies by \cite{Donnari21a} and \cite{Donnari21b}, where the authors
use a different prescription for identifying quenched centrals in Illustris
and TNG100. EAGLE breaks this pattern
in simulations by showing flatter curves more akin to observations, however
it fails to recover $f_Q \sim 1$ at high parameter values, hovering at
$f_Q \sim 0.6$ instead. Hence, in EAGLE there remains a~significant 
fraction of star forming systems at even the highest values of \mstar,
\mhalo\ and \mbh. We also notice that simulations, on average, tend
to transition into the quiescence-dominated regime later than the SDSS, 
reaching the grey shaded region at higher parameter values.
Table~\ref{tab:fq05} lists \mbh, \mstar\ and \mhalo\ values at which
all curves reach the $f_Q=0.5$ mark, clearly demonstrating this behaviour.

Fig.~\ref{fig:fq} also addresses the uncertainty associated with our choice
of black hole mass calibration in the leftmost bottom panel, where we compare
quenched fractions resulting from different inference methods.
Each curve is calculated for \mbh\ estimated from central
velocity dispersion, bulge mass or combination of the two, following
prescriptions in Eq.\ref{eq:S16}-\ref{eq:S16-ps}. In the case of
\cite{HaringRix04} calibration (labelled HR04 in the panel) the galaxy
sample is not subject to an inclination cut, as it is not relevant for
the measurement of \mbulge. All resulting $f_Q-M_{\rm BH}$ relations agree well with
one another with an RMS scatter around the mean quenched 
fraction of only $\sigma_{\rm RMS}=0.04$. This panel clearly
demonstrates that the relationship between $f_Q$ an \mbh\ is largely
invariant to the choice of calibration. This means that regardless 
of whether we use different scaling for early and late type galaxies,
choose a~different proxy (\vdisp\ or \mbulge) or introduce additional
sample selection criteria, our conclusions remain unaffected for
black hole mass in the SDSS.

Finally, we check how the measurement uncertainty present in observations would
demonstrate itself in the $f_Q$ relationships calculated for the simulations.
To this end we add a~random Gaussian noise to the measurements of
SFR, \mstar, \mhalo\ and \mbh\ to mimic the measurement uncertainties
in the SDSS. More specifically, for each simulated galaxy, we add 
a random draw from a~Gaussian centred on 0 with a~standard deviation
$\sigma=\sigma_X$ in each parameter of interest $X$. The $\sigma_X$
values were chosen to reflect the median random errors in the corresponding SDSS 
measurements and are listed in the rightmost middle panel in Fig.~\ref{fig:fq}.
The random error in SFR is only added for galaxies with log(sSFR/${\rm yr^{-1}}$) $>-11$,
since SFRs in the passive population in the SDSS are estimated using D4000
and hence treated as upper limits throughout our analysis. 
Hatched regions in the middle panel of Fig.~\ref{fig:fq} 
show the range of $f_Q$ curves resulting from 500 random realisations 
of simulated galaxies with scatter added as per the description above.
These regions demonstrate that the addition of random uncertainty to
the simulated data flattens $f_Q$ curves towards observation-like shapes.
The flattened curves, however, still reach $f_Q=0.5$ at higher parameter
values than in the SDSS, like it was previously seen in the raw results.

\begin{figure*}
	\includegraphics[width=\textwidth]{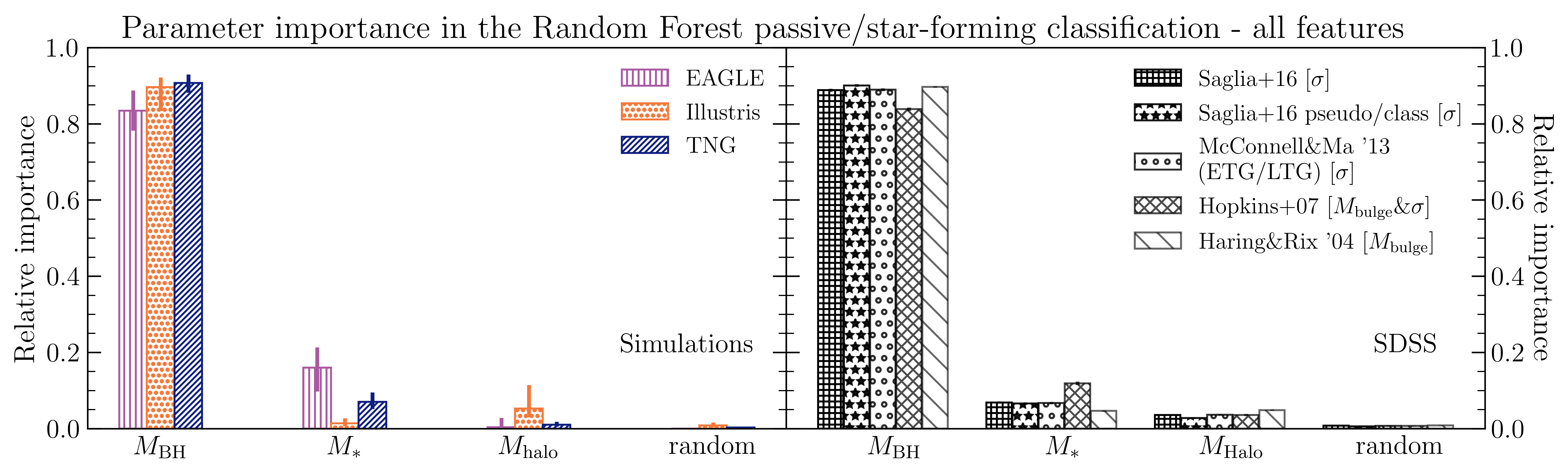} 

     \caption{Feature importances for \mbh, \mhalo\ and \mstar\ 
     					extracted from a~RF classification of central 
     					galaxies into star-forming and quenched categories. 
     					Bar heights represent 
     					the importance of each parameter for the classification
    		 			outcome in a~trained forest, and they sum up to one in each data set.
    		 			Error bars indicate the 5th and 95th 
    		 			percentiles of the distribution in parameter importance
    		 			drawn from 500 realisations of the experiment.  
    		 			Left panels: results in simulations. It is apparent that, irrespective of the model 
    		 			differences, all simulations predict \mbh\ to be the most important 
    		 			galactic property discriminating between the star-forming and quenched
    		 			populations. 
    		 			Right panels: results in the SDSS with \mbh\ estimated using calibrations
    		 			in Eq.~\ref{eq:S16}-\ref{eq:S16-ps}. It is clear that \mbh\ dwarfs
    		 			other parameters in its connection with quiescence, regardless of the choice
    		 			of calibration.}
\label{fig:rf-combined}
\end{figure*}

In Fig.~\ref{fig:ssfr-line} we take a~different look at how quenching 
progresses with increasing \mbh, \mhalo\ and \mstar. To this end
we show how the sSFR changes as a~function of each of those 
parameters, comparing the SDSS observations with simulation results.  
In order to fairly compare all data sets, we use EAGLE, Illustris and TNG
post-processed to account for random measurement errors in the 
observations, like we did in the middle panel of Fig.~\ref{fig:fq}.
The median curves in the SDSS include $1/V_{\rm max}$ and 
inclination corrections, which brings all rows in the figure to an 
equivalent footing. When we look at the observations alone, we
see that sSFR decreases with increasing \mhalo, \mbh\ and
\mstar, as expected from our analysis of quenched fractions. 
Halo mass shows the steepest evolution
in sSFR with the majority of \mhalo\ range covered by passive 
galaxies, following the quenching threshold at 
$\log(M_{\rm halo}/{\rm M_\odot}) \approx 12$ seen earlier in the $f_Q$ plots.

In the second row of Fig.~\ref{fig:ssfr-line} we compare 
EAGLE (coloured hatched regions) to the SDSS (light gray shaded 
regions) and notice that all curves in the suite 
begin 0.2-0.4~dex lower and have shallower slopes than the 
observations. The striking lack of quenched objects apparent 
earlier in Fig.~\ref{fig:fq} is also highlighted here by the median curves 
reaching a~lower limit of log(sSFR/${\rm yr^{-1}}$)  $\approx -11$, in contrast
with the other two suites. 
The Illustris and TNG curves
have steeper slopes than the observations in all variables
among \mbh, \mhalo\ and \mstar\ 
and clearly show that quenching occurs at higher parameter
values than in the SDSS. These suites, however, match the
observed curves at low values of \mstar, \mhalo\ and \mbh,
with \mstar\ showing good agreement over the broadest
range of values. When we compare the simulations among
each other, we find that TNG performs slightly better than 
Illustris and EAGLE, however none of the suites shows a~close
agreement with the SDSS.  

One important conclusion we draw from Fig.~\ref{fig:ssfr-line}
is that none of the three state-of-the-art cosmological
simulations are able to correctly predict the trends in sSFR with
\mbh, \mhalo\ and \mstar\ observed in the local Universe.
This statement is true for a fair comparison between simulations
and observations, where simulated data are augmented with
observational realism and the observed galaxy populations
are corrected for the effects of survey magnitude limits and
inclination cuts. The relationship for which predictions match
the SDSS most closely is \mstar\ and sSFR,
while for \mhalo\ and \mbh\ the onset of quenching is
predicted at much higher parameter values than it is observed
(in all simulations).
This interesting contrast shows how tuning the models to
recover observables like the galaxy stellar mass function
results in reasonable predictions for related secondary relationships,
e.g. the behaviour of sSFR as a function of \mstar. 
The lack of similar constraints from black hole mass
is also apparent, since the simulations struggle more to recover
the observed relations between \mbh\ and sSFR. 
Hence including these observational constraints from 
Fig.~\ref{fig:fq} in future model optimisation can potentially 
prove valuable for the development of the new generation of 
cosmological simulations. 
 
Figs.~\ref{fig:fq}~\&~\ref{fig:ssfr-line} together inform us 
about a~strong positive association between quenching and 
\mstar, \mhalo\ and \mbh\ in all simulations and in the observations.
When we compare the simulated universes to the observed Universe
in the SDSS, we find qualitatively similar trends in both which are
matched quantitatively with a~varying degree of success. When we focus
on \mstar\ trends only, the TNG result seems to be the closest match
with the observations, capturing the transition towards quiescence
with increasing parameter values most reliably. When we then look at
\mbh, it is the EAGLE suite which shows the closest trends to the SDSS.
However, EAGLE struggles to definitively quench galaxies at high black
hole masses. 

Although $f_Q$ and sSFR trends provide an interesting 
insight into the differences between EAGLE, Illustris and TNG, they 
do not indicate which among \mbh, \mhalo\ and \mstar, if any, 
are responsible for regulating galaxy quenching. We thus move 
on to using machine learning techniques in the next section,
in order to answer this fundamental question.

\subsection{Random forest classification}
\label{sec:rf}

In order to find out which galactic parameter among \mbh, \mhalo\
and \mstar\ is the most predictive in determining whether a galaxy
is star-forming or quenched, we perform a series of Random Forest
classifications described in detail in Sec.~\ref{sec:rf-method}. 
We conduct this machine learning experiment consistently in both 
simulations and the observations, repeating the training and testing 
sequence 500 times for each data set.
In this way we account for the random subsampling of the 
more numerous population between the star-forming and quiescent 
objects required to create an input sample of
50\% PA and 50\% SF galaxies. 
As a result we can compare the relationships between 
different galaxy parameters and quenching delivered by 
EAGLE, Illustris and TNG against the SDSS to see how different quenching 
mechanisms present in the simulations manifest themselves in the 
observables at hand.

In Fig.~\ref{fig:rf-combined} we show relative parameter importances
extracted from the Random Forest classifier for simulations
in the left panels and the SDSS on the right. Bar heights represent
the median importance for each parameter, while the error bars mark 
the 5th and 95th percentiles of the importance distribution created 
by repeating the experiment 500 times in each data set. The input 
features are listed in a decreasing order of importance from left 
to right in each panel and all importances
sum up to unity in each data set. Each experiment consists of 
four input features: \mbh, \mhalo, \mstar\ and a random draw
from a flat distribution between 0 and 1. The seemingly redundant
`random' feature allows us to check how much more important 
the highest ranking input feature is for assigning the PA/SF labels
than a random guess. In the SDSS we perform the classification
using all black hole mass calibrations from Eq.~\ref{eq:S16}-\ref{eq:S16-ps}, 
labelling the results with the parameters used to estimate \mbh. As 
discussed in the earlier parts of our analysis, the \cite{HaringRix04}
bulge mass calibration does not require a cut on galaxy inclination
and hence serves as a~consistency check for the influence of this selection
criterion on our classification result. 

We first focus on simulations in the left panel of Fig.~\ref{fig:rf-combined}. 
It is overwhelmingly apparent that the decision trees
grow almost solely using black hole mass as the criterion for 
splitting the sample. Even in EAGLE where \mstar\ picks up some residual 
importance, \mbh\ holds over 4 times more predictive power, as measured
by the Gini Importance calculated for the feature across all 
trees in the forest. In simulations, the primacy of black hole mass for predicting 
quenching is invariant under different implementations
of AGN feedback and baryonic processes. EAGLE, Illustris and TNG 
all predict black hole mass to be the most informative of quiescence,
despite a~multitude of differences in the subgrid prescriptions for the interaction
between AGN and the matter surrounding them. 

This unanimous theoretical prediction is met incredibly well in 
the observations (right panel in Fig.~\ref{fig:rf-combined}), 
where the relative importance of \mbh\ dwarfs the other two parameters. 
In the SDSS this prominent dominance of black hole mass is robust against 
the choice of calibration since the importances of \mbh\ agree with one 
another within $\pm 0.05$ for all adopted prescriptions. 
We also find a near-identical result when we force the RF to subsample
features at each split, further discussed in Appendix~\ref{sec:appendix:sqrt}.

We are also confident that our conclusions drawn with the SDSS data 
are not affected by any potential sources of bias. As we
explain in Appendix~\ref{sec:discussion:err}, the result is not 
driven by the differences in precision with which we measure
\mhalo\ and \mstar, as compared with \mbh. We also check 
that our sample selection does not influence our conclusions
by exploring different corrections and quality cuts in 
Appendix~\ref{sec:discussion:sample-cuts}. Finally, in order
to further convince ourselves that the use of calibrations
to estimate \mbh\ does not drive our conclusions we also
repeat the RF experiment in Appendix~\ref{sec:appendix:terrazas}
with a~sample of 90 central galaxies with dynamical measurements of 
black hole mass compiled by 
\cite{Terrazas17}. We find that out of the two
parameters available for the sample - \mstar\ and \mbh, 
black hole mass has a~significantly higher importance in the 
passive classification both when the decision trees randomly
sample a subset of all features and when all features are 
available for each split. This result is completely consistent 
with our general conclusions from this section, based on much
larger samples with indirect estimates of supermassive black masses.

\begin{figure}
	\includegraphics[width=\columnwidth]{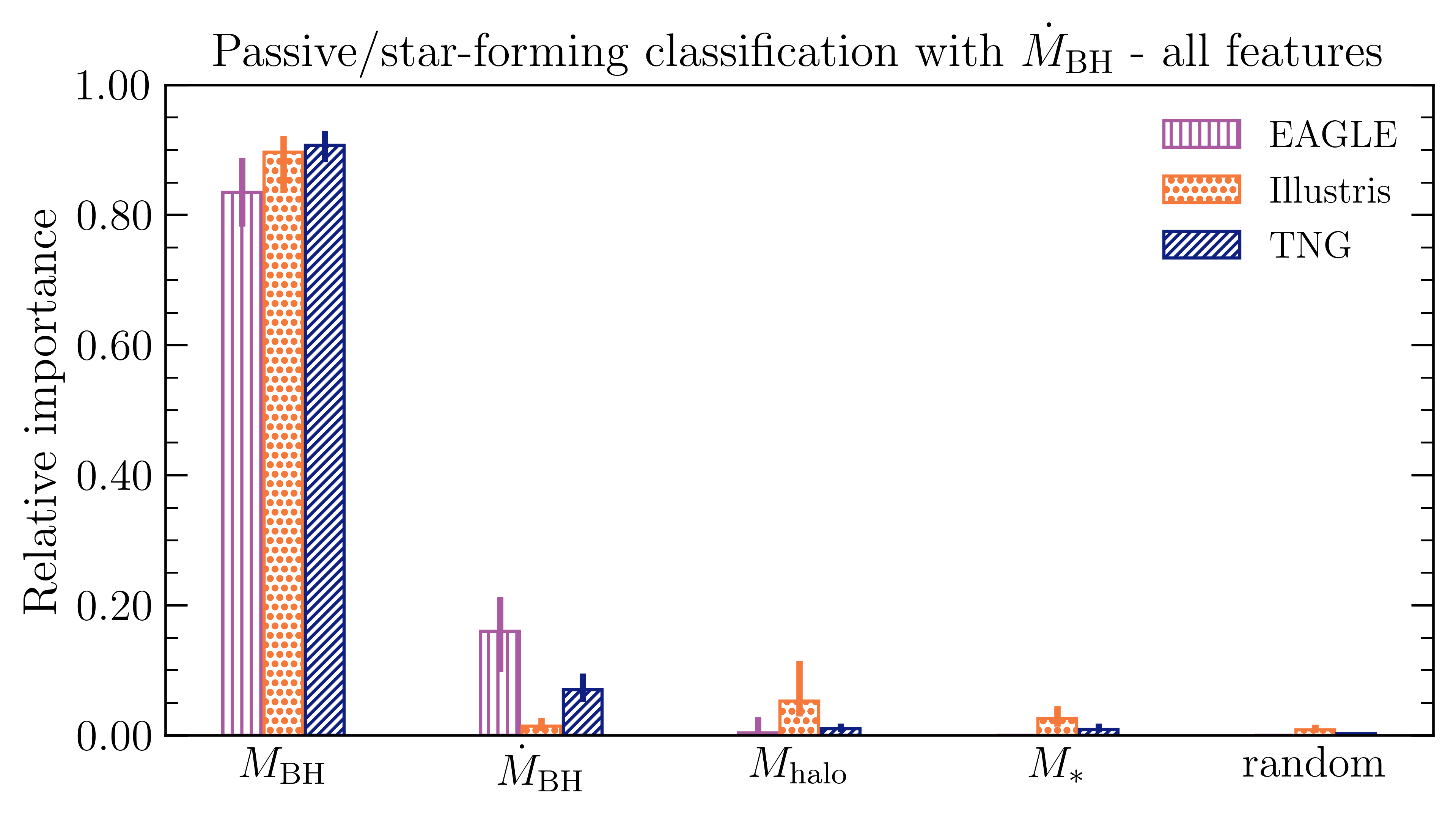}
	\caption{Same as the left panel in Fig.~\ref{fig:rf-combined} 
			 with black hole accretion
		 	 rate included as an additional parameter in the RF algorithm. It is apparent
			 that simulations predict a~negligible importance of redshift $z=0$ 
			 \mdot\ in quenching in comparison with \mbh\ and all remaining parameters.}
	\label{fig:rf-mdot}
\end{figure}

As a~final step in our Random Forest analysis we check how
well the current black hole accretion rate predicts
whether a~galaxy is quenched or star forming. 
Fig.~\ref{fig:rf-mdot} presents the results of our RF experiment
repeated with the \mdot\ parameter added to the input features 
for EAGLE, Illustris and TNG. The figure demonstrates that 
the simulations unanimously identify \mdot\ as holding
very little predictive power in contrast with \mbh\ in 
deciding whether a~galaxy is quenched or not. 
This result
shown in Fig.~\ref{fig:rf-mdot} carries important implications for 
the observational search for AGN quenching in action, discussed
further in Sec.~\ref{sec:L_AGN-mbh}.

The main conclusion we draw from the RF analysis is that all three 
different implementations of AGN feedback agree on a~strong connection 
between \mbh\ and quenching in central massive galaxies at redshift $z=0$.
At the same time, all three prescriptions for AGN feedback do not predict 
a strong relationship between quenching and the $z=0$ black hole accretion 
rate. This predicted importance of \mbh\ alone is very well reflected in 
the SDSS observations and holds true for all different \mbh\ calibrations tested. 
This striking agreement on \mbh\ dominance between simulations and observations 
suggests AGN feedback as a~viable mechanism behind transition towards quiescence
in massive central galaxies in the observable Universe. 


\subsection{Partial correlation analysis}
\label{sec:correlations}

We proceed with a~partial correlation analysis to explore 
connections among \mbh, \mhalo, \mstar\ and \mdot\ in the context 
of their correlations with sSFR. Partial correlations are 
a natural step down in complexity from the Random Forest analysis,
since the latter establishes relationships between the quenched/star-forming 
labels and all galactic parameters simultaneously while the former 
can only consider two variables at a time. The PCC calculation also
necessarily assumes that for any pair of variables one can be 
treated as a monotonic function of the other, while the RF makes no
a priori assumptions about the relationships between different features.
In other words, an RF can be 
thought of as an n-dimensional generalisation of a partial correlation,
which, due to its increased dimensionality, proves less straightforward
to visualise and interpret. We hence turn to the PCC analysis to further
understand and support our RF results with more conventional 
statistical tools.
 
Ultimately, there are two main scenarios we want to examine for the 
parameters at hand. Each of them can 
be either intrinsically tied to reduced sSFR and hence potentially 
causally connected to quenching, or parameters can be strongly 
correlated with one another and the predicted importance for quenching 
is a~consequence of this relationship. Partial correlation analysis allows 
us to differentiate between the two cases by comparing variables 
against each other. In a~given pair of parameters we can check 
how the correlation strength between one parameter and sSFR 
changes when the second parameter is controlled for. In this way 
we can cleanly separate out sSFR correlations of an incidental nature, 
since they disappear once partial correlation coefficients are calculated.
Additionally, this powerful tool
allows us to neatly visualise our results in two dimensions, showing 
how the quenching landscape changes in the plane defined by a~given 
pair of parameters \citep[e.g. see][]{Bluck19, Bluck20a, Piotrowska20}.

First, we calculate the Spearman's rank correlation coefficient 
($\rho_{\rm S}$) with sSFR for all variables among \mbh, \mdot, 
\mstar\ and \mhalo, checking to what extent specific SFR can be 
described as a~monotonic function of each of these parameters. 
As we explain above, the measurement of $\rho_{\rm S}$ for
each parameter can be affected by its correlation with another
galactic property, intrinsically connected with sSFR. In this case, 
a~parameter only marginally associated with sSFR can yield 
an inflated $\rho_{\rm S}$ owing to the correlation with said
fundamental property. In order to investigate this potential effect we 
divide our variables into three pairs. Each pair includes \mbh\ 
because it was identified as the most predictive quenching 
parameter in the RF analysis and hence is assumed as a variable
intrinsically connected to sSFR. In each pair we then calculate 
partial correlation coefficients (PCCs) and check to what extent 
the parameters  drive each other's $\rho_{\rm S}$ values.

A PCC measures the correlation strength between two variables 
(A\&B) at fixed third variable C. In so doing, it removes the impact 
on the correlation between A and B of the inter-correlation with C. 
Hence, for our purposes, we can, e.g. remove the dependence of 
sSFR on \mbh\ before assessing any residual dependence on \mstar\ 
and vice-versa. The PCC between quantities A and B, calculated when
controlling for C ($\rho_{\rm{ab|c}}$) is calculated by: 
\begin{equation}
\rho_{\rm{ab|c}}=\frac{\rho_{\rm{ab}}-\rho_{\rm{ac}}\cdot\rho_{\rm{bc}}}
{\sqrt{1-{\rho^2_{\rm{ac}}}}\sqrt{1-{\rho^2_{\rm{bc}}}}},
\end{equation}
where $\rho_{\rm{ab}}$ is the Spearman's rank correlation coefficient between 
A~and~B \citep{Kendall77}. In the case of the SDSS we also include $\omega_i$
weights from Eq.~\ref{eq:sdss-weight} in the calculation, in order to account 
for the Malmquist bias and the cut on disk-dominated galaxy inclination. 
We estimate uncertainties for all correlation coefficients by drawing 1000
bootstrap samples and selecting the 16th and 84th
percentiles of the resulting distributions. 

\begin{table} 
\caption{Orientation of the quenching vectors in Fig.~\ref{fig:corr-mdot}
			 through \ref{fig:corr-mhalo} as measured from the $y$-axis. 
			 All $\theta_{\rm Q}$ values are given in degrees $[^{\circ}]$.
			 In each row $\theta_{\rm Q}=90^{\circ}$ indicates quenching
			 solely connected to an increase in \mbh, while 
			 $\theta_{\rm Q}=0^{\circ}$ points towards quenching
			 only associated with an increase of \mdot, \mstar\ or \mhalo. }
\centering
\begin{tabular}{lcccc}
\toprule
 &   SDSS &  EAGLE &  Illustris &  TNG \\
\midrule 
\mbh--\mdot\ & -- & $116.2^{+1.4}_{-1.3}$ &  $103.0^{+1.2}_{-1.1}$  & $125.0^{+0.5}_{-0.5}$  \\
\rule{0pt}{4ex}   
\mbh--\mstar & $89.5^{+0.8}_{-0.8}$ & $63.9^{+6.9}_{-6.8}$  &  $120.2^{+0.6}_{-0.7}$ &  $84.4^{+3.1}_{-3.4}$  \\
\rule{0pt}{4ex}   
\mbh--\mhalo & $88.8^{+0.7}_{-0.7}$ & $109.8^{+2}_{-2}$ & $124.8^{+0.4}_{-0.4}$ & $117.2^{+0.6}_{-0.7}$ \\
\bottomrule
\end{tabular}
\label{tab:bluck-angles}
\end{table}

For each pair of parameters:  \mbh--\mdot, \mbh-\mstar\ and 
\mbh--\mhalo\ we also explore the variation in sSFR in the
two-dimensional plane defined by the parameters. By tracing
the direction of steepest decline in sSFR in the plane, we identify
an average `quenching vector' for each pair of variables. We calculate
its orientation following the prescription introduced
by \cite{Bluck20a}, treating both $\rho_{\rm pcc}$ values as vector 
components  in the parameter plane. The quenching vector is then 
inclined at $\theta_{\rm Q}$ to the $y$-axis:
\begin{equation}
\theta_{\rm Q} = \tan^{-1}\qty(\frac{\rho_{\rm xz|y}}{\rho_{\rm yz|x}}),
\end{equation}
where ${\rm x}=M_{\rm BH}$, ${\rm z}={\rm sSFR}$ and ${\rm y}$ is 
\mdot, \mstar\ and \mhalo\ in turn. We summarise all quenching vector 
orientations in Table~\ref{tab:bluck-angles} along with their uncertainties 
estimated in a~similar fashion to the errors on correlation coefficients.

\begin{figure*}[H]
\includegraphics[width=\textwidth]{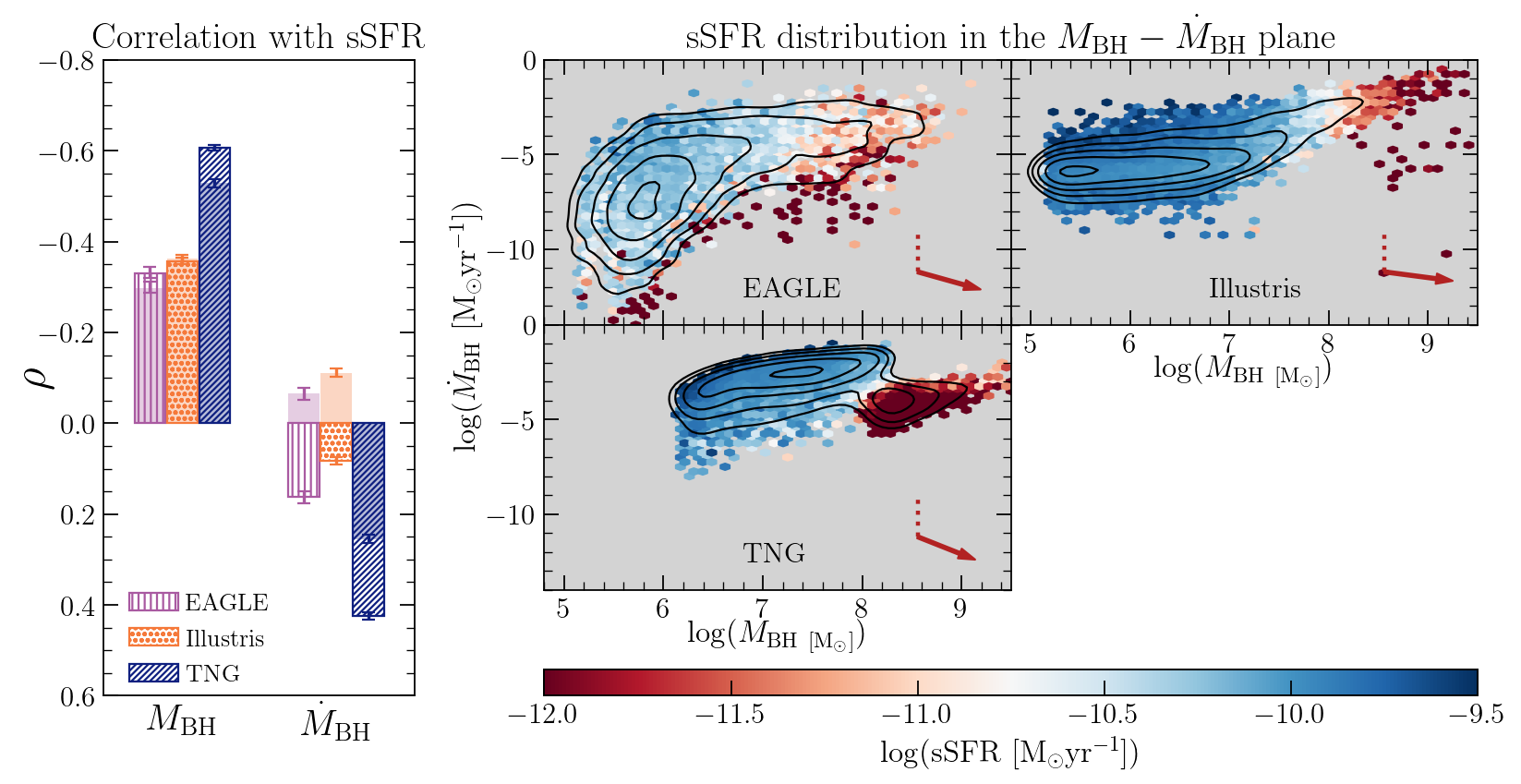} 
    \caption{Left: Spearman's rank correlation coefficient $\rho_{\rm S}$ 
    		(full bars) and partial correlation coefficient $\rho_{\rm pcc}$ 
    		(hatched bars) between sSFR, \mbh\ and \mdot. 
    		 It is apparent that \mdot\ 
    		anti-correlates with quenching in the simulations once its
    		dependence on \mbh\ is accounted for.
    		Right: \mbh--\mdot\ plane colour-coded by
    		the median sSFR in a~given hexagonal bin. Red arrows
    		indicate the average quenching direction and are oriented at 
    		$\theta_Q=116^\circ,\ 103^\circ$ and $125^\circ$ for EAGLE, 
    		Illustris and TNG respectively (see Table~\ref{tab:bluck-angles}). 
    		Black contours show the
    		density of objects in the plane. In all panels the sSFR gradient changes 
    		horizontally or slightly towards the bottom right corner, 
    		indicating that quenching requires primarily an increase in \mbh\ 
    		and a~slight decrease in \mdot.}
    \label{fig:corr-mdot}
\end{figure*}

\subsubsection{$L_{\rm AGN}$ cannot reveal AGN feedback quenching in action}
\label{sec:mdot-no-corr}

\begin{figure*}
\includegraphics[width=\textwidth]{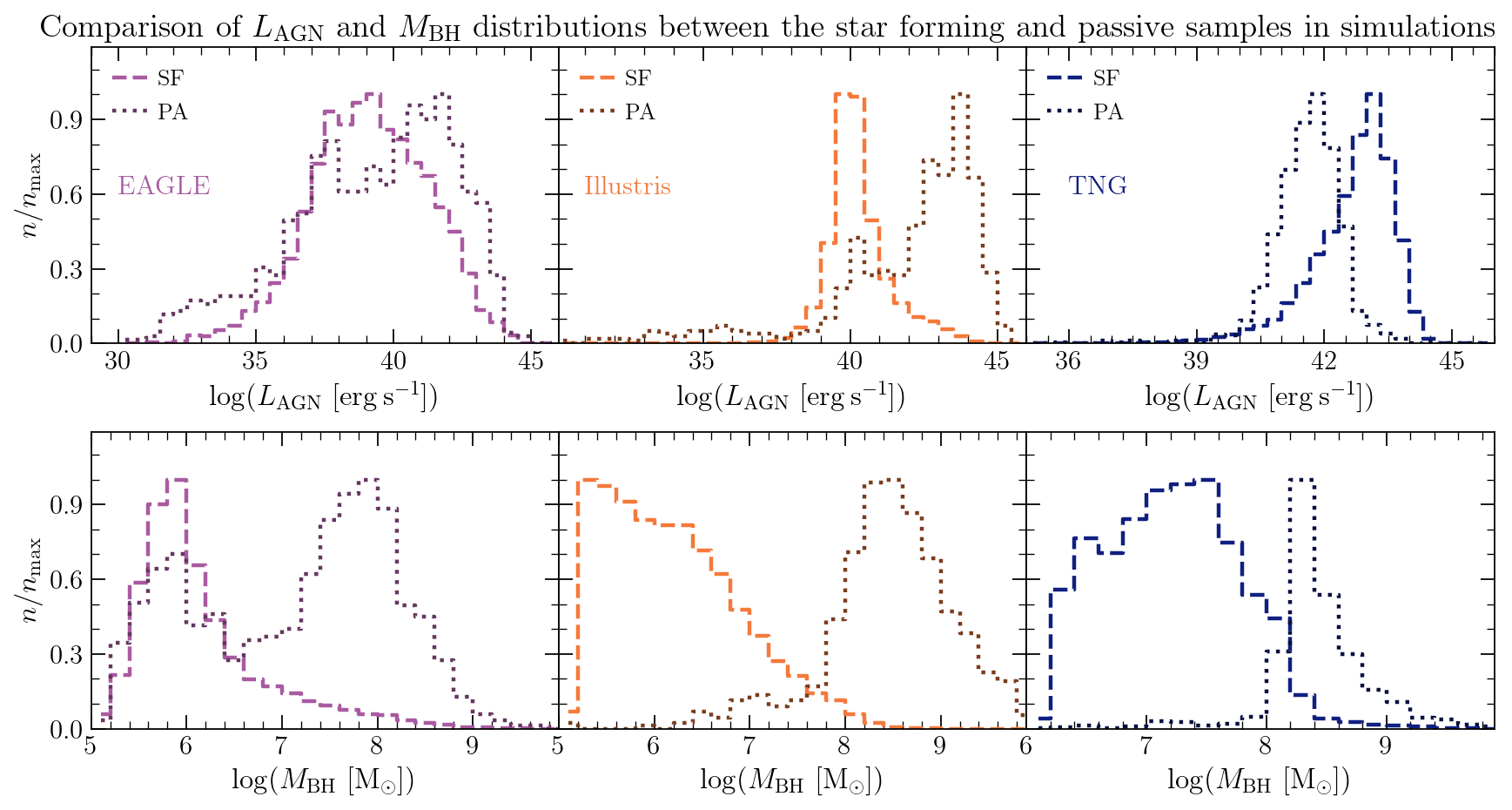}
	\caption{Comparison of distributions in bolometric AGN luminosity (top) and black hole mass
				  (bottom) between the star forming (SF) and passive (PA) galaxy populations
				  in the simulations. All bottom panels show a~clear contrast between \mbh\ distributions,
				  where PA systems have high \mbh, while their
				  SF counterparts host smaller black holes. 
				  In $L_{\rm AGN}$ the picture is different: in TNG higher AGN 
				  luminosities are linked to active star formation, Illustris shows an
				  exactly opposite trend, while in EAGLE $L_{\rm AGN}$ is insensitive to the
				  star formation state of the galactic host. }
	\label{fig:LAGN-MBH}
\end{figure*}

\begin{figure}
\centering
\includegraphics[width=0.9\columnwidth]{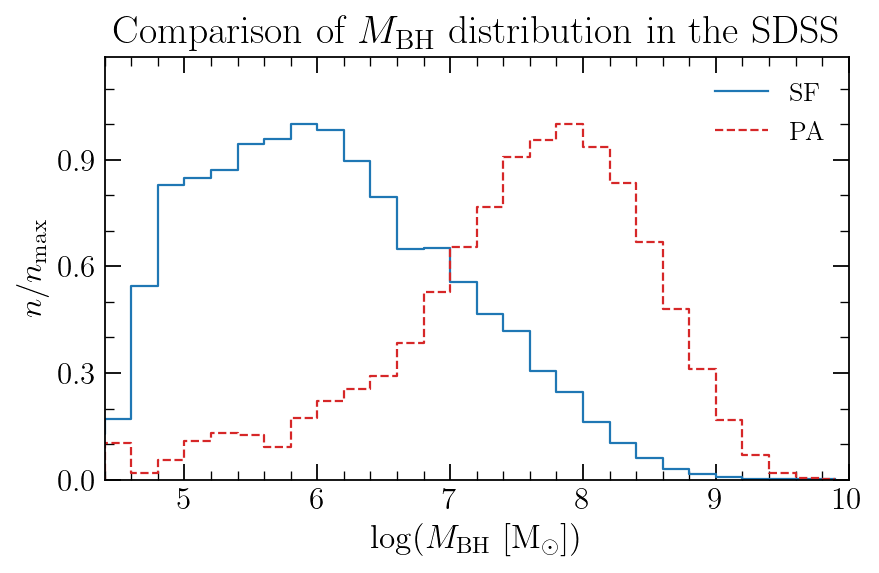}
	\caption{Comparison of black hole mass distributions
				  between the star forming (SF) and passive (PA) 
				  galaxy populations in the SDSS. The histograms
				  are weighted to account for the Malmquist bias
				  and the inclination correction. Passive galaxies
				  are clearly associated with more massive black
				  holes than their star-forming counterparts. }
	\label{fig:sdss-mbh}
\end{figure}

We first compare \mbh\ against \mdot\ in Fig.~\ref{fig:corr-mdot}.
Full bars in the left panel show the Spearman's rank correlation
coefficient between the parameters and sSFR, while the hatched ones
present their respective partial correlation coefficients, evaluated
at fixed values of the remaining parameter.
Please note that the vertical axis is inverted such that upright bars
reflect a~positive correlation with quenching (a negative correlation 
with sSFR). At first glance, Fig.~\ref{fig:corr-mdot} informs us about 
little to no measurable association between the current instantaneous
accretion rate of a~black hole and the quenching stage of its galactic host.
It is also apparent in all simulations that once the connection between 
\mdot\ and \mbh\ is accounted for, an \textit{increased 
accretion rate at given black hole mass} is associated with 
an \textit{increase in sSFR}, pointing towards a co-fuelling scenario
in which star formation and black hole growth coincide in time.

\begin{figure*}
\includegraphics[width=\textwidth]{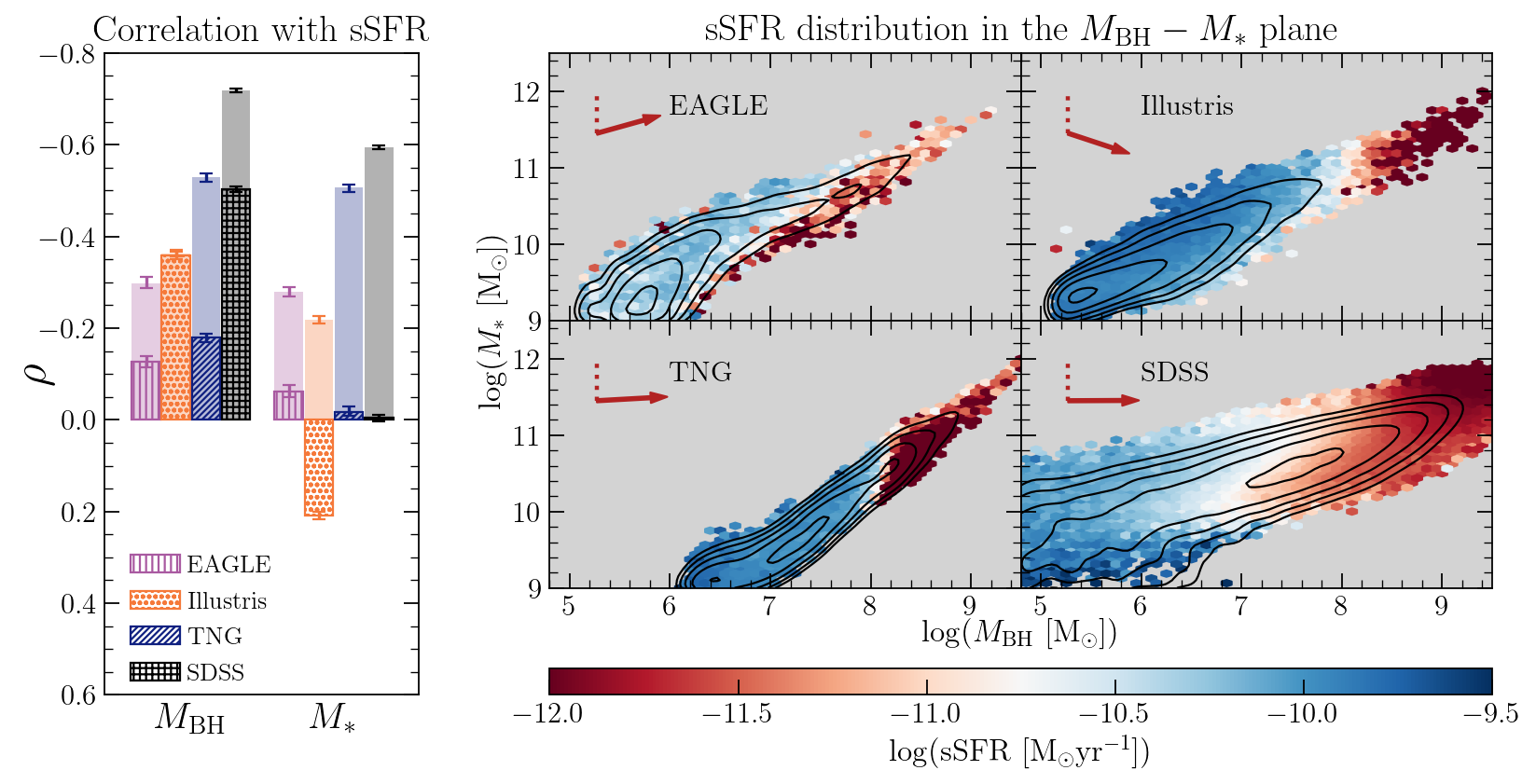} 
    \caption{Identical in structure to Fig.~\ref{fig:corr-mdot} but with \mstar\
    		 replacing \mdot. Left: The hatched bars demonstrate that \mstar\ is 
    		 barely associated with quenching once the \mbh--\mstar\ 
    		 relationship is controlled for in the partial correlation.
    		 Right: The gradient in sSFR is primarily horizontal in all panels, 
    		 illustrating that \mbh\ is more fundamentally linked to quenching
    		 than \mstar\ in both observations and simulations.}
    \label{fig:corr-mstar}
\end{figure*}

In the right panel of Fig.~\ref{fig:corr-mdot} we show the distribution of 
galaxies in the $M_{\rm BH}-\dot{M}_{\rm BH}$ plane, colour-coded by sSFR.
In all three panels a~negative gradient in sSFR is primarily driven by 
an increase in \mbh\ and, to a lesser extent, by a~decrease in \mdot. 
In Illustris and TNG low sSFRs are associated exclusively with high 
black hole masses, while in EAGLE there exists a small population
of low-\mbh\ quenched galaxies. The qualitative interpretation of 
sSFR gradients is confirmed 
by $\theta_{\rm Q}$ values of $116.2^{+1.4}_{-1.3}$ in EAGLE, 
$103.0^{+1.2}_{-1.1}$ in Illustris and $125.0^{+0.5}_{-0.5}$ in TNG, which orient
all quenching vectors towards the bottom right corners of their
respective plots. It is important to point out, however, that $\theta_{\rm Q}$ 
values are much closer to $\theta_{\rm Q}=90^{\circ}$ than 
$\theta_{\rm Q}=180^{\circ}$, quantitatively supporting the dominance of
\mbh\ over \mdot\, as seen previously in the RF experiment.

To better understand the observable consequences of AGN feedback 
we show $L_{\rm AGN}$ and \mbh\ distributions for passive (PA) and
star forming (SF) galaxies in Fig.~\ref{fig:LAGN-MBH}. We estimate
AGN luminosities from the black hole accretion rates via 
$L_{\rm AGN}= \eta c^2 \dot{M}_{\rm BH}$, where we take $\eta=0.1$, 
which is a~value commonly accepted in the literature as an average 
radiative efficiency of AGN in the local Universe 
\citep[e.g][]{Shankar09, Bluck11, Davis11, Wu13}. We also
include \mbh\ distributions in the SDSS 
in Fig.~\ref{fig:sdss-mbh}, for a comparison with the 
theoretical predictions.

In the bottom panels of Fig.~\ref{fig:LAGN-MBH} we see the same 
pattern repeated in all three simulation suites, which is also
present in Fig.~\ref{fig:sdss-mbh} in the SDSS. Namely, 
passive galaxies host more massive black holes than the star 
forming population, regardless of the subgrid prescription for 
AGN feedback in the simulations. This feature explains why \mbh\ is found to be
an efficient route to predicting a star forming/passive classification
in the Random Forest. 

In the top panels of Fig.~\ref{fig:LAGN-MBH} 
the situation is 
strikingly different, since all three simulations predict different 
behaviour in $L_{\rm AGN}$. In EAGLE the star forming and quiescent
galaxy populations show essentially no difference in the instantaneous 
AGN activity, as demonstrated by the overlapping $L_{\rm AGN}$ histograms. 
In Illustris passive galaxies are associated with higher $L_{\rm AGN}$, 
while TNG presents an entirely opposite trend where SF galaxies host 
more luminous AGN.
The stark contrast between the top and bottom panels in 
Fig.~\ref{fig:LAGN-MBH} once again demonstrates how different 
implementations of AGN feedback lead to a~ubiquitous signature in 
\mbh, while the prediction in \mdot\ ($L_{\rm AGN}$) is far less clear.
As demonstrated in Fig.~\ref{fig:sdss-mbh}, the SDSS observations
match the theoretical prediction in black hole mass, showing a clear
separation in \mbh\ distribution between the SF an PA galaxy
populations. However, the overlap between the two distributions 
is larger than in the simulations, most likely owing to the 
uncertainty associated with the \mbh\ estimation in the SDSS.

The comparisons we draw between \mbh\ and \mdot\ in 
Figs.~\ref{fig:corr-mdot},~\ref{fig:LAGN-MBH}~\&~\ref{fig:sdss-mbh} 
strengthen our conclusions 
reached in the RF analysis. The PCC calculation demonstrates how the connection
between quenching and black hole accretion rate at redshift $z=0$ 
is negligible in comparison to the relationship with black hole mass. 
This statement is true for all three different implementations of 
AGN feedback in EAGLE, Illustris and TNG, which vary in how much 
feedback energy is generated by the black holes and how that energy 
couples to their environment. This variation in subgrid prescriptions
leaves an imprint on the $L_{\rm AGN}$ distributions, visibly 
separating the models based on this observable. Most importantly, 
however, we need to remember that all these simulations quench 
their central galaxies solely via the AGN feedback.
Hence, our results demonstrate that a dependence of quenching on 
the instantaneous accretion rate at $z=0$ is not a necessary 
consequence of AGN quenching. In contrast, the strong
dependence on \mbh\ is necessary, given the ubiquitousness of 
this trend in three very different simulations.

\subsubsection{Connection to black hole mass drives high correlation with quiescence in \mhalo\ and \mstar}

\begin{figure*}
\includegraphics[width=\textwidth]{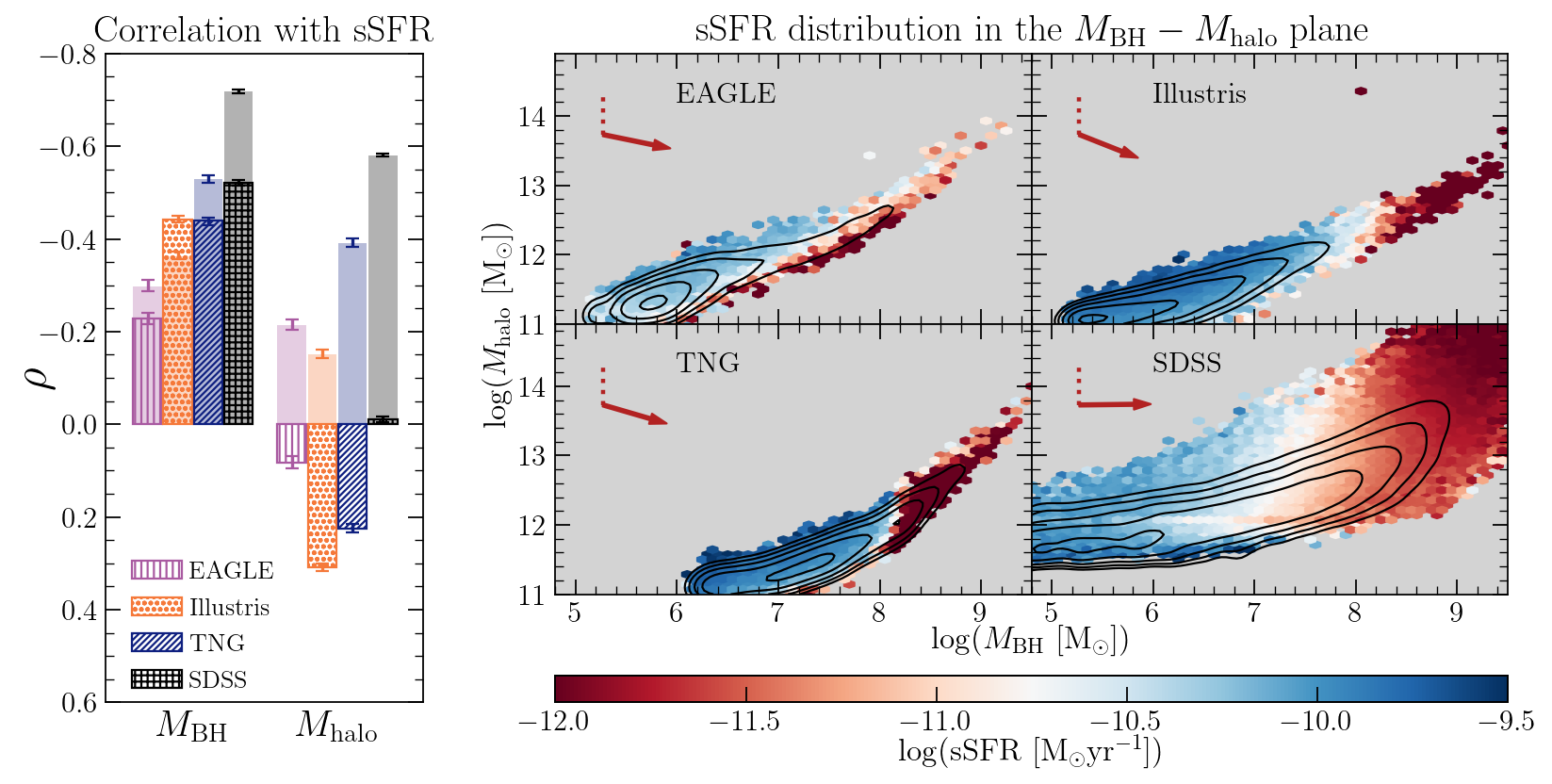} 
    \caption{Identical in structure to Fig.~\ref{fig:corr-mdot} but with \mhalo\ 
    		replacing \mdot. Left: Hatched bars show that more massive halos result 
    		in more star formation once the correlation between \mbh\ and \mhalo\ 
    		is taken into account in simulations. In the SDSS the result is consistent with no 
    		correlation, rendering the intrinsic connection between sSFR and 
    		\mhalo\ negligible once the inter-correlation with \mbh\ is controlled for.
    		Right: the sSFR gradient points towards the bottom right corner in
    		simulations and horizontally in the SDSS, indicating that \mbh\ and
    		not \mhalo\ is the fundamental quenching parameter.}
    	\label{fig:corr-mhalo}
\end{figure*}

In Fig.~\ref{fig:corr-mstar} we present a~comparison between \mbh\ and 
\mstar, following the structure previously seen in Fig.~\ref{fig:corr-mdot}. 
When we calculate the Spearman's rank correlation coefficient between 
sSFR and both \mbh\ and \mstar, we obtain similarly high correlation
values for both parameters in all data sets.
However, when we account for 
the \mstar--\mbh\ connection, the sSFR-\mbh\ correlation remains 
relatively high, while the sSFR-\mstar\ PCC is drastically changed. 
It becomes negligible in SDSS, EAGLE and TNG, while for Illustris the
correlation even changes sign (at fixed \mbh\ galaxies are more star-forming
with increasing \mstar\ in the suite). This behaviour is highlighted
in the parameter plane in the right panel, which shows a~negative
sSFR gradient primarily driven by an increase in \mbh\ across all data sets.
Like in the previous case of \mdot, the transition is most abrupt in TNG,
where all objects beyond $\log(M_{\rm BH}/{\rm M_\odot})\sim8$ have 
log(sSFR/${\rm yr^{-1}}$) $\sim -11$ and below. Our visual gradient inspection
is confirmed by the orientation of the quenching vectors, which
point at $\theta_{\rm Q}\sim 90^{\circ}$ in SDSS and TNG and within
$\pm 45^{\circ}$ from $\theta_{\rm Q}=90^{\circ}$ in Illustris and EAGLE.

This result, together with our previous RF analysis, confirms that the
observed connection between galaxy stellar mass and quenching
results largely from the black hole-galaxy co-evolution, rather than from 
quenching being a~function of the stellar mass of the system 
(e.g. due to supernova feedback). Quiescence in massive centrals 
is primarily caused by the black holes in simulations
and hence \mbh\ is the 
parameter determining whether a~given galaxy is likely to be quenched.
This result is identical to that found by \mbox{\cite{Bluck16}} 
for SDSS and Illustris with the use of area statistics in quenched fraction 
calculations. In this work we update the methodology by using different
statistical tests in tandem with machine learning analysis, as well as 
exploring a~broader range of theoretical predictions by including EAGLE
and TNG in our investigations.

Finally, we show a~comparison between \mbh\ and \mhalo\ in 
Fig.~\ref{fig:corr-mhalo}. Like in the case of \mstar, both parameters 
are similarly strongly correlated with sSFR, however the difference
in $\rho_{\rm S}$ is more pronounced, in favour of black hole mass.
When we account for the \mhalo--\mbh\ connection, the correlation
changes sign in all simulations and almost vanishes in the SDSS.
This implies that at a fixed black hole mass, increasing the dark matter 
halo mass contributes to an increase in star formation activity
of a~galaxy in the simulations. This is likely a consequence of
higher cooling rates associated with more massive haloes - the
original `problem' mitigated by AGN feedback which is still present
as a secondary effect, once the primary role of black holes is
accounted for. In this scenario increasing halo mass acts to resist 
quenching rather than aiding it - hence the virial shocks
are not effective at keeping halo gas hot and buoyant. 

In the observations our result suggest no contribution from \mhalo\ 
to quenching, however a weak trend in agreement with the simulations
is found in the SDSS by \cite{Bluck20a}. 
The difference between this work and our analysis is in the use of
$\Delta{\rm sSFR}$ in \cite{Bluck20a} instead of sSFR for the 
correlation calculations. Nonetheless this does not influence our main 
conclusion that \mbh\ is overwhelmingly more important
than \mhalo\ for the intrinsic galaxy quenching in the observed 
Universe, in agreement with cosmological simulations and this prior study.

The dominance of \mbh\ over \mhalo\ is further demonstrated in the 
parameter plane, where the negative gradient in sSFR is, on average, 
pointing towards the bottom right corner in all simulation panels. 
In the SDSS the gradient
appears more horizontal with a~hint of increase in sSFR for increasing
\mhalo\ at fixed \mbh. These qualitative observations are supported
by the calculated orientation of the quenching vectors, which have
$\theta_{\rm Q} \in (90^{\circ}, 135^{\circ})$ in all simulations. 
In the SDSS the arrow is almost precisely aligned with the $x$-axis, 
i.e. $\theta_{\rm Q}\approx 90^{\circ}$.

The conclusions we draw from Fig.~\ref{fig:corr-mhalo}, together 
with the RF analysis, reject one potential halo heating mechanism,
which was proposed as a~means to quench massive central galaxies.
This solution assumes that cold flows in massive haloes 
(\mhalo$\gtrsim 10^{12}{\rm M}_{\odot}$) would experience shocks
as they fall through the deep halo potential wells. They would then 
heat up and prevent cooling required for future star formation 
in a~process dubbed virial shock heating \citep{Dekel06}. 
In such a~scenario cold gas supply would be ceased for halo masses 
above a~critical shock mass and hence \mhalo\ could serve as an 
excellent classification criterion for quenched galaxy population. 
This is not the case according to our RF results, where the importance
of \mhalo\ is dwarfed by \mbh. 
Similarly, one would expect halo mass to completely determine
the quenching state of a~galaxy at fixed black hole mass, 
meanwhile our partial correlation analysis demonstrates that the 
opposite behaviour is the case, i.e.~that there is a very strong
dependence of quenching on \mbh\ at fixed \mhalo . This
result is most pronounced in the SDSS, where the \mhalo\ $\rho_{\rm pcc}$
almost vanishes. 

In all three simulations the PCC result suggests
that at a~fixed black hole mass more massive haloes experience 
elevated levels of star formation. This relationship could be potentially 
explained by increased Bremsstrahlung cooling rates in denser gas,
whose density increases with increasing halo mass \citep{Fabian12}.
Therefore we conclude that both in simulations and observations 
virial shock heating does not appear to be a~viable quenching 
mechanism in massive central galaxies. With this statement we
also confirm the results found previously by \cite{Bluck16} and 
\cite{Bluck20a}, reached with different 
statistical methods for Illustris and the SDSS.
More importantly, in this work we investigate theoretical
predictions for the partial correlations between \mbh, \mhalo\
and sSFR in two additional simulation suites: EAGLE \& TNG,
finding that these are in agreement with trends predicted
in Illustris.

In summary, throughout our correlation analysis and RF classification
we extract testable predictions for observable consequences of central
galaxy quenching in three state-of-the art cosmological simulations.
Since EAGLE, Illustris and TNG all quench their centrals using AGN, 
we effectively test predictions for black hole feedback quenching 
models against the observed Universe in the SDSS. We find that \mbh\ 
is the most predictive parameter for determining
whether a~galaxy is quenched or not, both in the simulations and the 
observations. Hence, so far in this study, we find 
a close agreement on the predicted observable consequences
of AGN feedback among \textit{all three simulations}, in which the
parameter most predictive of quenching in massive centrals
is the same, regardless of the implementation of AGN feedback.
We also find that \mbh\ shows the
strongest connection to quenching among all parameters, when explicitly 
compared with other variables through partial correlations. The vanishing
PCCs between sSFR, \mhalo\ and \mstar\ suggest that the transition 
to quiescence is not strongly influenced by virial shocks or
supernovae, both in the simulations and observations. 
Most strikingly, however, we learn that
cosmological simulations unanimously predict a weak dependence
of quenching on redshift $z=0$ \mdot, in contrast with its strong connection 
with \mbh.

\begin{figure*}
	\begin{subfigure}[b]{\textwidth}
		\centering
	 	\includegraphics[width=0.666\textwidth]{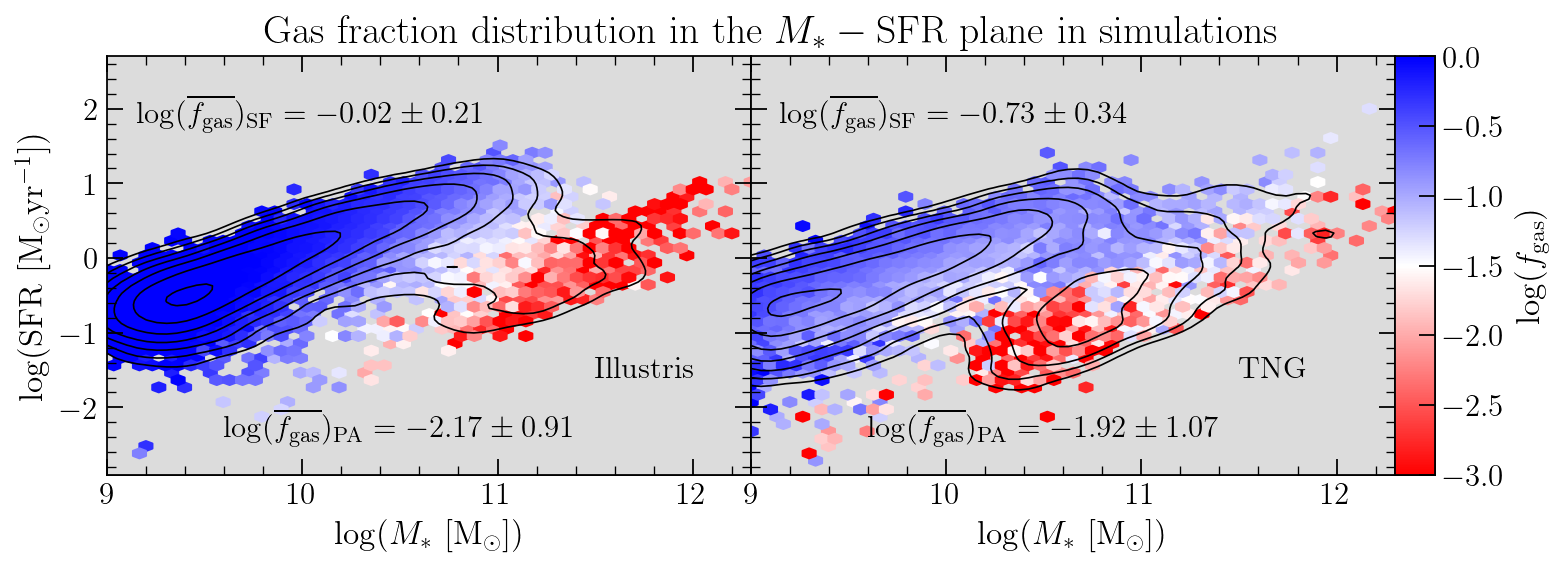} 
	 	\caption{raw result in the simulations}
	 	\label{fig:msfr-fgas-raw}
	\end{subfigure}
	\\
	\begin{subfigure}[b]{\textwidth}
		\includegraphics[width=\textwidth]{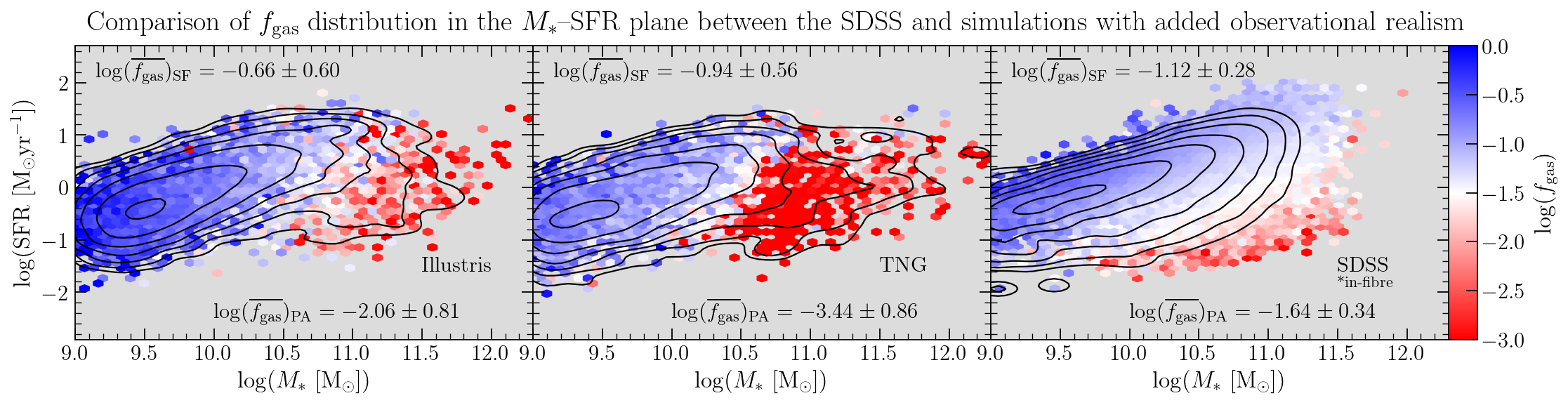} 
    	\caption{simulations with added observational realism}
  		\label{fig:msfr-fgas-sdss} 
  	\end{subfigure}
  	\caption{Comparison of \fgas\ distribution in the \msfr\
    				plane between simulations and the SDSS. 
    				$(\overline{f_{\rm gas}})_{\rm SF}$ and 
    				$(\overline{f_{\rm gas}})_{\rm PA}$
    				are median gas fractions calculated for the star-forming and 
    				passive galaxy populations. Black contours show data 
    				distribution in the plane and for SDSS they include 
    				$\flatfrac{1}{V_{\rm max}}$ and inclination corrections.   				
    				The raw values in (\subref{fig:msfr-fgas-raw})
    				show that $f_{\rm gas}$ is lower in passive galaxies for 
    				both simulations, however the magnitude of this difference 
    				is larger in Illustris. In (\subref{fig:msfr-fgas-sdss}) we mimic 
    				the combined effect of fibre apertures, S/N cut on the 
    				${\rm H}\alpha$ emission line and random scatter on
    				all presented observables in the simulations to allow for 
    				a~fairer comparison with observations. All panels show 
    				qualitatively similar trends in \fgas, which decreases 
    				between the star forming and passive populations. 
    				However, this transition is more abrupt and 
    				spans a~broader range of values in simulations than in the 
    				SDSS. Adding observational realism to simulated galaxies 
    				primarily softens the trend in \fgas\ for Illustris and causes 
    				the scatter in the main sequences to increase in both 
    				simulation suites.}
    \label{fig:msfr-fgas} 		 
\end{figure*}

\begin{table}
\caption{Median differences in \fgas\ and SFE between the 
		 PA and SF populations in Figs.~\ref{fig:msfr-fgas}~\&~\ref{fig:msfr-sfe}
		 in units of [dex].}
\centering
\begin{tabular}{m{15mm}cc}
\toprule
 & $\log(\overline{f_{\rm gas}})_{\rm PA} - \log(\overline{f_{\rm gas}})_{\rm SF} $ & $\log(\overline{\rm SFE})_{\rm PA} - \log(\overline{\rm SFE})_{\rm SF} $ \\
\midrule
\rule{0pt}{0ex}
SDSS & $-0.52 \pm 0.44$ & $-1.02 \pm 0.61$ \\
\rule{0pt}{3ex}   
Illustris raw & $-2.15 \pm 0.93$ & $0.07 \pm 0.84$ \\  
\rule{0pt}{0ex}
Illustris \newline + realism & $-1.40 \pm 1.01$ & $0.55 \pm 0.95$ \\
\rule{0pt}{3ex}   
TNG raw & $-1.19 \pm 1.12$ & $-1.83 \pm 0.38$ \\
\rule{0pt}{0ex}
TNG \newline + realism & $-2.50 \pm 1.03$ & $-2.01 \pm 0.55$ \\

\bottomrule
\end{tabular}
\label{tab:delta-pa}
\end{table}

\subsection{Gas content and quenching in the SDSS, Illustris and TNG}
\label{sec:gas}

\begin{figure*}
	\begin{subfigure}[b]{\textwidth}
		\centering
	 	\includegraphics[width=0.666\textwidth]{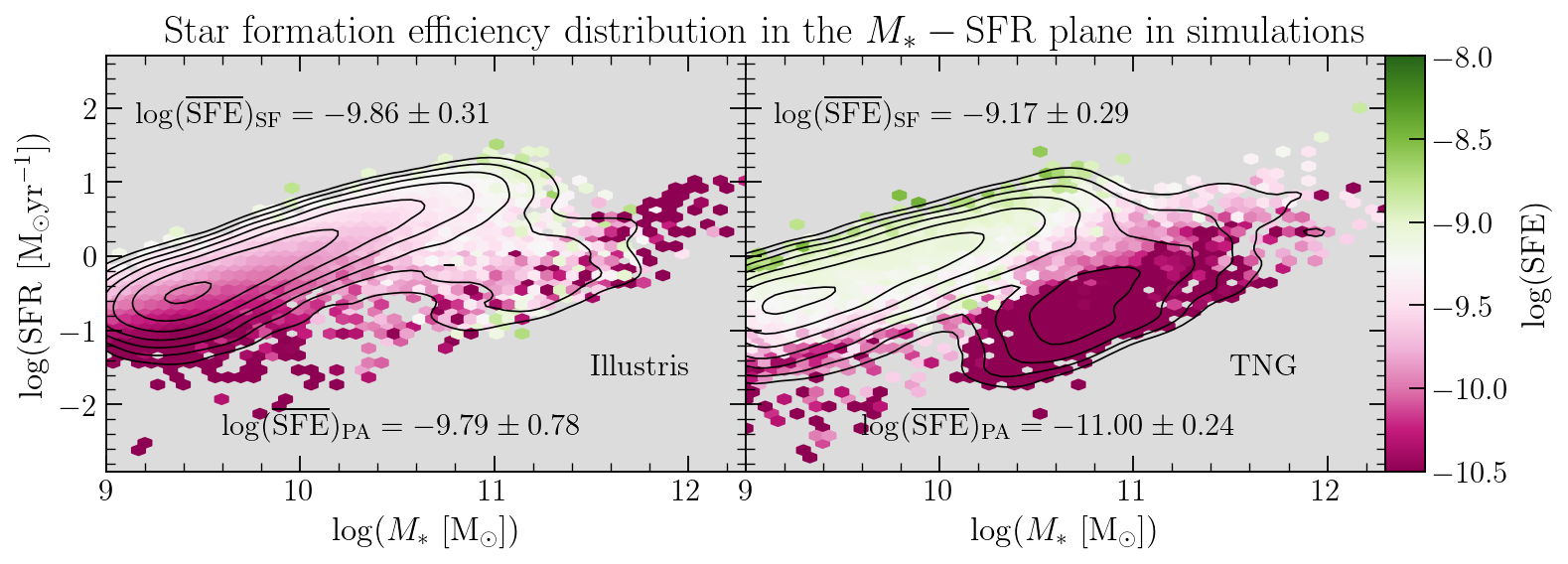} 
	 	\caption{raw result in the simulations}
	 	\label{fig:msfr-sfe-raw}
	\end{subfigure}
	\\
	\begin{subfigure}[b]{\textwidth}
		\includegraphics[width=\textwidth]{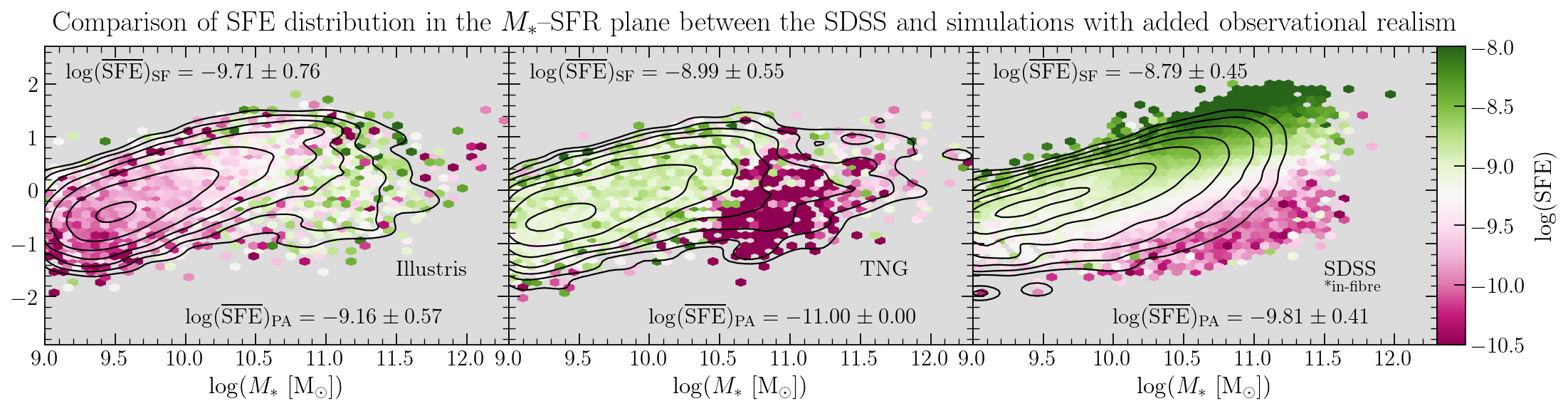} 
    	\caption{simulations with added observational realism}
  		\label{fig:msfr-sfe-sdss} 
  	\end{subfigure}
    \caption{Analogous to Fig.~\ref{fig:msfr-fgas}, 
    				colour-coded by the median SFE instead of $f_{\rm gas}$.
    				(\subref{fig:msfr-sfe-raw}) simulations disagree on
    				the change in SFE between the star forming and passive populations.
    				In Illustris star formation efficiency is slightly higher in the quenched objects
    				than on the main sequence, while in TNG the opposite is true. The
    				latter also shows a much more dramatic difference between the star-forming
    				and quenched populations, in which the median SFE drops by almost 2~dex. 
    				(\subref{fig:msfr-sfe-sdss}) The addition of observational realism
    				has the effect of further highlighting the opposite trends present in the simulations.
    				In SDSS the SFE mildly increases with increasing mass along 
    				the MS and decreases rapidly perpendicular to it. It is apparent
    				that the SFE trend present in TNG agrees qualitatively with the observations,
    				in contrast with Illustris, where it is inverted.}
    	\label{fig:msfr-sfe}
\end{figure*}

In this section we characterise the potential mode of operation 
of AGN feedback quenching, looking for its connection to the molecular
gas content of galaxies.We focus on Illustris, TNG and SDSS data sets 
only, omitting EAGLE due to the lack of publicly available molecular 
gas mass estimates in this suite. We also note that, as highlighted 
by the quenched fractions and sSFR relations in Sec.~\ref{sec:qf}, 
the quenching trends predicted by EAGLE disagree with 
those seen in the SDSS, especially at high \mbh. Hence, a detailed 
investigation of how this AGN feedback mode operates is not 
strictly necessary in our work. In the SDSS we limit our study to 
non-AGN dominated galaxies which meet our S/N criteria on 
emission line fluxes (see Sec.~\ref{sec:sample-selection}) 
in order to reliably estimate \hmol\ masses from dust reddening 
of optical spectra. In simulations we check for the potential
influence of non-AGN selection by removing objects with \mdot\
falling in the top 20\% of the whole population. We find that
this selection does not impact any trends in the simulations, 
however it significantly depletes the already limited sample of
quiescent galaxies in Illustris.
 
In our analysis we focus on two quantities of interest - 
molecular gas
fraction $(f_{\rm gas} = M_{\rm H2}/M_\ast)$ and star formation
efficiency (${\rm SFE} = \flatfrac{\rm SFR}{M_{\rm H2}}$). Multiplied
together, these two yield the specific star formation rate of 
a~galaxy:
\begin{equation}
{\rm sSFR} = \flatfrac{\rm SFR}{M_\ast}=\flatfrac{M_{\rm H2}}{M_\ast} \times \flatfrac{\rm SFR}{M_{\rm H2}}
= f_{\rm gas} \times {\rm SFE}
\end{equation}
and hence allow us to decompose the reduction in sSFR, i.e.~quenching, 
into fuel-driven and and efficiency-driven evolution. In order to 
decrease \fgas\ one needs to invoke processes
which act to reduce molecular gas reservoirs, like e.g.~removal 
through AGN-driven outflows or heating of the CGM to prevent 
replenishment. Conversely, in order to lower the SFE (or, equivalently, increase 
the molecular gas depletion time 
$\tau_{\rm dep}=\flatfrac{1}{\rm SFE}$) one can imagine a mechanism
which does not affect the amount of gas but rather reduces 
the efficiency with which it collapses to form new stars, like 
e.g.~injection of turbulence into the ISM \citep[e.g.][]{Krumholz05}. 

By examining the changes in \fgas\ and SFE we can thus
characterise the observational consequences of different AGN 
feedback modes as implemented in Illustris and TNG. A fair 
comparison between predicted trends and those observed in the SDSS
can then inform us about the potential physical mechanisms
responsible for quenching in the observed Universe and highlight 
potential areas for improvement in the implemented models.

\subsubsection{A universal trend: gas fractions decrease with increasing black hole mass}
\label{sec:gas:fgas}

In Fig.~\ref{fig:msfr-fgas} we present the evolution \fgas\
across the \msfr\ plane
in the simulations and observations. Galaxies are grouped into hexagonal
bins in the plane and each bin is colour-coded by its median \fgas, 
while black contours indicate the density of objects in each panel. 
The SDSS contours include $\flatfrac{1}{V_{\rm max}}$ and inclination 
corrections. The figure also lists the median \fgas\ in 
the passive and star-forming galaxy populations labelled as 
$(\overline{f_{\rm gas}})_{\rm PA}$ and $(\overline{f_{\rm gas}})_{\rm SF}$
respectively. Analogously to Fig.~\ref{fig:ms} very low SFR objects
are redistributed to locations corresponding to upper limits on SFR
from \cite{Brinchmann04}, in order for us to present all simulated
galaxies in the plane. We would also like to remind the reader that all
results presented in Sec.~\ref{sec:gas} use the \cite{Gnedin14} model
for the HI/H2 transition, out of the four models made publicly available
by the Illustris and TNG teams. Although we find differences in the 
${\rm H_2}$ content on a galaxy-by-galaxy basis, the statistical 
behaviour of the galaxy population is not affected by our choice
of transition model. We further describe the models and show the 
robustness of our result to this choice in Appendix~\ref{sec:appendix:h2-models}.

In Fig.~\ref{fig:msfr-fgas-raw} we focus on the raw result in the simulations first, 
characterising a theoretical prediction for idealised molecular gas observations covering
the whole spatial extent of galaxies. Both panels in Fig.~\ref{fig:msfr-fgas-raw} 
show a~clear bimodality in colour, where star forming (SF) galaxies are 
associated with higher gas fractions than their passive (PA) counterparts.
These differences are most striking in Illustris, as highlighted by the
colour ranges and median change in \fgas\ of 2.15~dex between the populations.
This suite also shows a~decreasing trend in \fgas\ with increasing
\mstar\ across the MS, which is very mild in comparison with the drop
in \fgas\ away from the MS. TNG shows similar qualitative trends 
in a~narrower range of values with a~median difference between
populations of 1.19~dex. The passive sequence also shows significant
scatter in \fgas\ at high \mstar, in part driven by low sample statistics.
Overall, we note that the two simulations are in good qualitative agreement 
with each other but that Illustris exhibits a~steeper declining trend
in gas fraction between the star forming and passive galaxy populations.

\begin{table}
\caption{Standard deviation $\sigma_X$ of random error in quantity
$X$ in the SDSS. }
\centering
\begin{tabular}{c|ccccccc}
\toprule
$X$ & $M_{\rm \ast, tot}$ & $M_{\rm \ast, fib}$ & ${\rm SFR}_{\rm tot}$ 
& ${\rm SFR}_{\rm fib}$ & $M_{\rm H2}$ & \mhalo & \mbh \\
\midrule
$\sigma_X$ & 0.15 & 0.10 & 0.30 & 0.30 & 0.40 & 0.50 & 0.50 \\
\bottomrule
\end{tabular}
\label{tab:sigmax}
\end{table}

In Fig.~\ref{fig:msfr-fgas-sdss} we compare the distribution of \fgas\
in the simulations against the SDSS. In order to bring all data sets to
a comparable footing, we apply several layers of observational realism
to galaxies in Illustris and TNG. As we explain in detail in 
Appendix~\ref{sec:sim-realism}, we mimic the effect of an SDSS fibre
aperture by focusing on central regions in the galaxies and integrating
$M_{\rm H2}$, \mstar\ and SFR profiles up to their assigned fibre radii.
Depending on the simulated galaxy mass, such mock fibre radius is 
randomly drawn from a~distribution of physical fibre sizes seen in a~given
stellar mass bin in the SDSS. We also imitate the way in which emission
line quality cuts influence the observed sample by removing a~number of
simulated galaxies with low SFRs. Removed objects are chosen at random
from the SFR distribution in the simulations, such that the fraction of
galaxies remaining in a~given SFR bin traces the SDSS sample completeness
as function of SFR after imposing the emission line S/N cuts.
Finally, we also account for the random measurement errors in 
all quantities of interest by adding random Gaussian noise to their values 
drawn from the simulations. In this step, previously described for quenched fractions
in Section~\ref{sec:qf}, for each quantity $X$ among the in-fibre \mstar, SFR 
and $M_{\rm H2}$, a~given simulated galaxy is assigned $X_{\rm new}$
with $\log(X_{\rm new}) = \log(X) + N(0, \sigma_X)$. The values of $\sigma_X$
are dictated by random error distributions in the SDSS and are listed in 
Table~\ref{tab:sigmax} for all relevant quantities. All observational
realism steps along with their influence on the results in 
Fig.~\ref{fig:msfr-fgas-raw} are described in detail in 
Appendix~\ref{sec:sim-realism}.

Fig.~\ref{fig:msfr-fgas-sdss} shows some common trends present 
in both simulations with added observational realism and the SDSS. 
Across all three panels, main sequence galaxies have higher gas fractions
than their quenched counterparts, with simulations showing significantly
more dramatic difference in median \fgas\ between the two populations.
In the SDSS this variation amounts to only 0.52~dex and includes 
all sample corrections in the calculation. We can also
see a~slightly decreasing trend in \fgas\ with increasing \mstar\ across the
MS in all panels, which is significantly milder than the evolution between MS and
the passive sequence. 

Apart from these similarities, Fig.~\ref{fig:msfr-fgas-sdss} also
shows a range of differences between both simulation suites and
the SDSS. While in the observations the lowest gas fractions are
only found in galaxies with the most extreme deviations away
from the Main Sequence, in both simulations these extreme
values of \fgas\ are distributed across the entire passive sequence.
This trend is slightly more extreme in TNG, where the galactic
centres of quenched galaxies are almost exclusively gas-poor.
It is also important to note that the gradient in \fgas\ in the
SDSS is approximately perpendicular to the MS, while in 
both simulation suites it points almost horizontally with
an abrupt drop in \fgas\ above $\sim 10^{10.8} {\rm M_\odot}$
in Illustris and $\sim  10^{10.5}  {\rm M_\odot}$ in TNG.
In TNG this dramatic change appears around a threshold mass
for which the AGN feedback mode switches from thermal to 
kinetic \citep{Weinberger17}.  

Despite the addition of random scatter in Illustris and TNG, the observations
still cover the largest area in the \msfr\ parameter space
and show the smoothest evolution in \fgas\ across the plane. The shortage
of simulated galaxies in the passive cloud is highlighted even further by
the imitation of S/N cut on emission lines, with only a~handful of objects
left to compare with the SDSS. The observed and simulated main
sequences also differ, with the SDSS contours showing a~much tighter 
relation than those seen in Illustris and TNG. This comparison 
suggests that the emerging MS relations in the simulations are
potentially too broad, given that the width of the SDSS Main Sequence 
can be attributed to measurement uncertainties to a~large extent.
However, we do recognise that our treatment of measurement 
uncertainties in simulated quantities is rather simplistic and
hence we only point out this MS feature as a~hint of a~potentially 
interesting result. An accurate translation between simulated observables
and their realistic measurement requires careful consideration of the 
electromagnetic radiation emitted by galaxies, which is beyond the scope
of this work.

Fig.~\ref{fig:msfr-fgas-raw} indicates that one potential impact 
black holes have on their galactic hosts is depleting dense 
gas available for star formation. This is
demonstrated by the dramatic decrease in \fgas\ between 
the star forming and passive galaxy populations drawn from the simulations. 
When we post-process the
simulations with observational realism and compare them against
the SDSS in Fig.~\ref{fig:msfr-fgas-sdss}, we find a qualitative
agreement across all panels, albeit the range of values we see
in observations is more modest (see Table~\ref{tab:delta-pa} for a 
As we point out earlier, the trends are also more dramatic
in the simulations than the observations, suggesting that although
cosmological simulations successfully recover the large-scale
effects quenching has on galaxy populations, the details 
of its implementation still leave room for further improvement.  

These results suggest that our current observations of molecular 
gas content of local, massive central galaxies can 
potentially be explained by AGN acting to
prevent the re-supply of fuel for future star formation. In 
Illustris and TNG, AGN achieve this primarily by preventing 
the accretion of new gas from hot galactic haloes. 
However, based on our comparison alone, we cannot reject 
the potential involvement of ejective forms of AGN feedback which would 
lead to similar observational signatures and be particularly relevant for 
the lower end of our investigated stellar mass range.

\subsubsection{Conflicting trends in star formation efficiency}
\label{sec:gas:sfe}

In Fig.~\ref{fig:msfr-sfe} we explore how star formation efficiency 
\mbox{(${\rm SFE} = \flatfrac{\rm SFR}{M_{\rm H2}}$)} changes across the \msfr\ plane.
Analogously to Sec.~\ref{sec:gas:fgas} we analyse
the `raw' result first and then add simulation realism to compare the simulations
with the SDSS. In all panels, objects with ${\rm SFR}=0$ in the simulations are assigned 
a nominal low value of log(SFE/${\rm yr^{-1}}$) $=-11$ and hexagonal bins 
dominated by these objects saturate the colour scale with dark magenta. 

In Fig.~\ref{fig:msfr-sfe-raw} Illustris and TNG do not agree on the 
trends in SFE, in contrast with the previously seen \fgas\ distributions. 
Illustris galaxies have slightly higher 
SFEs in the bulk of the passive sequence, with a~median \textit{increase} of only 
0.07~dex when compared to SF objects. The passive sequence in Illustris is 
also subject to variation owing to small number statistics.
In contrast, TNG experiences a~dramatic drop in median SFE of 1.83~dex and
its passive sequence is dominated by objects with ${\rm SFR}=0$. These are redistributed 
to higher sSFR values to allow a visual comparison in the \msfr\ plane 
and are assigned a nominal low value for SFE of $10^{-11} {\rm yr^{-1}}$. 

In Fig.~\ref{fig:msfr-sfe-sdss} we apply observational realism to the simulations
(see appendix~\ref{sec:sim-realism}) and compare them against the SDSS.
Adding observational realism results, on average, in 
higher SFE values in both simulation suites. 
This is a combined effect of removing low-sSFR simulated
galaxies to mimic S/N ratio cut in the SDSS and restricting galaxies
to their centres, which are more efficient at forming stars than the outskirts.
When we then focus on observations, we find a~slight increase in SFE 
with increasing \mstar\ on the Main Sequence along with 
a dramatic decrease of almost 2~dex in the direction perpendicular 
to the MS. 
The evolution of SFE in the SDSS is smooth and only shows 
pronounced scatter around the edge of the passive sequence owing 
to low number statistics. {These general trends are not
well reproduced in the simulations and hence all three panels show visible
differences among each other. In Illustris the direction of the trend
is essentially inverted, such that MS values of SFE are approximately 0.6~dex lower 
than those seen in the scarcely populated passive sequence. 
In TNG, although the quenched population shows lower SFE values
than the MS, this change is significantly more abrupt than observed
in the SDSS. Similarly to the trends seen in \fgas, the simulations
show steeper gradients in SFE than the observations, which primarily
point in the horizontal direction, rather than perpendicular to the
Main Sequence.
 Nonetheless, adding observational
realism once again has the effect of strengthening the trends visible 
in the raw simulation data, in this
case amplifying the contrast between Illustris and TNG.

Our analysis of Fig.~\ref{fig:msfr-sfe-raw} demonstrates that different
models predict contrasting behaviour in star formation efficiency in the
\msfr\ plane. Both Illustris and TNG quench their
massive centrals primarily via AGN feedback, however it is 
apparent in the colour gradients that the two suites predict two opposing 
trends in SFE. Interestingly, neither of the results is exactly reflected in
the observations when we compare them against the SDSS in 
Fig.~\ref{fig:msfr-sfe-sdss}. Adding observational realism further
highlights the juxtaposition of Illustris and TNG, yielding a~bimodal
distribution in SFE between passive and star forming galaxies in both
suites. Between the two theoretical predictions, there seems
to be no model clearly favoured by the observed Universe. However, 
given the trends seen in the SDSS, we would expect a successful
prediction to recover a significant decrease in both \fgas\ and
SFE in a direction perpendicular to the Main Sequence.

\subsection{The dependence of \fgas\ and SFE on \mbh}
\label{sec:gas-line}

\begin{figure}
\includegraphics[width=\columnwidth]{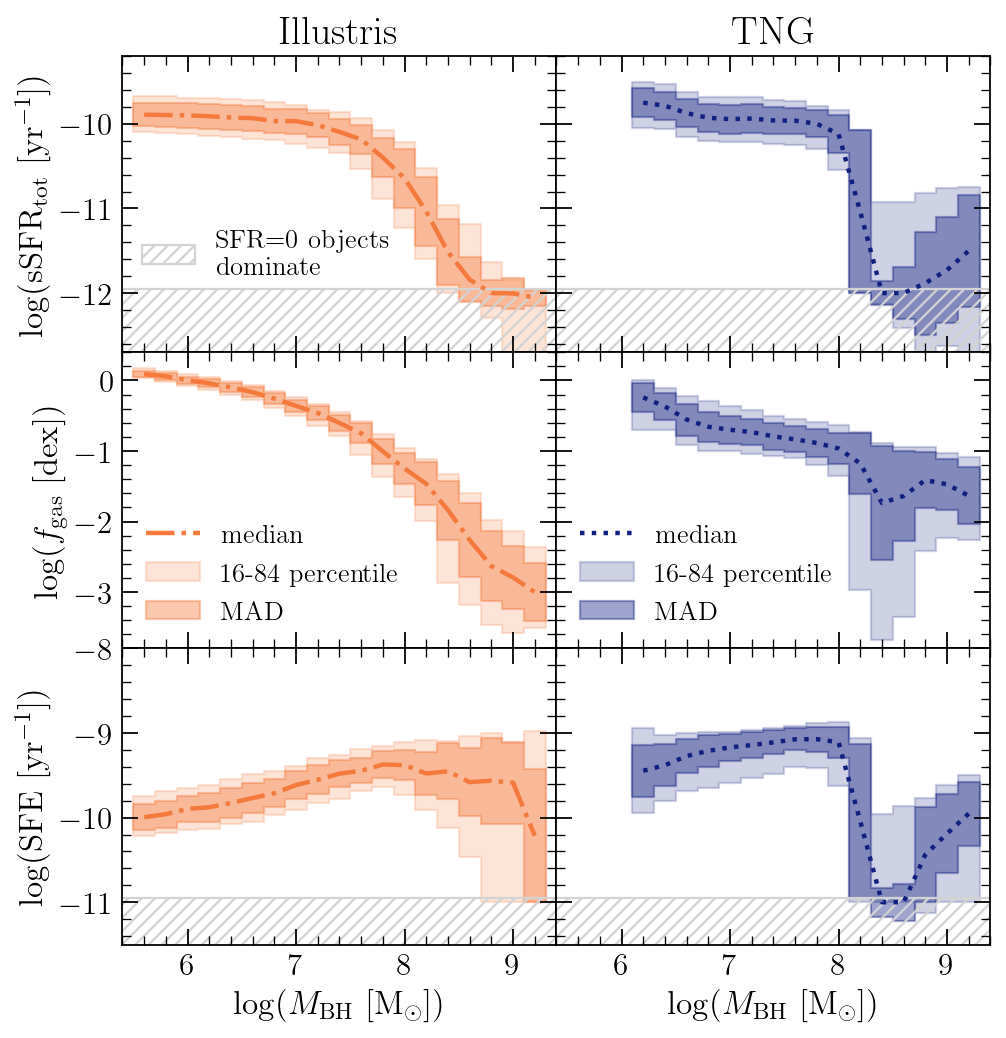}
	\caption{Median sSFR (top row), gas fraction (middle row) 
				and SFE (bottom row)
    				as function of \mbh\ in Illustris and TNG. Gray hatched 
   				regions indicate where the results are dominated by galaxies 
   				with SFR=0. Both simulations agree on the trends in sSFR and
   				\fgas, where both quantities decrease with increasing \mbh.
   				The variation in sSFR is more dramatic in TNG, showing 
   				a sharp drop around $\log(M_{\rm BH})\sim 8.1$. In \fgas\
   				Illustris galaxies show a~broader range in values, spanning
   				3~dex in logarithmic units. Simulations disagree about 
   				the behaviour of SFE, which steadily increases with \mbh\
   				in Illustris, while in TNG is initially roughly constant and
   				drops off dramatically at high black hole masses.}
\label{fig:line-plots-raw}
\end{figure}

\begin{figure*}
\includegraphics[width=\textwidth]{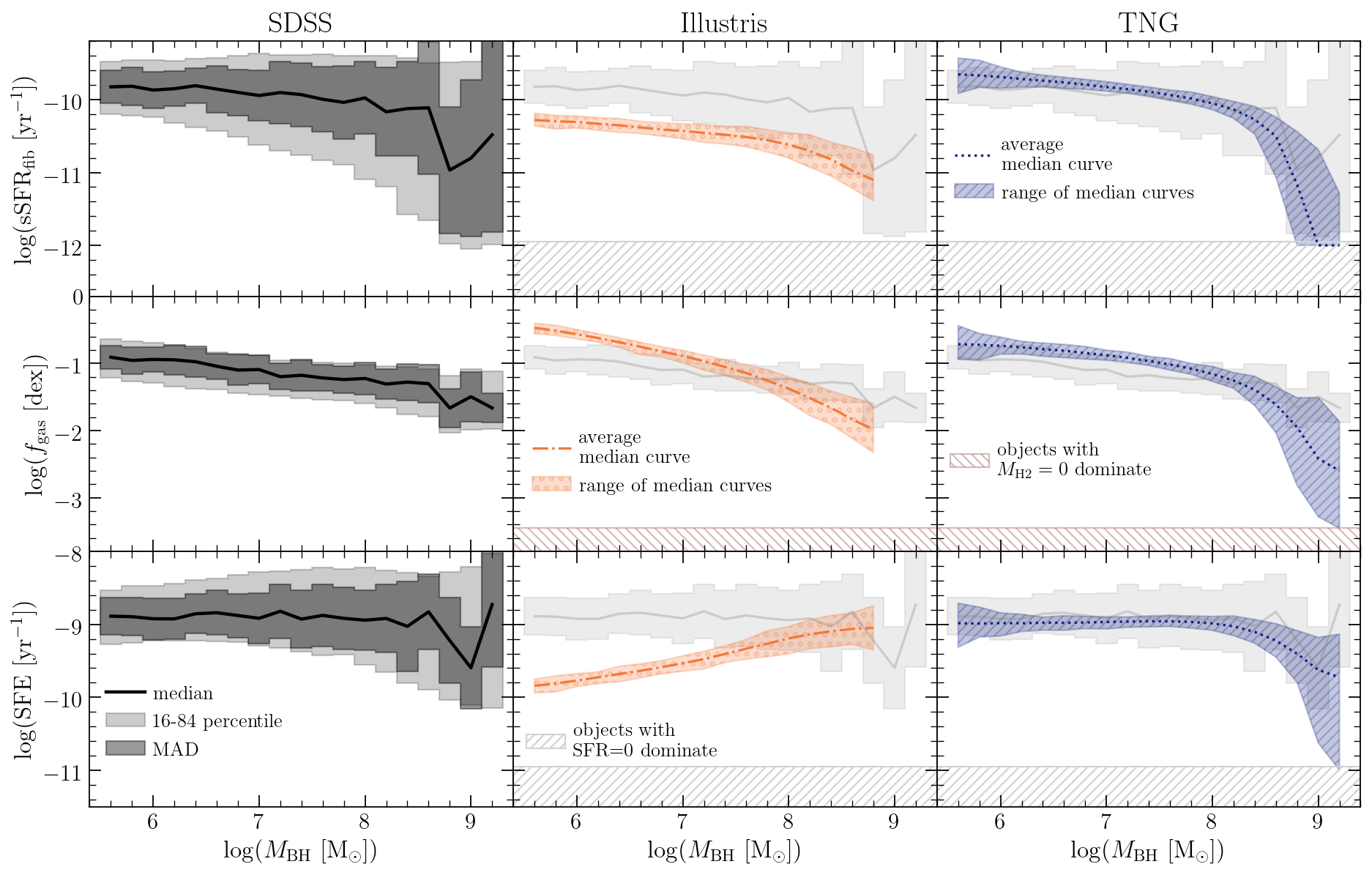} 
    \caption{Median sSFR (top row), gas fraction (middle row) 
				and SFE (bottom row) as function of \mbh\ for SDSS, Illustris
				and TNG (left to right panels respectively). In observations, all quantities are obtained from fibre
				spectroscopic measurements and the sample is restricted to galaxies
				with visible emission line signal. The simulation panels differ 
				from raw results in Fig.~\ref{fig:line-plots-raw} due to the application
				of observational realism described in appendix~\ref{sec:sim-realism}. 
				The SDSS curves include $1/V_{\rm max}$ and inclination 
				corrections placing all panels on equal footing.	
				Both simulations show decreasing trends in ${\rm sSFR_{fib}}$
				and \fgas\ in line with the observations, however TNG is a~better
				quantitative match with the SDSS. SFE in the observations shows
				a mild decrease at the highest \mbh\ and a similar trend is visible 
				in TNG. The curves in Illustris show an opposite behaviour, 
				increasing SFE with increasing \mbh, which is clearly discrepant with 
				the observations.}
    		\label{fig:line-plots-sdss}
\end{figure*}

In the final part of our analysis we look directly at possible connections
between \mbh\ and molecular gas content of galaxies to further investigate
the mode of operation of AGN feedback quenching.
We first analyse the simulations only and proceed by splitting the samples 
into 0.2~dex wide bins in \mbh\ to calculate median values of 
sSFR, \fgas\ and SFE in each bin. We plot the curves resulting from 
this calculation as dotted and dash-dotted lines in Fig.~\ref{fig:line-plots-raw}, 
indicating the median absolute deviation (MAD) and the 16-84 percentile 
range of values in a~given bin with shaded regions. 
In order to include all galaxies with ${\rm SFR}=0$, we assign 
them nominally low values for both sSFR and SFE of 
log(sSFR/${\rm yr^{-1}}$) $= -12$ and 
\mbox{log(SFE/${\rm yr^{-1}}$) $= -11$}, respectively. Gray hatched
regions in the top and bottom panel rows indicate bins dominated by 
the SFR=0 objects, which occur for 
$10^{8.2}{\rm M_\odot} \lesssim M_{\rm BH} \lesssim 10^{8.6} {\rm M_\odot}$ 
in TNG and $M_{\rm BH} \gtrsim 10^{8.8}{\rm M_\odot}$ in Illustris.

In the top two rows in Fig.~\ref{fig:line-plots-raw} we
see a~qualitative agreement between the trends in 
Illustris and TNG. Central galaxies in both simulations experience
decreasing sSFR and \fgas\ as a function of increasing \mbh. 
In Illustris the change in sSFR is more gradual, while in TNG
it resembles a step function around a characteristic black hole
mass of $\log(M_{\rm BH}/{\rm M_\odot}) \sim 8$, where the AGN feedback mode 
changes from `quasar' to `kinetic' \citep{Zinger20}. 
The gas fraction curve in TNG 
also shows a visible dip around the same characteristic \mbh, 
while in Illustris it is a smooth decreasing trend, spanning 
$3\ {\rm dex}$ in \fgas. The SFE curves in the bottom panel
show the most difference between the simulations. In TNG the
efficiency remains roughly constant for $M_{\rm BH} < 10^8 {\rm M_\odot}$ 
and drops dramatically beyond this critical black hole mass. 
In Illustris the the SFE behaviour seems to go against 
TNG -- the efficiency rises steadily with increasing
\mbh\ and then falls off at the highest black hole masses, where the
data shows significant scatter due to low number statistics.

In Fig.~\ref{fig:line-plots-sdss} we take these theoretical predictions,
add observational realism and compare the final results with the SDSS.
The observed sample is restricted to objects with reliable
emission line measurements according to the selection criteria in 
Section~\ref{sec:sample-selection} and the calculated curves
are weighted to correct for the Malmquist bias and the 
cut on disk inclination. The coloured shaded regions in the 
simulations show a~range
of median curves resulting from 500 random realisations of 
adding measurement error to the raw data post-processed to 
reflect aperture correction and emission line S/N ratio cuts.
In the process of mimicking an SDSS fibre aperture in the simulations
we arrive at a~population of galaxies which have $M_{\rm H2}=0$
and assign them with a~nominal low gas fraction value 
of $\log(f_{\rm gas})=-3.5$. 
All objects with both $M_{\rm H2}=0$ and SFR=0 are removed
from the calculations since their value of SFE is mathematically
ambiguous (without the knowledge of the functional forms 
of ${\rm SFR}(M_{\rm BH})$ and $M_{\rm H2}(M_{\rm BH})$ we are
unable to determine the finite value of the otherwise undefined
SFE in the limit). Hatched rectangular regions in
Fig.~\ref{fig:line-plots-sdss} indicate where objects devoid of 
star formation (gray hatch) and molecular gas (brown hatch) dominate. 

We first look at the in-fibre sSFR (${\rm sSFR_{fib}}$) 
in the top left corner in Fig.~\ref{fig:line-plots-sdss}. 
The decrease towards low sSFR values in the fibre occurs at 
significantly higher \mbh\ than we previously saw in the case 
of total sSFR. This is a direct result of the
emission line S/N ratio cuts biasing the sample towards more star-forming 
systems. The shaded regions are also visibly broader than before, owing to 
a~fivefold decrease in sample size. This dramatic change between the 
in-fibre and total result illustrates how critical it is to apply similar
selection criteria to both the simulations and observations for 
a successful comparative analysis. 
Standard procedures necessary in the observations
(e.g.~emission line quality cuts or restricting the field of view to 
galactic centres) need to be applied to theoretical predictions
in order to ensure a fair comparison between the models and the observed 
Universe.

Both simulations in Fig.~\ref{fig:line-plots-sdss} show a~decreasing 
trend in the aperture-restricted sSFR, qualitatively consistent with 
the observations. TNG shows a~better quantitative match, tracing the 
SDSS curve reasonably well across the entire range in \mbh. Gas fraction 
in the middle row of Fig.~\ref{fig:line-plots-sdss} shows a significant, 
steady decline of $\sim 1\ {\rm dex}$ in the SDSS. Similarly to sSFR, 
both simulation suites predict trends qualitatively similar to the 
observations. Galaxies with low \mbh\ in Illustris have higher initial 
\fgas\ than the observations, which then decreases more rapidly to become 
lower than in the SDSS at highest \mbh.
The \fgas\ curve in TNG traces the SDSS median values within their MADs 
until $M_{\rm BH} \sim 10^{8.4} {\rm M_\odot}$, where it rapidly drops 
to values around 1 dex lower than the SDSS.

The most striking result presented in Fig.~\ref{fig:line-plots-sdss} 
is the comparison of star formation efficiency trends in the bottom row. 
In the SDSS, the SFE remains approximately constant until 
$M_{\rm BH} \sim 10^{8.5} {\rm M_\odot}$, where it falls off by a~total 
of 0.5 dex to then pick up in the highest \mbh\ bin. Galaxies in Illustris 
show a different trend, where their median SFE rises by 1 dex with 
increasing \mbh\ to reach the initial SDSS value of 
${\rm SFE} \approx 10^{-9}{\rm yr^{-1}}$ at 
$M_{\rm BH} \sim 10^{8.8} {\rm M_\odot}$. 
In contrast, TNG seems to follow the SDSS curve
across a broad range in black hole mass, with the exception of the 
highest \mbh\ bin. 

Comparing Figs.~\ref{fig:line-plots-raw}~\&~\ref{fig:line-plots-sdss}
allows us to characterise the impact of adding observational
realism on the inferred theoretical predictions.
Across all three variables: sSFR, \fgas\ and SFE mimicking
measurement uncertainty smoothens the gradients present
in both simulations. It is particularly visible in TNG, where
the step function behaviour is removed from the curves.
The aperture correction in Illustris acts to reduce sSFR and \fgas,
and to increase SFE, primarily at the highest \mbh. In contrast,
in TNG galactic centres are more star-forming at low \mbh\ and
devoid of both gas and star-formation at 
$\log(M_{\rm BH}/{\rm M_\odot}) \gtrsim 8$. Finally, mimicking
a S/N ratio cut results in removing high-\mbh\ objects in both
simulations, since low SFRs are primarily associated with 
massive black holes in these two model implementations.
Combined together, our post-processing steps primarily act
to highlight the raw trends identified earlier in 
Fig.~\ref{fig:line-plots-raw}.
We see how Illustris and TNG agree on the behaviour of sSFR 
and \fgas\ as a function of \mbh\ and how they clearly disagree 
on the trends in SFE.
For a detailed discussion on the impact of all observational
realism steps we refer the interested reader to 
Appendix~\ref{sec:sim-realism}.

Overall, our analysis of molecular gas trends predicted
by the simulations shows that neither model perfectly 
reflects the observed Universe. We note, however, that the 
SFE trends seen in the SDSS are in a better agreement with those
predicted by TNG than Illustris, especially when we consider 
galactic centres with full observational realism.

\section{Discussion}
\label{sec:discussion}

\subsection{Quenching modus operandi in massive central galaxies}
\label{sec:quenching-mo}

The result of our Random Forest classification, together with 
the PCC analysis, clearly indicate that black hole mass is the 
most successful parameter at determining the 
quenching state of massive central galaxies at redshift $z=0$. 
We find this statement about the \mbh\ dominance to be true 
regardless of the detailed implementation of the feedback model 
in the simulations or the choice of \mbh\
calibration in the observations. Throughout this paper we 
also show that \mbh\ is superior to other galactic parameters 
like \mhalo\ and \mstar, a~conclusion previously reached with 
a~different statistical approach by \cite{Bluck16} in the SDSS 
and in a~sample of direct black hole mass measurements 
in \cite{Terrazas16, Terrazas17}. We improve on these previous 
works by extending the set of cosmological simulations to EAGLE 
and IllustrisTNG, enriching our analysis with molecular gas 
content within the galaxies, as well as using a more sophisticated
machine learning based analysis method.

In this work we limit our research to central galaxies in order 
to characterise the intrinsic quenching mechanisms only. Since 
centrals reside at the minima of their dark matter halo potentials, 
we expect the effect of environment on ceasing star formation to be 
minimal, avoiding e.g. ram pressure stripping of molecular gas 
\citep[e.g.][]{Peng12, Bluck16, Bluck20b}. Our results 
demonstrate that all three cosmological simulations: EAGLE,
Illustris and TNG, share a common theoretical prediction for
the observable consequences of AGN feedback quenching
and that this prediction is met overwhelmingly well in the 
Universe as observed in the SDSS.   

More specifically, we find that the current star
formation state of a~galaxy is determined by the 
\textit{integrated history of AGN feedback} (encoded in \mbh) rather 
than the instantaneous AGN luminosity at redshift $z \sim 0$
(as inferred from \mdot). Our findings confirm conclusions reached 
for small samples of direct black hole mass and accretion rate 
measurements in the local Universe \citep{Terrazas16, Terrazas17} 
as well as individual analyses of cosmological simulations 
\citep{Davies19, Terrazas20, Zinger20}. Our study is unique, however, 
in that it treats all simulations consistently throughout the whole 
analysis. We apply the same statistical tests and Random Forest 
classification to reveal a unanimous conclusion in all simulation suites
-- at redshift $z = 0$ \mbh\ is the key parameter determining 
galaxy quiescence, while the instantaneous \mdot\ holds very little
predictive power once its connection to \mbh\ is accounted for.
The theoretical prediction of \mbh\ dominance is also recovered
clearly in the SDSS observations, where we subject the data to
an identical analysis.

Although the importance of \mbh\ and AGN quenching are at the 
focus of this paper, our analysis delivers a useful discussion 
of other quenching avenues potentially available to massive central 
galaxies. Our Random Forest experiment, together with the PCC analysis, 
show that there indeed exists a~connection between quiescence and 
both \mstar\ and \mhalo\ individually. However, we interpret these
relations as secondary in nature and stemming from the 
inter-correlations between stellar, halo and black hole mass.
Once \mbh\ is identified as the most important parameter among \mbh, 
\mhalo\ and \mstar\ and its numerical association with stellar and 
halo mass is accounted for, a~picture emerges in which the previously 
recognised correlations are incidental, rather than intrinsic. 

These results neatly illustrate that correlation does not imply 
causation, regardless of how high the correlation measurement is. 
By showing how the correlations between sSFR, \mhalo\ and \mstar\ 
vanish at a fixed \mbh\ we unambiguously demonstrate
that these correlations are spurious, i.e. not causal in origin.
Hence there is no doubt that  more massive galaxies are more likely 
to be quiescent as demonstrated by e.g. \cite{Baldry06}, \cite{Peng10} 
and \cite{Peng12}, however this relationship is primarily a~consequence 
of \mstar\ being tied to \mbh\ in the black hole - galaxy 
co-evolution. Similarly, massive haloes are populated with quenched 
objects, in agreement with e.g. \cite{Woo13} and \cite{Woo15}, however 
it is likely not due to the influence of virial shock heating 
\citep[as suggested by][]{Dekel06}, but rather driven 
by the connection between the black hole and halo masses. 

As a~final remark we would like to stress that AGN feedback is unlikely 
to be the only mechanism responsible for galaxy transition to quiescence 
in its entirety. In fact, given the wealth of evidence for a~connection 
between quenching and other galactic parameters, as well as the measurement 
uncertainty associated with all variables of interest, one would 
expect multiple processes to be at play simultaneously, bringing star 
formation to a halt together. 

A good example of such process would be the occurrence of stabilising 
torques from the central bulge which could prevent gas from collapsing
to form new stars via `morphological quenching' \citep{Martig09}. This
scenario would decrease in the efficiency with which molecular
gas is forged into news stars, leading to the observed
low SFEs in the SDSS. This mechanism on its own, however, cannot 
explain the reduced gas fractions experienced by passive galaxies 
and hence could not be solely responsible for galactic transition 
to quiescence in the observations. The resolution limit in 
EAGLE, Illustris and TNG unfortunately does not allow us to investigate 
the theoretical predictions for morphological quenching potentially
working in tandem with AGN feedback. However, recent simulations 
of idealised galaxies suggest that the presence of a central bulge
is associated with decreased star formation activity in models 
which do not include prescriptions for black hole feedback 
\citep{Gensior20}.

\subsubsection{How do AGN prevent galaxies from forming stars?}
\label{sec:AGN-models}

In this work we investigate theoretical predictions for the 
observable consequences of AGN feedback quenching in three 
different feedback model implementations. Numerical simulations 
of galaxy evolution typically divide the interaction of supermassive
black holes with their environment into two categories: the thermal,  
`quasar mode'  feedback operating at high accretion rates and the 
preventative, `radio mode' feedback associated with less efficient 
accretion 
\cite[e.g.][]{Bower06, Croton06, Sijacki07, Booth11, Vogelsberger14b, Somerville15, Weinberger18}.
In the quasar mode, AGN demonstrate their presence through powerful 
radiation across the whole electromagnetic spectrum, feeding on cold, 
dense gas to launch high-velocity winds from the black hole accretion 
disks \citep[e.g.][]{Crenshaw10, Rupke17}. 
The radiatively inefficient radio (`mechanical') AGN mode, 
on the other hand, is associated with synchrotron emission from charged 
particles in AGN jets \citep[e.g.][]{McNamara07, Fabian12}.  
These two modes differ in both the location of energy injection 
(local vs at large distances) and the method by which this energy 
is deposited into the gas (radiative vs. mechanical).

The `kinetic bubble' model implemented in Illustris
(see Sec.~\ref{sec:cosmosims}) is designed to mimic the effect
of interactions between AGN jets and the CGM at large distances
away from a galaxy. By placing bubbles at random locations
within the CGM, AGN in Illustris evacuate the gas residing in 
galactic haloes and, to a lesser extent, heat up the leftover
gas reservoirs. This prescription is extremely efficient at depleting
hot gas haloes around massive galaxies, leaving them devoid of fuel 
for subsequent cooling and future star formation. Although this 
AGN feedback model successfully quenches central galaxies, it does 
so by removing hot gas haloes around them. This predicted outcome is
inconsistent with observations, which commonly show large quantities 
of hot, ionised gas surrounding massive centrals \citep[e.g.][]{Fabian12}. 

AGN feedback at low Eddington ratios in TNG aims to improve 
on the Illustris design by implementing a kinetic mode, which would 
keep massive galaxies quenched without stripping their hot gas haloes.
This kinetic feedback mode is designed to imitate the potential effect 
of interaction between the galaxy ISM and AGN winds at small distances
around the AGN and proves very successful at preventing excessive gas
removal from around the galaxy.
By introducing momentum kicks in random directions, the model
aims at capturing nucleus-scale interactions which over time percolate
outside the galaxy, increasing the entropy of gas in
the CGM as well as the ISM. As shown 
in \cite{Terrazas20} this kinetic feedback mode also results
in cold gas being pushed out of galaxies, since the wind energies
exceed the binding energy of gas within the hosts.

When we look at the connection between \mbh\ and different 
galactic properties compared across all simulations and the SDSS, 
we see that the observations show trends {which are not
perfectly captured by the theoretical predictions from Illustris
and TNG. Nonetheless, our analysis shows promising qualitative 
agreement between the predicted and observed trends, which seems
more pronounced in TNG out of the two simulation suites. 
It is apparent in Sec.~\ref{sec:qf} 
that although transition to quiescence happens at higher values of 
\mbh, \mhalo\ and \mstar\ than in the SDSS for all simulations, the trends 
in TNG show a promising resemblance both in terms of slopes 
and their behaviour in the parameter value limits, once observation-like scatter is added
to the data. The EAGLE suite stands out as well, showing the shallowest 
slopes among all simulations. More importantly, however, quenched fractions in EAGLE
point towards insufficient strength of its AGN quenching mechanism,  
which results in a~quenched fraction ceiling of $f_Q \sim 0.6$ and 
a~sSFR floor of log(sSFR/${\rm yr^{-1}}$) $\sim -11$ (i.e. barely quenched at all). 

The analysis of molecular gas content in Illustris, TNG and the SDSS in 
Sec.~\ref{sec:gas} allows us to directly compare the 
effect of two different feedback implementations on the AGN host galaxies.
The mechanical bubble model implemented in Illustris successfully 
removes fuel available for star formation by keeping the halo gas hot 
and preventing its collapse on the galaxy to replenish cold gas reservoirs. 
In this fashion, the simulation efficiently shuts down star formation in the 
outskirts of massive galaxies, leaving behind galactic centres 
with enough gas to fuel both the central black hole and the star 
formation in the central region. 

We then contrast this picture with TNG, where the preventative 
feedback in the form of momentum injection around the AGN leads
to a~more balanced scenario. In a fashion similar to that seen in the
SDSS, both SFE and \fgas\ decrease, together resulting in decreasing 
sSFR with increasing \mbh. The efficiency is mostly insensitive to 
the change in black  hole mass until around 
$M_{\rm BH} \sim 10^8 {\rm M_\odot}$, where it falls off, dropping by 
almost $1\ {\rm dex}$ in the simulation. As discussed in \cite{Zinger20}
and \cite{Terrazas20}, once the kinetic feedback turns on 
around this critical \mbh, the momentum kicks: 
a) increase turbulence in the ISM to prevent gravitational 
collapse of cold gas; b) increase the entropy and temperature
of the CGM to prevent cooling; and c) push cold gas out of the
galaxies, depleting fuel available for star formation.

When we compare the two `observation-like' results in the
simulations against the SDSS, we see that the observable
Universe does not strongly favour either of the 
theoretical predictions. However, we point out that
the TNG trends in all three variables: sSFR, \fgas\ and SFE 
show a~closer qualitative match to the SDSS than Illustris,
when one focuses their
analysis on central, massive galaxies at redshift $z \sim 0$.
The difference is most apparent in SFE, where Illustris goes
against the observations, showing a~significant increase in star
formation efficiency with increasing black hole mass. 
At the same time, the trends predicted by TNG,
although qualitatively consistent, are significantly steeper
than those seen in the SDSS. The change in gas content
between star-forming and passive galaxies seems too abrupt
due to the dichotomy between AGN feedback modes, which 
is likely an oversimplification of the physics as a known 
problem in the simulation. 
We also stress that the success of the theoretical
prediction in the TNG is only valid in the central regions, imitating
the SDSS fibre coverage. The overall black hole mass threshold 
at which quenching occurs is still too high in comparison with 
observations and this effect cannot be alleviated by mimicking
the observational measurement errors.

\begin{figure}
\centering	
\includegraphics[width=0.7\columnwidth]{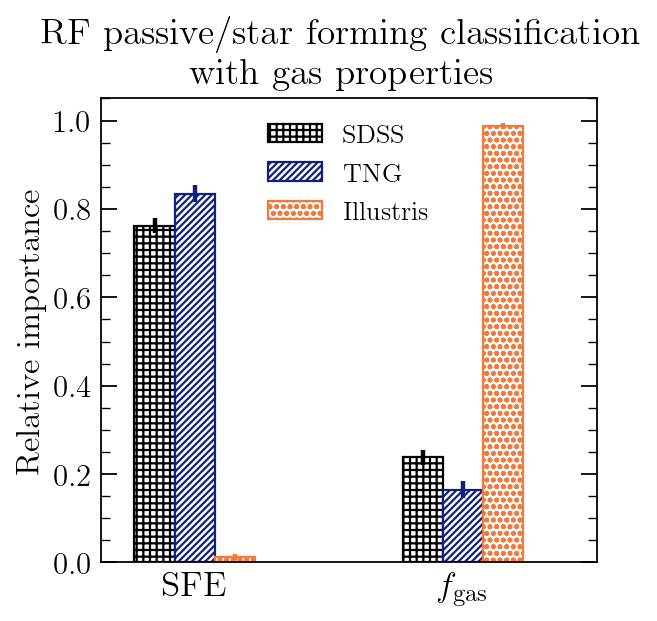} 
    \caption{Feature importance in the Random Forest classification into star 
    		 forming and passive galaxies, using \fgas\ and SFE as input features.
    		 The algorithm is allowed to choose from all features at each split.
    		 SFE clearly dominates in the observations and holds over three times 
    		 more predictive power than \fgas. In Illustris quiescence is more 
    		 accurately predicted by \fgas, while in TNG SFE holds significantly 
    		 more predictive power, in agreement with the SDSS observations.}
    	\label{fig:rf-gas}
\end{figure}

\subsubsection{RF classification with \fgas\ and SFE}

As a final consideration in this discussion, we perform a random 
forest classification on Illustris, TNG and the SDSS, using \fgas\ 
and SFE as input features for the algorithm. In Fig.~\ref{fig:rf-gas}
we show median feature importances extracted from 500 realisations of 
a forest for each data set, marking the 5th and 95th percentiles of
the importance distribution with the vertical error bars. 

It is apparent in Fig.~\ref{fig:rf-gas} that SFE holds more 
predictive power than \fgas\ in the SDSS, where its relative importance 
is over twice that of gas fraction (in agreement with \cite{Ellison20}). 
As expected from our previous 
analysis in Sec.~\ref{sec:gas}, the TNG result makes a prediction 
similar to the observed trends, where the cumulative information 
gain provided by the SFE in the classification is higher than that of 
\fgas\ by over 3 times. In Illustris the predictions also
summarise our previous analysis very well. The suite predicts 
\fgas\ to hold almost all predictive power in determining whether 
a given galaxy is star-forming or quenched. This result
further illustrates how the AGN feedback in Illustris primarily
acts to deplete molecular gas reservoirs to prevent future star 
formation, without substantial changes in star formation efficiency.

Our consistent comparison between observations and theoretical 
predictions from the simulations suggests that AGN can 
potentially affect their galaxies through a~combination of 
thermal energy and momentum injection into the surrounding gas. 
In this scenario, the central black hole will both maintain high 
temperatures in the halo gas, not allowing reservoir replenishment; 
and increase turbulence in the ISM, resisting gravitational collapse.
Our RF experiment with molecular gas properties shows 
that TNG agrees with the observations, while Illustris does not. 
This result points towards a success of the kinetic mode over 
the mechanical bubble implementation in the AGN feedback
model development within the Illustris framework.
 
Following the encouraging qualitative agreement between the 
SDSS and the simulations, we would expect massive enough black holes
to cause and maintain low star formation rates in their galactic
hosts. The AGN could achieve that through a radiatively inefficient 
feedback channel, injecting turbulence into the ISM as well as increasing 
the temperature and entropy of the CGM. Such an AGN feedback model
would result in both a decrease in gas fractions and star formation
efficiency, preventing gravitational collapse and subsequent star 
formation. 
In this paradigm, the AGN feedback could be observed in action 
through direct measurements of molecular gas content and kinematics 
across the galaxy coupled with a direct measurement of black hole mass. 
Unfortunately, these three kinds of observations are not simultaneously 
available for large galaxy samples at the moment and hence the 
detailed tests of this theoretical prediction can only be completed 
in the future.

\subsubsection{$L_{\rm AGN}$ and \mbh\ -- consequences for direct AGN feedback observations}
\label{sec:L_AGN-mbh}

Our analysis in Sec.~\ref{sec:mdot-no-corr} carries 
important implications for the observational interpretation 
of AGN quenching. Since black hole accretion rate is responsible 
for the measured bolometric luminosity of an AGN, the 
\textit{negligible importance of instantaneous \mdot\ implies little 
significance of AGN detection and classification for the study 
of black hole quenching in action.} In fact, this prediction explains 
why a~substantial body of research in the field delivers contradicting 
results, with some authors finding a~relationship between the current 
AGN activity and quenching \citep[e.g.][]{Nandra07, Bundy08, Georgakakis08}, 
while others find no evidence of such a relationship 
\citep[e.g.][]{Rosario13b, Aird12} for different samples of AGN hosts.

The negligible importance of \mdot\ in the simulations, 
however, does not imply that black holes are not responsible for 
quenching of star-formation in massive central galaxies. As we explain 
in Section~\ref{sec:correlations}, it is the fossil record of
the integrated AGN activity, traced by \mbh, rather than its instantaneous 
power injection at $z=0$ traced by \mdot, which can successfully predict 
which galaxies are no longer star forming. This statement is true for 
all three different prescriptions for AGN feedback implemented in EAGLE, 
Illustris and TNG and does not make any assumptions about the functional
form of $\dot{M}_{\rm BH}(t)$. 

The redshift $z=0$ measurement of $M_{\rm BH} = \int \dot{M}_{\rm BH}\, {\rm dt}$ 
in the simulations is agnostic to whether e.g. a single violent 
event was responsible for the bulk of the growth of \mbh\ or whether it 
was a slow growth through constant low-\mdot\ over time. Hence, it would
be interesting to investigate the importance of peak black hole 
accretion rates for the star-forming/quenched classification to
compare them against \mbh\ measurements. Unfortunately, owing to the
nature of black holes and the `no-hair theorem' \citep{Misner73} it is impossible to 
infer their exact accretion history from contemporaneous observations. Being limited
to simulations only, we would not be able to make a comparison
between theoretical predictions and the observable Universe, which is
the key objective of this work.

As we show in Sec.~\ref{sec:mdot-no-corr}, all three simulation 
suites predict different distributions in $L_{\rm AGN}$ between 
the star forming and passive galaxy populations, stemming from their 
different prescriptions for AGN feedback. The model 
favoured most in comparison with observations -- TNG -- predicts 
higher AGN luminosities in star forming galaxies. This observational 
signature is expected in a~cold mode accretion scenario where there 
is abundant dense gas available to feed both the black hole accretion 
disk in the galactic centre as well as the star formation within the 
galactic disk. Once the accretion becomes inefficient and the 
feedback mode changes, black holes are still accreting hot gas, 
which is sufficient to provide AGN feedback of the mechanical kind. 

As we show in our analysis, this mechanism most strongly affects the 
efficiency of star formation, hence causing the association of quenched 
galaxies with primarily low accretion rates (and hence AGN luminosities). 
These results suggest that the most violent processes observed 
in the AGN during their quasar phase are associated with episodes of 
thriving star formation rather than galactic quiescence, and it is the 
inefficient accretion of hot gas onto the black hole environment 
which ultimately maintains galaxies in their very low star formation state. 
In fact, such trends have already been reported in the observational 
literature by e.g. \cite{Hickox09}, \cite{Xue10}, \cite{Heckman14} 
and \cite{Trump15}.

Even if we refrain from identifying the model most compatible 
with the observable Universe, there are further reasons for 
which a~measurement of $L_{\rm AGN}$ in large studies of
currently active AGN is non-trivial to interpret in the 
context of quenching. All three simulations analysed in this 
study invoke supermassive black hole feedback in order to 
successfully quench their massive galaxies, however they implement
different subgrid treatments to achieve this goal. As we see 
in our analysis all three prescriptions predict three different AGN
luminosity trends between the passive and star-forming galaxy 
populations. In contrast, EAGLE, Illustris and TNG unanimously 
predict an association between high black hole masses and quiescence, 
regardless of the model assumptions. Therefore \mbh\ proves to 
be a robust tracer of AGN feedback quenching, insensitive to its
exact mode of operation, whereas the connection between $L_{\rm AGN}$
and quenching is highly model-dependent.

The robustness of black hole mass is not only restricted 
to predictions drawn from numerical simulations. Ultimately, 
a condition necessary for a quenching mechanism to succeed is 
being able to offset the cooling rates within the halo. 
Without enough heat continuously provided into the CGM, 
the halo gas would cool and accrete onto a~galaxy, leading 
to an ever-active star formation as a consequence of 
a~`cooling catastrophe'. As shown analytically in 
Appendix~B by \cite{Bluck20a}, the energy required for an 
AGN to offset cooling in the halo is directly proportional 
to \mbh, which is the integral of \mdot\ over time. These 
considerations do not make any assumptions about the microphysics 
of AGN heating and, instead, universally show how the instantaneous 
behaviour of accretion has no immediate impact on the galactic 
evolution cycle. From this simple argument one would yet again
expect to observe quenching signatures as a function of 
black hole mass rather than luminosity of active galactic 
nuclei, as is indeed confirmed in three leading hydrodynamical 
models here, with differing feedback prescriptions.

\subsubsection{Lessons learned from observational realism}

In the process of comparing data on equal footing, we 
had a~unique chance to look at simulation results from 
the perspective of observational realism. As a~result we find
that cosmological simulations produce smaller passive 
populations than the observed Universe. This issue is 
particularly visible in Illustris, {where quiescent 
central galaxies account for only $\sim 12\%\, (\sim 35\%)$ 
of the whole population 
with $M_\ast > 10^{10}\, (M_ast > 10^{10.6})$. The situation 
in EAGLE is less dramatic since the suite manages to produce 
a~quiescent population of $\sim 25\%\, (\sim 42\%)$ among massive 
centrals in these mass ranges. Finally, TNG produces the largest
passive population amounting to $\sim 34\%\, (\sim 67\%)$, 
however this fraction is still lower than the $\sim 50\%\  
(\sim 68\%)$ we see in the SDSS (calculated with volume 
corrections to account for the Malmquist bias). We also learn 
that if the intrinsic behaviour of different 
galactic parameters 
as a~function of \mbh\ is almost a~step function, the current 
measurement uncertainties in the observations may very well hide 
it, producing very smooth curves like in the case of TNG. 

Finally, the simulated galaxy trends seem to predict relations 
observed in the Universe rather well without accounting for 
the measurement uncertainty. One obvious example of this property 
is the simulated Main Sequence which is potentially too 
wide across all data sets. EAGLE, Illustris and TNG all
reproduce the observed width of the MS very well, however when 
we account for more realistic observations by adding random scatter 
informed by the observations, the \msfr\ relations become 
significantly too spread out. This result suggests that the 
intrinsic Main Sequence relation is quite tight and, more 
importantly, tighter than predicted by the simulations. This may 
in turn point towards MS regularisation processes that are 
overlooked in the currently available models. 

We do, however, want to note that our realism treatment here is rather 
simplistic - we do not perform full radiative transfer 
calculations to obtain simulated galaxy spectra and hence we
do not use the same techniques to estimate relevant galactic
parameters in both simulations and observations.
Nonetheless, regardless of their simplicity our realism steps 
do account for
the three main caveats associated with the SDSS observations 
- fibre aperture correction, S/N ratio cuts on emission line 
fluxes and the random measurement errors.

\section{Summary and conclusions}
\label{sec:SUMMARY}

In this work we investigate the physical mechanisms responsible 
for quenching in massive central galaxies by analysing multiple 
sets of theoretical predictions and testing them against
observations of the local Universe. More specifically, we make 
use of the publicly available data from three state-of-the-art 
cosmological simulations: EAGLE \citep{Schaye15, Crain15}, 
Illustris \citep{Genel14, Vogelsberger14a, Vogelsberger14b, Sijacki15} 
and TNG\citep{Marinacci18, Naiman18, Nelson18, Springel18, Pillepich18b}, 
to understand how star formation is quenched in these 
approximations of our Universe. We then compare the detailed 
predictions from these three simulations to observational 
measurements from the SDSS. Because all three simulations 
invoke AGN feedback as their primary tool to shut down star 
formation in massive centrals, we have a~perfect opportunity 
to thoroughly explore this quenching avenue in the local Universe.

We first perform a~Random Forest experiment to determine which 
parameter among \mbh, \mstar, \mhalo\ and \mdot\ 
(in the case of simulations) holds the most 
predictive power in determining whether a~given galaxy is star 
forming or quenched. We then follow up the outcome of machine 
learning with a~partial correlation analysis, to investigate 
the relationships between different variables and check whether 
their connection with quiescence is intrinsic or incidental. 
As a~final step we analyse the behaviour of gas content of 
galaxies during their transition towards quiescence. In order
to do this we infer ${\rm H_2}$ masses from optical extinction 
in the SDSS spectra and compare the observations to the 
published HI/${\rm H_2}$ catalogues in Illustris and TNG.
Our primary results are as follows:

\begin{enumerate}

\item the Random Forest analysis reveals black hole mass as the most important parameter for classifying a~galaxy into a~quenched or star forming category. This result is a~unanimous prediction of all three simulations and thus independent of the exact implementation of the AGN feedback model. It is also true in the SDSS, where it is invariant under different choices of \mbh\ calibration used in lieu of dynamical measurements of black hole mass. Our comparison between simulations and observations puts forward the dominance of black hole mass as the most robust observable consequence of AGN feedback quenching in massive central galaxies. This prediction is met exceptionally well in the SDSS which suggests that supermassive black hole feedback is the most likely quenching avenue in observed central galaxies
(Figs.\ref{fig:rf-combined}~\&~\ref{fig:rf-mdot}). \\

\item when we calculate the partial correlation coefficients we find that other galactic 
parameters like \mstar\ and \mhalo\ are both predicted and observed to 
carry little intrinsic correlation with a~reduction in sSFR. Instead, their
their connection to quenching found previously through a direct 
correlation analysis proves incidental once the \mbh-\mstar\ and
\mbh-\mhalo\ relationships are accounted for 
(Figs.~\ref{fig:corr-mstar}~\&~\ref{fig:corr-mhalo}). \\

\item simulations predict a~negligible importance of \mdot\ in comparison with \mbh. This finding suggests that the observational search for AGN feedback quenching should focus on the integrated history of the AGN feedback (encoded in \mbh) rather than the instantaneous activity of the AGN (as captured by $L_{\rm AGN}$) (Figs.~\ref{fig:corr-mdot}~\&~\ref{fig:LAGN-MBH})\\

\item moving on to the evolution in gas content and quiescence, 
we find that quenching is associated with both a~reduction in SFE 
and \fgas, estimated within the SDSS fibre in the observations
(Figs.~\ref{fig:msfr-fgas}~\&~\ref{fig:msfr-sfe}). We also 
find that it is SFE rather than \fgas\ which holds the most 
predictive power in determining the star-forming / passive 
classification when only these two parameters are considered
by the RF algorithm (Fig.~\ref{fig:rf-gas}).\\

\item when we compare theoretical predictions with the SDSS, we find that
TNG predicts trends more consistent with observations than Illustris. We find
the most visible disagreement in SFE, which rises with increasing black hole
mass in Illustris and decreases in TNG (as it does in the SDSS). Once 
we add observational realism to the data, these differences are highlighted 
even more: quiescent galaxies in Illustris form stars efficiently at their centres, 
while their TNG counterparts remain quiescent at their cores. At the same time 
we stress that although TNG shows good qualitative predictions for the observed 
trends, observations do not match them in the highest \mbh\ 
bins, where gas fractions and in-fibre sSFRs are significantly lower 
in TNG than in the SDSS 
(Figs.\ref{fig:line-plots-raw}~\&~\ref{fig:line-plots-sdss}).
More importantly, however, TNG also  significantly overestimates the 
quenching thresholds in 
\mhalo\ and \mbh\ when compared with the SDSS (Fig.~\ref{fig:fq}).
\end{enumerate}	
	

Our Random Forest experiment, together with the partial correlation 
analysis show results consistent with an AGN quenching scenario and
inconsistent with other quenching mechanism like e.g.~supernova 
feedback or virial shock heating. All three simulations predict a~negligible 
association between \mdot\ (and, by extension $L_{\rm AGN}$) and 
quiescence, regardless of their subgrid treatment of AGN feedback.
This finding might explain why previous studies aimed 
at linking elevated AGN activity with quenching delivered 
ambiguous results in which different authors were finding both
positive and negative correlations between $L_{\rm AGN}$ and 
quiescence or even a lack thereof. The difference in timescales
between the change in SFR and the change in \mdot\ (the AGN
duty cycle) is crucial for the potential appearance of these correlations
in the observable Universe, as argued by e.g. \cite{Hickox09} and
\cite{Harrison19}. Hence, in this study a~clear picture emerges in which
it is the \textit{the total power output of the AGN integrated over time,
i.e energy, which dictates the fate of a~galaxy} rather than its 
instantaneous activity. The simulations clearly show how different 
prescriptions for AGN feedback implementation result in contradictory 
trends in the $L_{\rm AGN}$ observable, while the picture in 
\mbh\ is consistent across the suites. We therefore conclude 
that black hole mass is a more robust tracer 
of AGN feedback than \mdot, boasting independence of the model 
prescription for the feedback mode of operation.

Finally, through our analysis of molecular gas content we find 
the most probable mode of operation of the AGN. We show 
how the central black hole has influence on both a) depleting 
the star forming gas reservoirs and b) decreasing the efficiency 
with which the remaining gas is collapsing to form new stars. 
We learn that a~successful AGN feedback model most likely 
\textit{involves a~combination of turbulence injection and heating}, 
which leads to the observed depletion of molecular gas and 
decreased star formation efficiency upon transition to quiescence.

\section*{Acknowledgements}

We would like to sincerely thank an anonymous referee
for their insightful comments which greatly improved this publication.
We would also like to express our deepest gratitude to the Illustris, 
IllustrisTNG  and EAGLE teams
for making their data publicly available and straightforward to use,
along with the excellent online documentation.
JMP and YP gratefully acknowledge funding from the MERAC Foundation.
AFLB and RM gratefully acknowledge ERC Advanced Grant 695671
”QUENCH”, and support from the Science and Technology Facilities
Council (STFC). YP acknowledges National Science Foundation of 
China (NSFC) Grant No. 12125301, 11773001 and the science 
research grants from the China Manned Space Project with No. CMS-CSST-2021-A07.
The flagship simulations of the IllustrisTNG have been run on the HazelHen
Cray XC40-system at the High Performance Computing Center
Stuttgart as part of project GCS-ILLU of the Gauss centres for Supercomputing (GCS). 
Illustris Simulations were run on the Harvard Odyssey and CfA/ITC 
clusters, the Ranger and Stampede supercomputers at the Texas 
Advanced Computing Center as part of XSEDE, the Kraken supercomputer 
at Oak Ridge National Laboratory as part of XSEDE, the CURIE supercomputer 
at CEA/France as part of PRACE project RA0844 and
the SuperMUC computer at the Leibniz Computing Centre, Germany, 
as part of project pr85je.

This work makes use of data from SDSS-I \& SDSS-IV. Funding for
the SDSS has been provided by the Alfred P. Sloan Foundation, the
Participating Institutions, the National Science Foundation, the U.S.
Department of Energy, the National Aeronautics and Space Administration, 
the Japanese Monbukagakusho, the Max Planck Society,
and the Higher Education Funding Council for England. Additional
funding towards SDSS-IV has been provided by the U.S. Department 
of Energy Office of Science. SDSS-IV acknowledges support and
resources from the Center for High-Performance Computing at the
University of Utah. The SDSS website is: \href{www.sdss.org}{www.sdss.org}.

The SDSS is managed by the Astrophysical Research Consortium 
for the Participating Institutions of the SDSS Collaboration. 
For SDSS-IV this includes the Brazilian Participation Group,
the Carnegie Institution for Science, Carnegie Mellon University,
the Chilean Participation Group, the French Participation Group,
Harvard-Smithsonian Center for Astrophysics, Instituto de 
Astrofisica de Canarias, The Johns Hopkins University, Kavli Institute
for the Physics and Mathematics of the Universe (IPMU) / University
of Tokyo, Lawrence Berkeley National Laboratory, Leibniz Institut
fur Astrophysik Potsdam (AIP), Max-Planck-Institut fur Astronomie
(MPIA Heidelberg), Max-Planck-Institut fur Astrophysik (MPA
Garching), Max-Planck-Institut fur Extraterrestrische Physik (MPE),
National Astronomical Observatory of China, New Mexico State
University, New York University, University of Notre Dame, 
Observatario Nacional / MCTI, The Ohio State University, 
Pennsylvania State University, Shanghai Astronomical Observatory, 
United Kingdom Participation Group, Universidad Nacional Autonoma de
Mexico, University of Arizona, University of Colorado Boulder,
University of Oxford, University of Portsmouth, University of Utah,
University of Virginia, University of Washington, University of
Wisconsin, Vanderbilt University, and Yale University.

\section*{Data availability}

All data used in this study have been previously published
and are available at the following online locations:

\renewcommand\labelitemi{---}

\begin{itemize}
\item EAGLE: \href{http://virgodb.dur.ac.uk/}{virgodb.dur.ac.uk/}
\item Illustris: \href{https://www.illustris-project.org/}{www.illustris-project.org/}
\item TNG: \href{https://www.tng-project.org/}{https://www.tng-project.org/}
\item MPA-JHU release of spectrum measurements: \\ \href{https://wwwmpa.mpa-garching.mpg.de/SDSS/DR7/}{wwwmpa.mpa-garching.mpg.de/SDSS/DR7/}
\item SDSS Group Catalogue: \href{https://gax.sjtu.edu.cn/data/Group.html}{gax.sjtu.edu.cn/data/Group.html}
\item NYU Galaxy Value Added Catalogue: \\ \href{http://sdss.physics.nyu.edu/vagc/}{sdss.physics.nyu.edu/vagc/}
\item morphological catalogues: \\ \href{https://doi.org/10.1088/0067-0049/196/1/11}{doi.org/10.1088/0067-0049/196/1/11} \\ \href{https://doi.org//10.26093/cds/vizier.22100003}{doi.org//10.26093/cds/vizier.22100003}
\end{itemize}




\bibliographystyle{mnras}
\bibliography{bibliography}

\begin{thebibliography}{}
\makeatletter
\relax
\def\mn@urlcharsother{\let\do\@makeother \do\$\do\&\do\#\do\^\do\_\do\%\do\~}
\def\mn@doi{\begingroup\mn@urlcharsother \@ifnextchar [ {\mn@doi@}
  {\mn@doi@[]}}
\def\mn@doi@[#1]#2{\def\@tempa{#1}\ifx\@tempa\@empty \href
  {http://dx.doi.org/#2} {doi:#2}\else \href {http://dx.doi.org/#2} {#1}\fi
  \endgroup}
\def\mn@eprint#1#2{\mn@eprint@#1:#2::\@nil}
\def\mn@eprint@arXiv#1{\href {http://arxiv.org/abs/#1} {{\tt arXiv:#1}}}
\def\mn@eprint@dblp#1{\href {http://dblp.uni-trier.de/rec/bibtex/#1.xml}
  {dblp:#1}}
\def\mn@eprint@#1:#2:#3:#4\@nil{\def\@tempa {#1}\def\@tempb {#2}\def\@tempc
  {#3}\ifx \@tempc \@empty \let \@tempc \@tempb \let \@tempb \@tempa \fi \ifx
  \@tempb \@empty \def\@tempb {arXiv}\fi \@ifundefined
  {mn@eprint@\@tempb}{\@tempb:\@tempc}{\expandafter \expandafter \csname
  mn@eprint@\@tempb\endcsname \expandafter{\@tempc}}}

\bibitem[\protect\citeauthoryear{{Abazajian} et~al.,}{{Abazajian}
  et~al.}{2009}]{Abazajian09}
{Abazajian} K.~N.,  et~al., 2009, \mn@doi [\apjs]
  {10.1088/0067-0049/182/2/543}, \href
  {https://ui.adsabs.harvard.edu/abs/2009ApJS..182..543A} {182, 543}

\bibitem[\protect\citeauthoryear{{Aird} et~al.,}{{Aird} et~al.}{2012}]{Aird12}
{Aird} J.,  et~al., 2012, \mn@doi [\apj] {10.1088/0004-637X/746/1/90}, \href
  {https://ui.adsabs.harvard.edu/abs/2012ApJ...746...90A} {746, 90}

\bibitem[\protect\citeauthoryear{{Aravena} et~al.,}{{Aravena}
  et~al.}{2019}]{Aravena19}
{Aravena} M.,  et~al., 2019, \mn@doi [\apj] {10.3847/1538-4357/ab30df}, \href
  {https://ui.adsabs.harvard.edu/abs/2019ApJ...882..136A} {882, 136}

\bibitem[\protect\citeauthoryear{{Baldry}, {Glazebrook}, {Brinkmann},
  {Ivezi{\'c}}, {Lupton}, {Nichol}  \& {Szalay}}{{Baldry}
  et~al.}{2004}]{Baldry04}
{Baldry} I.~K.,  {Glazebrook} K.,  {Brinkmann} J.,  {Ivezi{\'c}} {\v{Z}}.,
  {Lupton} R.~H.,  {Nichol} R.~C.,   {Szalay} A.~S.,  2004, \mn@doi [\apj]
  {10.1086/380092}, \href
  {https://ui.adsabs.harvard.edu/abs/2004ApJ...600..681B} {600, 681}

\bibitem[\protect\citeauthoryear{{Baldry}, {Balogh}, {Bower}, {Glazebrook},
  {Nichol}, {Bamford}  \& {Budavari}}{{Baldry} et~al.}{2006}]{Baldry06}
{Baldry} I.~K.,  {Balogh} M.~L.,  {Bower} R.~G.,  {Glazebrook} K.,  {Nichol}
  R.~C.,  {Bamford} S.~P.,   {Budavari} T.,  2006, \mn@doi [\mnras]
  {10.1111/j.1365-2966.2006.11081.x}, \href
  {https://ui.adsabs.harvard.edu/abs/2006MNRAS.373..469B} {373, 469}

\bibitem[\protect\citeauthoryear{{Baldwin}, {Phillips}  \&
  {Terlevich}}{{Baldwin} et~al.}{1981}]{BPT81}
{Baldwin} J.~A.,  {Phillips} M.~M.,   {Terlevich} R.,  1981, \mn@doi [\pasp]
  {10.1086/130766}, \href
  {https://ui.adsabs.harvard.edu/abs/1981PASP...93....5B} {93, 5}

\bibitem[\protect\citeauthoryear{{Bell} et~al.,}{{Bell} et~al.}{2004}]{Bell04}
{Bell} E.~F.,  et~al., 2004, \mn@doi [\apj] {10.1086/420778}, \href
  {https://ui.adsabs.harvard.edu/abs/2004ApJ...608..752B} {608, 752}

\bibitem[\protect\citeauthoryear{{Bell} et~al.,}{{Bell} et~al.}{2012}]{Bell12}
{Bell} E.~F.,  et~al., 2012, \mn@doi [\apj] {10.1088/0004-637X/753/2/167},
  \href {https://ui.adsabs.harvard.edu/abs/2012ApJ...753..167B} {753, 167}

\bibitem[\protect\citeauthoryear{{Belli} et~al.,}{{Belli}
  et~al.}{2021}]{Belli21}
{Belli} S.,  et~al., 2021, arXiv e-prints, \href
  {https://ui.adsabs.harvard.edu/abs/2021arXiv210207881B} {p. arXiv:2102.07881}

\bibitem[\protect\citeauthoryear{{Binney}}{{Binney}}{2004}]{Binney04}
{Binney} J.,  2004, \mn@doi [\mnras] {10.1111/j.1365-2966.2004.07277.x}, \href
  {https://ui.adsabs.harvard.edu/abs/2004MNRAS.347.1093B} {347, 1093}

\bibitem[\protect\citeauthoryear{{B{\^\i}rzan}, {Rafferty}, {McNamara}, {Wise}
  \& {Nulsen}}{{B{\^\i}rzan} et~al.}{2004}]{Birzan04}
{B{\^\i}rzan} L.,  {Rafferty} D.~A.,  {McNamara} B.~R.,  {Wise} M.~W.,
  {Nulsen} P.~E.~J.,  2004, \mn@doi [\apj] {10.1086/383519}, \href
  {https://ui.adsabs.harvard.edu/abs/2004ApJ...607..800B} {607, 800}

\bibitem[\protect\citeauthoryear{{Blanton} et~al.,}{{Blanton}
  et~al.}{2005}]{Blanton05}
{Blanton} M.~R.,  et~al., 2005, \mn@doi [\aj] {10.1086/429803}, \href
  {https://ui.adsabs.harvard.edu/abs/2005AJ....129.2562B} {129, 2562}

\bibitem[\protect\citeauthoryear{{Bluck}, {Conselice}, {Almaini}, {Laird},
  {Nandra}  \& {Gr{\"u}tzbauch}}{{Bluck} et~al.}{2011}]{Bluck11}
{Bluck} A. F.~L.,  {Conselice} C.~J.,  {Almaini} O.,  {Laird} E.~S.,  {Nandra}
  K.,   {Gr{\"u}tzbauch} R.,  2011, \mn@doi [\mnras]
  {10.1111/j.1365-2966.2010.17521.x}, \href
  {https://ui.adsabs.harvard.edu/abs/2011MNRAS.410.1174B} {410, 1174}

\bibitem[\protect\citeauthoryear{{Bluck}, {Ellison}, {Patton}, {Simard},
  {Mendel}, {Teimoorinia}, {Moreno}  \& {Starkenburg}}{{Bluck}
  et~al.}{2014}]{Bluck14}
{Bluck} A. F.~L.,  {Ellison} S.~L.,  {Patton} D.~R.,  {Simard} L.,  {Mendel}
  J.~T.,  {Teimoorinia} H.,  {Moreno} J.,   {Starkenburg} E.,  2014, arXiv
  e-prints, \href {https://ui.adsabs.harvard.edu/abs/2014arXiv1412.3862B} {p.
  arXiv:1412.3862}

\bibitem[\protect\citeauthoryear{{Bluck} et~al.,}{{Bluck}
  et~al.}{2016}]{Bluck16}
{Bluck} A. F.~L.,  et~al., 2016, \mn@doi [\mnras] {10.1093/mnras/stw1665},
  \href {https://ui.adsabs.harvard.edu/abs/2016MNRAS.462.2559B} {462, 2559}

\bibitem[\protect\citeauthoryear{{Bluck} et~al.,}{{Bluck}
  et~al.}{2019}]{Bluck19}
{Bluck} A. F.~L.,  et~al., 2019, \mn@doi [\mnras] {10.1093/mnras/stz363}, \href
  {https://ui.adsabs.harvard.edu/abs/2019MNRAS.485..666B} {485, 666}

\bibitem[\protect\citeauthoryear{{Bluck}, {Maiolino}, {S{\'a}nchez}, {Ellison},
  {Thorp}, {Piotrowska}, {Teimoorinia}  \& {Bundy}}{{Bluck}
  et~al.}{2020a}]{Bluck20a}
{Bluck} A. F.~L.,  {Maiolino} R.,  {S{\'a}nchez} S.~F.,  {Ellison} S.~L.,
  {Thorp} M.~D.,  {Piotrowska} J.~M.,  {Teimoorinia} H.,   {Bundy} K.~A.,
  2020a, \mn@doi [\mnras] {10.1093/mnras/stz3264}, \href
  {https://ui.adsabs.harvard.edu/abs/2020MNRAS.492...96B} {492, 96}

\bibitem[\protect\citeauthoryear{{Bluck} et~al.,}{{Bluck}
  et~al.}{2020b}]{Bluck20b}
{Bluck} A. F.~L.,  et~al., 2020b, \mn@doi [\mnras] {10.1093/mnras/staa2806},
  \href {https://ui.adsabs.harvard.edu/abs/2020MNRAS.499..230B} {499, 230}

\bibitem[\protect\citeauthoryear{{Bluck}, {Maiolino}, {Brownson}, {Conselice},
  {Ellison}, {Piotrowska}  \& {Thorp}}{{Bluck} et~al.}{2021}]{Bluck21}
{Bluck} A. F.~L.,  {Maiolino} R.,  {Brownson} S.,  {Conselice} C.~J.,
  {Ellison} S.~L.,  {Piotrowska} J.~M.,   {Thorp} M.~D.,  submitted, 2021,
  \mnras

\bibitem[\protect\citeauthoryear{{Bolatto} et~al.,}{{Bolatto}
  et~al.}{2017}]{Bolatto17}
{Bolatto} A.~D.,  et~al., 2017, \mn@doi [\apj] {10.3847/1538-4357/aa86aa},
  \href {https://ui.adsabs.harvard.edu/abs/2017ApJ...846..159B} {846, 159}

\bibitem[\protect\citeauthoryear{{Bondi}}{{Bondi}}{1952}]{Bondi52}
{Bondi} H.,  1952, \mn@doi [\mnras] {10.1093/mnras/112.2.195}, \href
  {https://ui.adsabs.harvard.edu/abs/1952MNRAS.112..195B} {112, 195}

\bibitem[\protect\citeauthoryear{{Bondi} \& {Hoyle}}{{Bondi} \&
  {Hoyle}}{1944}]{Bondi44}
{Bondi} H.,  {Hoyle} F.,  1944, \mn@doi [\mnras] {10.1093/mnras/104.5.273},
  \href {https://ui.adsabs.harvard.edu/abs/1944MNRAS.104..273B} {104, 273}

\bibitem[\protect\citeauthoryear{{Booth} \& {Schaye}}{{Booth} \&
  {Schaye}}{2009}]{Booth09}
{Booth} C.~M.,  {Schaye} J.,  2009, \mn@doi [\mnras]
  {10.1111/j.1365-2966.2009.15043.x}, \href
  {https://ui.adsabs.harvard.edu/abs/2009MNRAS.398...53B} {398, 53}

\bibitem[\protect\citeauthoryear{{Booth} \& {Schaye}}{{Booth} \&
  {Schaye}}{2011}]{Booth11}
{Booth} C.~M.,  {Schaye} J.,  2011, \mn@doi [\mnras]
  {10.1111/j.1365-2966.2011.18203.x}, \href
  {https://ui.adsabs.harvard.edu/abs/2011MNRAS.413.1158B} {413, 1158}

\bibitem[\protect\citeauthoryear{{Bower}, {Benson}, {Malbon}, {Helly}, {Frenk},
  {Baugh}, {Cole}  \& {Lacey}}{{Bower} et~al.}{2006}]{Bower06}
{Bower} R.~G.,  {Benson} A.~J.,  {Malbon} R.,  {Helly} J.~C.,  {Frenk} C.~S.,
  {Baugh} C.~M.,  {Cole} S.,   {Lacey} C.~G.,  2006, \mn@doi [\mnras]
  {10.1111/j.1365-2966.2006.10519.x}, \href
  {https://ui.adsabs.harvard.edu/abs/2006MNRAS.370..645B} {370, 645}

\bibitem[\protect\citeauthoryear{{Brammer} et~al.,}{{Brammer}
  et~al.}{2009}]{Brammer09}
{Brammer} G.~B.,  et~al., 2009, \mn@doi [\apjl] {10.1088/0004-637X/706/1/L173},
  \href {https://ui.adsabs.harvard.edu/abs/2009ApJ...706L.173B} {706, L173}

\bibitem[\protect\citeauthoryear{{Breiman}}{{Breiman}}{2001}]{Breiman01}
{Breiman} L.,  2001, \mn@doi [{Machine Learning}] {10.1023/A:1010933404324},
  45, 5

\bibitem[\protect\citeauthoryear{{Brinchmann}, {Charlot}, {White}, {Tremonti},
  {Kauffmann}, {Heckman}  \& {Brinkmann}}{{Brinchmann}
  et~al.}{2004}]{Brinchmann04}
{Brinchmann} J.,  {Charlot} S.,  {White} S.~D.~M.,  {Tremonti} C.,  {Kauffmann}
  G.,  {Heckman} T.,   {Brinkmann} J.,  2004, \mn@doi [\mnras]
  {10.1111/j.1365-2966.2004.07881.x}, \href
  {https://ui.adsabs.harvard.edu/abs/2004MNRAS.351.1151B} {351, 1151}

\bibitem[\protect\citeauthoryear{{Bruzual} \& {Charlot}}{{Bruzual} \&
  {Charlot}}{2003}]{Bruzual03}
{Bruzual} G.,  {Charlot} S.,  2003, \mn@doi [\mnras]
  {10.1046/j.1365-8711.2003.06897.x}, \href
  {https://ui.adsabs.harvard.edu/abs/2003MNRAS.344.1000B} {344, 1000}

\bibitem[\protect\citeauthoryear{{Bundy} et~al.,}{{Bundy}
  et~al.}{2008}]{Bundy08}
{Bundy} K.,  et~al., 2008, \mn@doi [\apj] {10.1086/588719}, \href
  {https://ui.adsabs.harvard.edu/abs/2008ApJ...681..931B} {681, 931}

\bibitem[\protect\citeauthoryear{{Cameron} \& {Driver}}{{Cameron} \&
  {Driver}}{2009}]{Cameron09a}
{Cameron} E.,  {Driver} S.~P.,  2009, \mn@doi [\aap]
  {10.1051/0004-6361:20078558}, \href
  {https://ui.adsabs.harvard.edu/abs/2009A&A...493..489C} {493, 489}

\bibitem[\protect\citeauthoryear{{Cameron}, {Driver}, {Graham}  \&
  {Liske}}{{Cameron} et~al.}{2009}]{Cameron09b}
{Cameron} E.,  {Driver} S.~P.,  {Graham} A.~W.,   {Liske} J.,  2009, \mn@doi
  [\apj] {10.1088/0004-637X/699/1/105}, \href
  {https://ui.adsabs.harvard.edu/abs/2009ApJ...699..105C} {699, 105}

\bibitem[\protect\citeauthoryear{{Cardelli}, {Clayton}  \& {Mathis}}{{Cardelli}
  et~al.}{1989}]{Cardelli89}
{Cardelli} J.~A.,  {Clayton} G.~C.,   {Mathis} J.~S.,  1989, \mn@doi [\apj]
  {10.1086/167900}, \href
  {https://ui.adsabs.harvard.edu/abs/1989ApJ...345..245C} {345, 245}

\bibitem[\protect\citeauthoryear{{Carniani} et~al.,}{{Carniani}
  et~al.}{2015}]{Carniani15}
{Carniani} S.,  et~al., 2015, \mn@doi [\aap] {10.1051/0004-6361/201526557},
  \href {https://ui.adsabs.harvard.edu/abs/2015A&A...580A.102C} {580, A102}

\bibitem[\protect\citeauthoryear{{Colombo} et~al.,}{{Colombo}
  et~al.}{2020}]{Colombo20}
{Colombo} D.,  et~al., 2020, \mn@doi [\aap] {10.1051/0004-6361/202039005},
  \href {https://ui.adsabs.harvard.edu/abs/2020A&A...644A..97C} {644, A97}

\bibitem[\protect\citeauthoryear{{Crain} et~al.,}{{Crain}
  et~al.}{2015}]{Crain15}
{Crain} R.~A.,  et~al., 2015, \mn@doi [\mnras] {10.1093/mnras/stv725}, \href
  {https://ui.adsabs.harvard.edu/abs/2015MNRAS.450.1937C} {450, 1937}

\bibitem[\protect\citeauthoryear{{Crain} et~al.,}{{Crain}
  et~al.}{2017}]{Crain17}
{Crain} R.~A.,  et~al., 2017, \mn@doi [\mnras] {10.1093/mnras/stw2586}, \href
  {https://ui.adsabs.harvard.edu/abs/2017MNRAS.464.4204C} {464, 4204}

\bibitem[\protect\citeauthoryear{{Crenshaw}, {Schmitt}, {Kraemer}, {Mushotzky}
  \& {Dunn}}{{Crenshaw} et~al.}{2010}]{Crenshaw10}
{Crenshaw} D.~M.,  {Schmitt} H.~R.,  {Kraemer} S.~B.,  {Mushotzky} R.~F.,
  {Dunn} J.~P.,  2010, \mn@doi [\apj] {10.1088/0004-637X/708/1/419}, \href
  {https://ui.adsabs.harvard.edu/abs/2010ApJ...708..419C} {708, 419}

\bibitem[\protect\citeauthoryear{{Croton} et~al.,}{{Croton}
  et~al.}{2006}]{Croton06}
{Croton} D.~J.,  et~al., 2006, \mn@doi [\mnras]
  {10.1111/j.1365-2966.2005.09675.x}, \href
  {https://ui.adsabs.harvard.edu/abs/2006MNRAS.365...11C} {365, 11}

\bibitem[\protect\citeauthoryear{{Davies}, {Crain}, {McCarthy}, {Oppenheimer},
  {Schaye}, {Schaller}  \& {McAlpine}}{{Davies} et~al.}{2019}]{Davies19}
{Davies} J.~J.,  {Crain} R.~A.,  {McCarthy} I.~G.,  {Oppenheimer} B.~D.,
  {Schaye} J.,  {Schaller} M.,   {McAlpine} S.,  2019, \mn@doi [\mnras]
  {10.1093/mnras/stz635}, \href
  {https://ui.adsabs.harvard.edu/abs/2019MNRAS.485.3783D} {485, 3783}

\bibitem[\protect\citeauthoryear{{Davis} \& {Laor}}{{Davis} \&
  {Laor}}{2011}]{Davis11}
{Davis} S.~W.,  {Laor} A.,  2011, \mn@doi [\apj] {10.1088/0004-637X/728/2/98},
  \href {https://ui.adsabs.harvard.edu/abs/2011ApJ...728...98D} {728, 98}

\bibitem[\protect\citeauthoryear{{Dekel} \& {Birnboim}}{{Dekel} \&
  {Birnboim}}{2006}]{Dekel06}
{Dekel} A.,  {Birnboim} Y.,  2006, \mn@doi [\mnras]
  {10.1111/j.1365-2966.2006.10145.x}, \href
  {https://ui.adsabs.harvard.edu/abs/2006MNRAS.368....2D} {368, 2}

\bibitem[\protect\citeauthoryear{{Diemer} et~al.,}{{Diemer}
  et~al.}{2018}]{Diemer18}
{Diemer} B.,  et~al., 2018, \mn@doi [\apjs] {10.3847/1538-4365/aae387}, \href
  {https://ui.adsabs.harvard.edu/abs/2018ApJS..238...33D} {238, 33}

\bibitem[\protect\citeauthoryear{{Donnari} et~al.,}{{Donnari}
  et~al.}{2021a}]{Donnari21a}
{Donnari} M.,  et~al., 2021a, \mn@doi [\mnras] {10.1093/mnras/staa3006}, \href
  {https://ui.adsabs.harvard.edu/abs/2021MNRAS.500.4004D} {500, 4004}

\bibitem[\protect\citeauthoryear{{Donnari}, {Pillepich}, {Nelson}, {Marinacci},
  {Vogelsberger}  \& {Hernquist}}{{Donnari} et~al.}{2021b}]{Donnari21b}
{Donnari} M.,  {Pillepich} A.,  {Nelson} D.,  {Marinacci} F.,  {Vogelsberger}
  M.,   {Hernquist} L.,  2021b, \mn@doi [\mnras] {10.1093/mnras/stab1950},
  \href {https://ui.adsabs.harvard.edu/abs/2021MNRAS.506.4760D} {506, 4760}

\bibitem[\protect\citeauthoryear{{Dou} et~al.,}{{Dou} et~al.}{2021a}]{Dou21a}
{Dou} J.,  et~al., 2021a, \mn@doi [\apj] {10.3847/1538-4357/abd17c}, \href
  {https://ui.adsabs.harvard.edu/abs/2021ApJ...907..114D} {907, 114}

\bibitem[\protect\citeauthoryear{{Dou} et~al.,}{{Dou} et~al.}{2021b}]{Dou21b}
{Dou} J.,  et~al., 2021b, \mn@doi [\apj] {10.3847/1538-4357/abfaf7}, \href
  {https://ui.adsabs.harvard.edu/abs/2021ApJ...915...94D} {915, 94}

\bibitem[\protect\citeauthoryear{{Dubois}, {Peirani}, {Pichon}, {Devriendt},
  {Gavazzi}, {Welker}  \& {Volonteri}}{{Dubois} et~al.}{2016}]{Dubois16}
{Dubois} Y.,  {Peirani} S.,  {Pichon} C.,  {Devriendt} J.,  {Gavazzi} R.,
  {Welker} C.,   {Volonteri} M.,  2016, \mn@doi [\mnras]
  {10.1093/mnras/stw2265}, \href
  {https://ui.adsabs.harvard.edu/abs/2016MNRAS.463.3948D} {463, 3948}

\bibitem[\protect\citeauthoryear{{Ellison} et~al.,}{{Ellison}
  et~al.}{2020}]{Ellison20}
{Ellison} S.~L.,  et~al., 2020, \mn@doi [\mnras] {10.1093/mnrasl/slz179}, \href
  {https://ui.adsabs.harvard.edu/abs/2020MNRAS.493L..39E} {493, L39}

\bibitem[\protect\citeauthoryear{{Ellison}, {Lin}, {Thorp}, {Pan},
  {S{\'a}nchez}, {Bluck}  \& {Belfiore}}{{Ellison} et~al.}{2021}]{Ellison21}
{Ellison} S.~L.,  {Lin} L.,  {Thorp} M.~D.,  {Pan} H.-A.,  {S{\'a}nchez} S.~F.,
   {Bluck} A. F.~L.,   {Belfiore} F.,  2021, \mn@doi [\mnras]
  {10.1093/mnrasl/slaa199}, \href
  {https://ui.adsabs.harvard.edu/abs/2021MNRAS.502L...6E} {502, L6}

\bibitem[\protect\citeauthoryear{{Faber} et~al.,}{{Faber}
  et~al.}{2007}]{Faber07}
{Faber} S.~M.,  et~al., 2007, \mn@doi [\apj] {10.1086/519294}, \href
  {https://ui.adsabs.harvard.edu/abs/2007ApJ...665..265F} {665, 265}

\bibitem[\protect\citeauthoryear{{Fabian}}{{Fabian}}{2012}]{Fabian12}
{Fabian} A.~C.,  2012, \mn@doi [\araa] {10.1146/annurev-astro-081811-125521},
  \href {https://ui.adsabs.harvard.edu/abs/2012ARA&A..50..455F} {50, 455}

\bibitem[\protect\citeauthoryear{Fawcett}{Fawcett}{2006}]{Fawcett06}
Fawcett T.,  2006, \mn@doi [Pattern Recognition Letters]
  {https://doi.org/10.1016/j.patrec.2005.10.010}, 27, 861

\bibitem[\protect\citeauthoryear{{Federrath} \& {Klessen}}{{Federrath} \&
  {Klessen}}{2012}]{Federrath12}
{Federrath} C.,  {Klessen} R.~S.,  2012, \mn@doi [\apj]
  {10.1088/0004-637X/761/2/156}, \href
  {https://ui.adsabs.harvard.edu/abs/2012ApJ...761..156F} {761, 156}

\bibitem[\protect\citeauthoryear{{Feruglio}, {Maiolino}, {Piconcelli}, {Menci},
  {Aussel}, {Lamastra}  \& {Fiore}}{{Feruglio} et~al.}{2010}]{Feruglio10}
{Feruglio} C.,  {Maiolino} R.,  {Piconcelli} E.,  {Menci} N.,  {Aussel} H.,
  {Lamastra} A.,   {Fiore} F.,  2010, \mn@doi [\aap]
  {10.1051/0004-6361/201015164}, \href
  {https://ui.adsabs.harvard.edu/abs/2010A&A...518L.155F} {518, L155}

\bibitem[\protect\citeauthoryear{{Fluetsch} et~al.,}{{Fluetsch}
  et~al.}{2019}]{Fluetsch19}
{Fluetsch} A.,  et~al., 2019, \mn@doi [\mnras] {10.1093/mnras/sty3449}, \href
  {https://ui.adsabs.harvard.edu/abs/2019MNRAS.483.4586F} {483, 4586}

\bibitem[\protect\citeauthoryear{{Fluetsch} et~al.,}{{Fluetsch}
  et~al.}{2020}]{Fluetsch20}
{Fluetsch} A.,  et~al., 2020, arXiv e-prints, \href
  {https://ui.adsabs.harvard.edu/abs/2020arXiv200613232F} {p. arXiv:2006.13232}

\bibitem[\protect\citeauthoryear{{Fukazawa}, {Botoya-Nonesa}, {Pu}, {Ohto}  \&
  {Kawano}}{{Fukazawa} et~al.}{2006}]{Fukazawa06}
{Fukazawa} Y.,  {Botoya-Nonesa} J.~G.,  {Pu} J.,  {Ohto} A.,   {Kawano} N.,
  2006, \mn@doi [\apj] {10.1086/498081}, \href
  {https://ui.adsabs.harvard.edu/abs/2006ApJ...636..698F} {636, 698}

\bibitem[\protect\citeauthoryear{{Furlong} et~al.,}{{Furlong}
  et~al.}{2015}]{Furlong15}
{Furlong} M.,  et~al., 2015, \mn@doi [\mnras] {10.1093/mnras/stv852}, \href
  {https://ui.adsabs.harvard.edu/abs/2015MNRAS.450.4486F} {450, 4486}

\bibitem[\protect\citeauthoryear{{Genel} et~al.,}{{Genel}
  et~al.}{2014}]{Genel14}
{Genel} S.,  et~al., 2014, \mn@doi [\mnras] {10.1093/mnras/stu1654}, \href
  {https://ui.adsabs.harvard.edu/abs/2014MNRAS.445..175G} {445, 175}

\bibitem[\protect\citeauthoryear{{Genel} et~al.,}{{Genel}
  et~al.}{2018}]{Genel18}
{Genel} S.,  et~al., 2018, \mn@doi [\mnras] {10.1093/mnras/stx3078}, \href
  {https://ui.adsabs.harvard.edu/abs/2018MNRAS.474.3976G} {474, 3976}

\bibitem[\protect\citeauthoryear{{Gensior}, {Kruijssen}  \& {Keller}}{{Gensior}
  et~al.}{2020}]{Gensior20}
{Gensior} J.,  {Kruijssen} J.~M.~D.,   {Keller} B.~W.,  2020, \mn@doi [\mnras]
  {10.1093/mnras/staa1184}, \href
  {https://ui.adsabs.harvard.edu/abs/2020MNRAS.495..199G} {495, 199}

\bibitem[\protect\citeauthoryear{{Genzel} et~al.,}{{Genzel}
  et~al.}{2015}]{Genzel15}
{Genzel} R.,  et~al., 2015, \mn@doi [\apj] {10.1088/0004-637X/800/1/20}, \href
  {https://ui.adsabs.harvard.edu/abs/2015ApJ...800...20G} {800, 20}

\bibitem[\protect\citeauthoryear{{Georgakakis} et~al.,}{{Georgakakis}
  et~al.}{2008}]{Georgakakis08}
{Georgakakis} A.,  et~al., 2008, \mn@doi [\mnras]
  {10.1111/j.1365-2966.2008.12962.x}, \href
  {https://ui.adsabs.harvard.edu/abs/2008MNRAS.385.2049G} {385, 2049}

\bibitem[\protect\citeauthoryear{{Giallongo}, {Salimbeni}, {Menci}, {Zamorani},
  {Fontana}, {Dickinson}, {Cristiani}  \& {Pozzetti}}{{Giallongo}
  et~al.}{2005}]{Giallongo05}
{Giallongo} E.,  {Salimbeni} S.,  {Menci} N.,  {Zamorani} G.,  {Fontana} A.,
  {Dickinson} M.,  {Cristiani} S.,   {Pozzetti} L.,  2005, \mn@doi [\apj]
  {10.1086/427819}, \href
  {https://ui.adsabs.harvard.edu/abs/2005ApJ...622..116G} {622, 116}

\bibitem[\protect\citeauthoryear{{Gnedin} \& {Draine}}{{Gnedin} \&
  {Draine}}{2014}]{Gnedin14}
{Gnedin} N.~Y.,  {Draine} B.~T.,  2014, \mn@doi [\apj]
  {10.1088/0004-637X/795/1/37}, \href
  {https://ui.adsabs.harvard.edu/abs/2014ApJ...795...37G} {795, 37}

\bibitem[\protect\citeauthoryear{{Gnedin} \& {Kravtsov}}{{Gnedin} \&
  {Kravtsov}}{2011}]{Gnedin11}
{Gnedin} N.~Y.,  {Kravtsov} A.~V.,  2011, \mn@doi [\apj]
  {10.1088/0004-637X/728/2/88}, \href
  {https://ui.adsabs.harvard.edu/abs/2011ApJ...728...88G} {728, 88}

\bibitem[\protect\citeauthoryear{{Gudennavar}, {Bubbly}, {Preethi}  \&
  {Murthy}}{{Gudennavar} et~al.}{2012}]{Gudennavar12}
{Gudennavar} S.~B.,  {Bubbly} S.~G.,  {Preethi} K.,   {Murthy} J.,  2012,
  \mn@doi [\apjs] {10.1088/0067-0049/199/1/8}, \href
  {https://ui.adsabs.harvard.edu/abs/2012ApJS..199....8G} {199, 8}

\bibitem[\protect\citeauthoryear{{H{\"a}ring} \& {Rix}}{{H{\"a}ring} \&
  {Rix}}{2004}]{HaringRix04}
{H{\"a}ring} N.,  {Rix} H.-W.,  2004, \mn@doi [\apjl] {10.1086/383567}, \href
  {https://ui.adsabs.harvard.edu/abs/2004ApJ...604L..89H} {604, L89}

\bibitem[\protect\citeauthoryear{{Harrison}, {Alexander}, {Mullaney}  \&
  {Swinbank}}{{Harrison} et~al.}{2014}]{Harrison14}
{Harrison} C.~M.,  {Alexander} D.~M.,  {Mullaney} J.~R.,   {Swinbank} A.~M.,
  2014, \mn@doi [\mnras] {10.1093/mnras/stu515}, \href
  {https://ui.adsabs.harvard.edu/abs/2014MNRAS.441.3306H} {441, 3306}

\bibitem[\protect\citeauthoryear{{Harrison}, {Alexander}, {Rosario}, {Scholtz}
  \& {Stanley}}{{Harrison} et~al.}{2019}]{Harrison19}
{Harrison} C.~M.,  {Alexander} D.~M.,  {Rosario} D.~J.,  {Scholtz} J.,
  {Stanley} F.,  2019, arXiv e-prints, \href
  {https://ui.adsabs.harvard.edu/abs/2019arXiv191201020H} {p. arXiv:1912.01020}

\bibitem[\protect\citeauthoryear{{Hastie}, {Tibshirani}  \&
  {Friedman}}{{Hastie} et~al.}{2009}]{Hastie09}
{Hastie} T.,  {Tibshirani} R.,   {Friedman} J.,  2009, {The Elements of
  Statistical Learning (2nd edition}).
{Springer-Verlag}

\bibitem[\protect\citeauthoryear{{Heckman} \& {Best}}{{Heckman} \&
  {Best}}{2014}]{Heckman14}
{Heckman} T.~M.,  {Best} P.~N.,  2014, \mn@doi [\araa]
  {10.1146/annurev-astro-081913-035722}, \href
  {https://ui.adsabs.harvard.edu/abs/2014ARA&A..52..589H} {52, 589}

\bibitem[\protect\citeauthoryear{{Heinz}, {Reynolds}  \& {Begelman}}{{Heinz}
  et~al.}{1998}]{Heinz98}
{Heinz} S.,  {Reynolds} C.~S.,   {Begelman} M.~C.,  1998, \mn@doi [\apj]
  {10.1086/305807}, \href
  {https://ui.adsabs.harvard.edu/abs/1998ApJ...501..126H} {501, 126}

\bibitem[\protect\citeauthoryear{{Henriques}, {White}, {Lilly}, {Bell}, {Bluck}
   \& {Terrazas}}{{Henriques} et~al.}{2019}]{Henriques19}
{Henriques} B. M.~B.,  {White} S. D.~M.,  {Lilly} S.~J.,  {Bell} E.~F.,
  {Bluck} A. F.~L.,   {Terrazas} B.~A.,  2019, \mn@doi [\mnras]
  {10.1093/mnras/stz577}, \href
  {https://ui.adsabs.harvard.edu/abs/2019MNRAS.485.3446H} {485, 3446}

\bibitem[\protect\citeauthoryear{{Hickox} et~al.,}{{Hickox}
  et~al.}{2009}]{Hickox09}
{Hickox} R.~C.,  et~al., 2009, \mn@doi [\apj] {10.1088/0004-637X/696/1/891},
  \href {https://ui.adsabs.harvard.edu/abs/2009ApJ...696..891H} {696, 891}

\bibitem[\protect\citeauthoryear{{Hinshaw} et~al.,}{{Hinshaw}
  et~al.}{2013}]{Hinshaw13}
{Hinshaw} G.,  et~al., 2013, \mn@doi [\apjs] {10.1088/0067-0049/208/2/19},
  \href {https://ui.adsabs.harvard.edu/abs/2013ApJS..208...19H} {208, 19}

\bibitem[\protect\citeauthoryear{{Hlavacek-Larrondo}, {Fabian}, {Edge},
  {Ebeling}, {Sanders}, {Hogan}  \& {Taylor}}{{Hlavacek-Larrondo}
  et~al.}{2012}]{Hlavacek-Larrondo12}
{Hlavacek-Larrondo} J.,  {Fabian} A.~C.,  {Edge} A.~C.,  {Ebeling} H.,
  {Sanders} J.~S.,  {Hogan} M.~T.,   {Taylor} G.~B.,  2012, \mn@doi [\mnras]
  {10.1111/j.1365-2966.2011.20405.x}, \href
  {https://ui.adsabs.harvard.edu/abs/2012MNRAS.421.1360H} {421, 1360}

\bibitem[\protect\citeauthoryear{{Hlavacek-Larrondo}
  et~al.,}{{Hlavacek-Larrondo} et~al.}{2015}]{Hlavacek-Larrondo15}
{Hlavacek-Larrondo} J.,  et~al., 2015, \mn@doi [\apj]
  {10.1088/0004-637X/805/1/35}, \href
  {https://ui.adsabs.harvard.edu/abs/2015ApJ...805...35H} {805, 35}

\bibitem[\protect\citeauthoryear{{Hopkins} \& {Elvis}}{{Hopkins} \&
  {Elvis}}{2010}]{Hopkins10}
{Hopkins} P.~F.,  {Elvis} M.,  2010, \mn@doi [\mnras]
  {10.1111/j.1365-2966.2009.15643.x}, \href
  {https://ui.adsabs.harvard.edu/abs/2010MNRAS.401....7H} {401, 7}

\bibitem[\protect\citeauthoryear{{Hopkins}, {Hernquist}, {Cox}, {Robertson}  \&
  {Krause}}{{Hopkins} et~al.}{2007}]{Hopkins07}
{Hopkins} P.~F.,  {Hernquist} L.,  {Cox} T.~J.,  {Robertson} B.,   {Krause} E.,
   2007, \mn@doi [\apj] {10.1086/521601}, \href
  {https://ui.adsabs.harvard.edu/abs/2007ApJ...669...67H} {669, 67}

\bibitem[\protect\citeauthoryear{{Hopkins}, {Hernquist}, {Cox}  \&
  {Kere{\v{s}}}}{{Hopkins} et~al.}{2008}]{Hopkins08}
{Hopkins} P.~F.,  {Hernquist} L.,  {Cox} T.~J.,   {Kere{\v{s}}} D.,  2008,
  \mn@doi [\apjs] {10.1086/524362}, \href
  {https://ui.adsabs.harvard.edu/abs/2008ApJS..175..356H} {175, 356}

\bibitem[\protect\citeauthoryear{{Hopkins}, {Quataert}  \& {Murray}}{{Hopkins}
  et~al.}{2011}]{Hopkins11}
{Hopkins} P.~F.,  {Quataert} E.,   {Murray} N.,  2011, \mn@doi [\mnras]
  {10.1111/j.1365-2966.2011.19306.x}, \href
  {https://ui.adsabs.harvard.edu/abs/2011MNRAS.417..950H} {417, 950}

\bibitem[\protect\citeauthoryear{{Hoyle} \& {Lyttleton}}{{Hoyle} \&
  {Lyttleton}}{1939}]{Hoyle39}
{Hoyle} F.,  {Lyttleton} R.~A.,  1939, \mn@doi [Proceedings of the Cambridge
  Philosophical Society] {10.1017/S0305004100021150}, \href
  {https://ui.adsabs.harvard.edu/abs/1939PCPS...35..405H} {35, 405}

\bibitem[\protect\citeauthoryear{{Hummer} \& {Storey}}{{Hummer} \&
  {Storey}}{1987}]{Hummer87}
{Hummer} D.~G.,  {Storey} P.~J.,  1987, \mn@doi [\mnras]
  {10.1093/mnras/224.3.801}, \href
  {https://ui.adsabs.harvard.edu/abs/1987MNRAS.224..801H} {224, 801}

\bibitem[\protect\citeauthoryear{{Humphrey}, {Buote}, {Canizares}, {Fabian}  \&
  {Miller}}{{Humphrey} et~al.}{2011}]{Humphrey11}
{Humphrey} P.~J.,  {Buote} D.~A.,  {Canizares} C.~R.,  {Fabian} A.~C.,
  {Miller} J.~M.,  2011, \mn@doi [\apj] {10.1088/0004-637X/729/1/53}, \href
  {https://ui.adsabs.harvard.edu/abs/2011ApJ...729...53H} {729, 53}

\bibitem[\protect\citeauthoryear{{Jorgensen}, {Franx}  \&
  {Kjaergaard}}{{Jorgensen} et~al.}{1995}]{Jorgensen95}
{Jorgensen} I.,  {Franx} M.,   {Kjaergaard} P.,  1995, \mn@doi [\mnras]
  {10.1093/mnras/276.4.1341}, \href
  {https://ui.adsabs.harvard.edu/abs/1995MNRAS.276.1341J} {276, 1341}

\bibitem[\protect\citeauthoryear{{Kauffmann} et~al.,}{{Kauffmann}
  et~al.}{2003}]{Kauffmann03}
{Kauffmann} G.,  et~al., 2003, \mn@doi [\mnras]
  {10.1046/j.1365-8711.2003.06291.x}, \href
  {https://ui.adsabs.harvard.edu/abs/2003MNRAS.341...33K} {341, 33}

\bibitem[\protect\citeauthoryear{{Kay}, {Pearce}, {Frenk}  \& {Jenkins}}{{Kay}
  et~al.}{2002}]{Kay02}
{Kay} S.~T.,  {Pearce} F.~R.,  {Frenk} C.~S.,   {Jenkins} A.,  2002, \mn@doi
  [\mnras] {10.1046/j.1365-8711.2002.05070.x}, \href
  {https://ui.adsabs.harvard.edu/abs/2002MNRAS.330..113K} {330, 113}

\bibitem[\protect\citeauthoryear{{Kendall} \& {Stuart}}{{Kendall} \&
  {Stuart}}{1977}]{Kendall77}
{Kendall} M.~G.,  {Stuart} A.,  1977, {The Advanced Theory of Statistics:
  Inference and Relationship. (vol 2)}.
London : C. Griffin

\bibitem[\protect\citeauthoryear{{Kormendy} \& {Ho}}{{Kormendy} \&
  {Ho}}{2013}]{KormendyHo13}
{Kormendy} J.,  {Ho} L.~C.,  2013, \mn@doi [\araa]
  {10.1146/annurev-astro-082708-101811}, \href
  {https://ui.adsabs.harvard.edu/abs/2013ARA&A..51..511K} {51, 511}

\bibitem[\protect\citeauthoryear{{Kormendy}, {Bender}  \& {Cornell}}{{Kormendy}
  et~al.}{2011}]{Kormendy11}
{Kormendy} J.,  {Bender} R.,   {Cornell} M.~E.,  2011, \mn@doi [\nat]
  {10.1038/nature09694}, \href
  {https://ui.adsabs.harvard.edu/abs/2011Natur.469..374K} {469, 374}

\bibitem[\protect\citeauthoryear{{Krumholz}}{{Krumholz}}{2013}]{Krumholz13}
{Krumholz} M.~R.,  2013, \mn@doi [\mnras] {10.1093/mnras/stt1780}, \href
  {https://ui.adsabs.harvard.edu/abs/2013MNRAS.436.2747K} {436, 2747}

\bibitem[\protect\citeauthoryear{{Krumholz} \& {McKee}}{{Krumholz} \&
  {McKee}}{2005}]{Krumholz05}
{Krumholz} M.~R.,  {McKee} C.~F.,  2005, \mn@doi [\apj] {10.1086/431734}, \href
  {https://ui.adsabs.harvard.edu/abs/2005ApJ...630..250K} {630, 250}

\bibitem[\protect\citeauthoryear{{Kumari}, {Maiolino}, {Belfiore}  \&
  {Curti}}{{Kumari} et~al.}{2019}]{Kumari19}
{Kumari} N.,  {Maiolino} R.,  {Belfiore} F.,   {Curti} M.,  2019, \mn@doi
  [\mnras] {10.1093/mnras/stz366}, \href
  {https://ui.adsabs.harvard.edu/abs/2019MNRAS.485..367K} {485, 367}

\bibitem[\protect\citeauthoryear{{Lim}, {Mo}, {Lan}  \& {M{\'e}nard}}{{Lim}
  et~al.}{2017}]{Lim17}
{Lim} S.~H.,  {Mo} H.~J.,  {Lan} T.~W.,   {M{\'e}nard} B.,  2017, \mn@doi
  [\mnras] {10.1093/mnras/stw2553}, \href
  {https://ui.adsabs.harvard.edu/abs/2017MNRAS.464.3256L} {464, 3256}

\bibitem[\protect\citeauthoryear{{Lin} et~al.,}{{Lin} et~al.}{2017}]{Lin17}
{Lin} L.,  et~al., 2017, \mn@doi [\apj] {10.3847/1538-4357/aa96ae}, \href
  {https://ui.adsabs.harvard.edu/abs/2017ApJ...851...18L} {851, 18}

\bibitem[\protect\citeauthoryear{{Lin} et~al.,}{{Lin} et~al.}{2020}]{Lin20}
{Lin} L.,  et~al., 2020, \mn@doi [\apj] {10.3847/1538-4357/abba3a}, \href
  {https://ui.adsabs.harvard.edu/abs/2020ApJ...903..145L} {903, 145}

\bibitem[\protect\citeauthoryear{{Liu}, {Hao}, {Wang}  \& {Yang}}{{Liu}
  et~al.}{2019}]{Liu19}
{Liu} C.,  {Hao} L.,  {Wang} H.,   {Yang} X.,  2019, \mn@doi [\apj]
  {10.3847/1538-4357/ab1ea0}, \href
  {https://ui.adsabs.harvard.edu/abs/2019ApJ...878...69L} {878, 69}

\bibitem[\protect\citeauthoryear{{Louppe}}{{Louppe}}{2014}]{Loupe14}
{Louppe} G.,  2014, arXiv e-prints, \href
  {https://ui.adsabs.harvard.edu/abs/2014arXiv1407.7502L} {p. arXiv:1407.7502}

\bibitem[\protect\citeauthoryear{{Maiolino} et~al.,}{{Maiolino}
  et~al.}{2012}]{Maiolino12}
{Maiolino} R.,  et~al., 2012, \mn@doi [\mnras]
  {10.1111/j.1745-3933.2012.01303.x}, \href
  {https://ui.adsabs.harvard.edu/abs/2012MNRAS.425L..66M} {425, L66}

\bibitem[\protect\citeauthoryear{{Malmquist}}{{Malmquist}}{1922}]{Malmquist1922}
{Malmquist} K.~G.,  1922, Meddelanden fran Lunds Astronomiska Observatorium
  Serie I, \href {https://ui.adsabs.harvard.edu/abs/1922MeLuF.100....1M} {100,
  1}

\bibitem[\protect\citeauthoryear{{Maraston} \& {Str{\"o}mb{\"a}ck}}{{Maraston}
  \& {Str{\"o}mb{\"a}ck}}{2011}]{Maraston11}
{Maraston} C.,  {Str{\"o}mb{\"a}ck} G.,  2011, \mn@doi [\mnras]
  {10.1111/j.1365-2966.2011.19738.x}, \href
  {https://ui.adsabs.harvard.edu/abs/2011MNRAS.418.2785M} {418, 2785}

\bibitem[\protect\citeauthoryear{{Marinacci} et~al.,}{{Marinacci}
  et~al.}{2018}]{Marinacci18}
{Marinacci} F.,  et~al., 2018, \mn@doi [\mnras] {10.1093/mnras/sty2206}, \href
  {https://ui.adsabs.harvard.edu/abs/2018MNRAS.480.5113M} {480, 5113}

\bibitem[\protect\citeauthoryear{{Marri} \& {White}}{{Marri} \&
  {White}}{2003}]{Marri03}
{Marri} S.,  {White} S.~D.~M.,  2003, \mn@doi [\mnras]
  {10.1046/j.1365-8711.2003.06984.x}, \href
  {https://ui.adsabs.harvard.edu/abs/2003MNRAS.345..561M} {345, 561}

\bibitem[\protect\citeauthoryear{{Martig}, {Bournaud}, {Teyssier}  \&
  {Dekel}}{{Martig} et~al.}{2009}]{Martig09}
{Martig} M.,  {Bournaud} F.,  {Teyssier} R.,   {Dekel} A.,  2009, \mn@doi
  [\apj] {10.1088/0004-637X/707/1/250}, \href
  {https://ui.adsabs.harvard.edu/abs/2009ApJ...707..250M} {707, 250}

\bibitem[\protect\citeauthoryear{{Mart{\'\i}n-Navarro}, {Brodie}, {Romanowsky},
  {Ruiz-Lara}  \& {van de Ven}}{{Mart{\'\i}n-Navarro}
  et~al.}{2018}]{Martin-Navarro18}
{Mart{\'\i}n-Navarro} I.,  {Brodie} J.~P.,  {Romanowsky} A.~J.,  {Ruiz-Lara}
  T.,   {van de Ven} G.,  2018, \mn@doi [\nat] {10.1038/nature24999}, \href
  {https://ui.adsabs.harvard.edu/abs/2018Natur.553..307M} {553, 307}

\bibitem[\protect\citeauthoryear{{Mathur}, {Fields}, {Peterson}  \&
  {Grupe}}{{Mathur} et~al.}{2012}]{Mathur12}
{Mathur} S.,  {Fields} D.,  {Peterson} B.~M.,   {Grupe} D.,  2012, \mn@doi
  [\apj] {10.1088/0004-637X/754/2/146}, \href
  {https://ui.adsabs.harvard.edu/abs/2012ApJ...754..146M} {754, 146}

\bibitem[\protect\citeauthoryear{{McAlpine} et~al.,}{{McAlpine}
  et~al.}{2016}]{McAlpine16}
{McAlpine} S.,  et~al., 2016, \mn@doi [Astronomy and Computing]
  {10.1016/j.ascom.2016.02.004}, \href
  {https://ui.adsabs.harvard.edu/abs/2016A&C....15...72M} {15, 72}

\bibitem[\protect\citeauthoryear{{McCarthy}, {Schaye}, {Bower}, {Ponman},
  {Booth}, {Dalla Vecchia}  \& {Springel}}{{McCarthy}
  et~al.}{2011}]{McCarthy11}
{McCarthy} I.~G.,  {Schaye} J.,  {Bower} R.~G.,  {Ponman} T.~J.,  {Booth}
  C.~M.,  {Dalla Vecchia} C.,   {Springel} V.,  2011, \mn@doi [\mnras]
  {10.1111/j.1365-2966.2010.18033.x}, \href
  {https://ui.adsabs.harvard.edu/abs/2011MNRAS.412.1965M} {412, 1965}

\bibitem[\protect\citeauthoryear{{McConnell} \& {Ma}}{{McConnell} \&
  {Ma}}{2013}]{MM13}
{McConnell} N.~J.,  {Ma} C.-P.,  2013, \mn@doi [\apj]
  {10.1088/0004-637X/764/2/184}, \href
  {https://ui.adsabs.harvard.edu/abs/2013ApJ...764..184M} {764, 184}

\bibitem[\protect\citeauthoryear{{McNamara} \& {Nulsen}}{{McNamara} \&
  {Nulsen}}{2007}]{McNamara07}
{McNamara} B.~R.,  {Nulsen} P.~E.~J.,  2007, \mn@doi [\araa]
  {10.1146/annurev.astro.45.051806.110625}, \href
  {https://ui.adsabs.harvard.edu/abs/2007ARA&A..45..117M} {45, 117}

\bibitem[\protect\citeauthoryear{{McNamara} et~al.,}{{McNamara}
  et~al.}{2000}]{McNamara00}
{McNamara} B.~R.,  et~al., 2000, \mn@doi [\apjl] {10.1086/312662}, \href
  {https://ui.adsabs.harvard.edu/abs/2000ApJ...534L.135M} {534, L135}

\bibitem[\protect\citeauthoryear{{Mendel}, {Simard}, {Palmer}, {Ellison}  \&
  {Patton}}{{Mendel} et~al.}{2014}]{Mendel14}
{Mendel} J.~T.,  {Simard} L.,  {Palmer} M.,  {Ellison} S.~L.,   {Patton} D.~R.,
   2014, \mn@doi [\apjs] {10.1088/0067-0049/210/1/3}, \href
  {https://ui.adsabs.harvard.edu/abs/2014ApJS..210....3M} {210, 3}

\bibitem[\protect\citeauthoryear{{Misner}, {Thorne}  \& {Wheeler}}{{Misner}
  et~al.}{1973}]{Misner73}
{Misner} C.~W.,  {Thorne} K.~S.,   {Wheeler} J.~A.,  1973, {Gravitation}

\bibitem[\protect\citeauthoryear{{Naiman} et~al.,}{{Naiman}
  et~al.}{2018}]{Naiman18}
{Naiman} J.~P.,  et~al., 2018, \mn@doi [\mnras] {10.1093/mnras/sty618}, \href
  {https://ui.adsabs.harvard.edu/abs/2018MNRAS.477.1206N} {477, 1206}

\bibitem[\protect\citeauthoryear{{Nandra} et~al.,}{{Nandra}
  et~al.}{2007}]{Nandra07}
{Nandra} K.,  et~al., 2007, \mn@doi [\apjl] {10.1086/517918}, \href
  {https://ui.adsabs.harvard.edu/abs/2007ApJ...660L..11N} {660, L11}

\bibitem[\protect\citeauthoryear{{Nelson} et~al.,}{{Nelson}
  et~al.}{2015}]{Nelson15}
{Nelson} D.,  et~al., 2015, \mn@doi [Astronomy and Computing]
  {10.1016/j.ascom.2015.09.003}, \href
  {https://ui.adsabs.harvard.edu/abs/2015A&C....13...12N} {13, 12}

\bibitem[\protect\citeauthoryear{{Nelson} et~al.,}{{Nelson}
  et~al.}{2018}]{Nelson18}
{Nelson} D.,  et~al., 2018, \mn@doi [\mnras] {10.1093/mnras/stx3040}, \href
  {https://ui.adsabs.harvard.edu/abs/2018MNRAS.475..624N} {475, 624}

\bibitem[\protect\citeauthoryear{{Nelson} et~al.,}{{Nelson}
  et~al.}{2019}]{Nelson19}
{Nelson} D.,  et~al., 2019, \mn@doi [Computational Astrophysics and Cosmology]
  {10.1186/s40668-019-0028-x}, \href
  {https://ui.adsabs.harvard.edu/abs/2019ComAC...6....2N} {6, 2}

\bibitem[\protect\citeauthoryear{{Nelson} et~al.,}{{Nelson}
  et~al.}{2021}]{Nelson21}
{Nelson} E.~J.,  et~al., 2021, \mn@doi [\mnras] {10.1093/mnras/stab2131}, \href
  {https://ui.adsabs.harvard.edu/abs/2021MNRAS.tmp.2068N} {}

\bibitem[\protect\citeauthoryear{{Noeske} et~al.,}{{Noeske}
  et~al.}{2007}]{Noeske07}
{Noeske} K.~G.,  et~al., 2007, \mn@doi [\apjl] {10.1086/517926}, \href
  {https://ui.adsabs.harvard.edu/abs/2007ApJ...660L..43N} {660, L43}

\bibitem[\protect\citeauthoryear{{Omand}, {Balogh}  \& {Poggianti}}{{Omand}
  et~al.}{2014}]{Omand14}
{Omand} C. M.~B.,  {Balogh} M.~L.,   {Poggianti} B.~M.,  2014, \mn@doi [\mnras]
  {10.1093/mnras/stu331}, \href
  {https://ui.adsabs.harvard.edu/abs/2014MNRAS.440..843O} {440, 843}

\bibitem[\protect\citeauthoryear{{Paolillo}, {Fabbiano}, {Peres}  \&
  {Kim}}{{Paolillo} et~al.}{2002}]{Paolillo02}
{Paolillo} M.,  {Fabbiano} G.,  {Peres} G.,   {Kim} D.~W.,  2002, \mn@doi
  [\apj] {10.1086/337919}, \href
  {https://ui.adsabs.harvard.edu/abs/2002ApJ...565..883P} {565, 883}

\bibitem[\protect\citeauthoryear{{Pawlowski} \& {Kroupa}}{{Pawlowski} \&
  {Kroupa}}{2020}]{Pawlowski20}
{Pawlowski} M.~S.,  {Kroupa} P.,  2020, \mn@doi [\mnras]
  {10.1093/mnras/stz3163}, \href
  {https://ui.adsabs.harvard.edu/abs/2020MNRAS.491.3042P} {491, 3042}

\bibitem[\protect\citeauthoryear{Pedregosa et~al.,}{Pedregosa
  et~al.}{2011}]{scikit-learn}
Pedregosa F.,  et~al., 2011, Journal of Machine Learning Research, 12, 2825

\bibitem[\protect\citeauthoryear{{Peng} \& {Renzini}}{{Peng} \&
  {Renzini}}{2020}]{Peng20}
{Peng} Y.-j.,  {Renzini} A.,  2020, \mn@doi [\mnras] {10.1093/mnrasl/slz163},
  \href {https://ui.adsabs.harvard.edu/abs/2020MNRAS.491L..51P} {491, L51}

\bibitem[\protect\citeauthoryear{{Peng} et~al.,}{{Peng} et~al.}{2010}]{Peng10}
{Peng} Y.-j.,  et~al., 2010, \mn@doi [\apj] {10.1088/0004-637X/721/1/193},
  \href {https://ui.adsabs.harvard.edu/abs/2010ApJ...721..193P} {721, 193}

\bibitem[\protect\citeauthoryear{{Peng}, {Lilly}, {Renzini}  \&
  {Carollo}}{{Peng} et~al.}{2012}]{Peng12}
{Peng} Y.-j.,  {Lilly} S.~J.,  {Renzini} A.,   {Carollo} M.,  2012, \mn@doi
  [\apj] {10.1088/0004-637X/757/1/4}, \href
  {https://ui.adsabs.harvard.edu/abs/2012ApJ...757....4P} {757, 4}

\bibitem[\protect\citeauthoryear{{Pillepich} et~al.,}{{Pillepich}
  et~al.}{2018a}]{Pillepich18a}
{Pillepich} A.,  et~al., 2018a, \mn@doi [\mnras] {10.1093/mnras/stx2656}, \href
  {https://ui.adsabs.harvard.edu/abs/2018MNRAS.473.4077P} {473, 4077}

\bibitem[\protect\citeauthoryear{{Pillepich} et~al.,}{{Pillepich}
  et~al.}{2018b}]{Pillepich18b}
{Pillepich} A.,  et~al., 2018b, \mn@doi [\mnras] {10.1093/mnras/stx3112}, \href
  {https://ui.adsabs.harvard.edu/abs/2018MNRAS.475..648P} {475, 648}

\bibitem[\protect\citeauthoryear{{Piotrowska}, {Bluck}, {Maiolino}, {Concas}
  \& {Peng}}{{Piotrowska} et~al.}{2020}]{Piotrowska20}
{Piotrowska} J.~M.,  {Bluck} A. F.~L.,  {Maiolino} R.,  {Concas} A.,   {Peng}
  Y.,  2020, \mn@doi [\mnras] {10.1093/mnrasl/slz172}, 492, L6

\bibitem[\protect\citeauthoryear{{Planck Collaboration} et~al.,}{{Planck
  Collaboration} et~al.}{2014}]{Planck14}
{Planck Collaboration} et~al., 2014, \mn@doi [\aap]
  {10.1051/0004-6361/201321529}, \href
  {https://ui.adsabs.harvard.edu/abs/2014A&A...571A...1P} {571, A1}

\bibitem[\protect\citeauthoryear{{Pontzen}, {Tremmel}, {Roth}, {Peiris},
  {Saintonge}, {Volonteri}, {Quinn}  \& {Governato}}{{Pontzen}
  et~al.}{2017}]{Pontzen17}
{Pontzen} A.,  {Tremmel} M.,  {Roth} N.,  {Peiris} H.~V.,  {Saintonge} A.,
  {Volonteri} M.,  {Quinn} T.,   {Governato} F.,  2017, \mn@doi [\mnras]
  {10.1093/mnras/stw2627}, \href
  {https://ui.adsabs.harvard.edu/abs/2017MNRAS.465..547P} {465, 547}

\bibitem[\protect\citeauthoryear{{Reines} \& {Volonteri}}{{Reines} \&
  {Volonteri}}{2015}]{Reines15}
{Reines} A.~E.,  {Volonteri} M.,  2015, \mn@doi [\apj]
  {10.1088/0004-637X/813/2/82}, \href
  {https://ui.adsabs.harvard.edu/abs/2015ApJ...813...82R} {813, 82}

\bibitem[\protect\citeauthoryear{{Renzini}}{{Renzini}}{2020}]{Renzini20}
{Renzini} A.,  2020, \mn@doi [\mnras] {10.1093/mnrasl/slaa054}, \href
  {https://ui.adsabs.harvard.edu/abs/2020MNRAS.495L..42R} {495, L42}

\bibitem[\protect\citeauthoryear{{Renzini} \& {Peng}}{{Renzini} \&
  {Peng}}{2015}]{Renzini15}
{Renzini} A.,  {Peng} Y.-j.,  2015, \mn@doi [\apjl]
  {10.1088/2041-8205/801/2/L29}, \href
  {https://ui.adsabs.harvard.edu/abs/2015ApJ...801L..29R} {801, L29}

\bibitem[\protect\citeauthoryear{{Rosario} et~al.,}{{Rosario}
  et~al.}{2013}]{Rosario13b}
{Rosario} D.~J.,  et~al., 2013, \mn@doi [\apj] {10.1088/0004-637X/771/1/63},
  \href {https://ui.adsabs.harvard.edu/abs/2013ApJ...771...63R} {771, 63}

\bibitem[\protect\citeauthoryear{{Rosas-Guevara} et~al.,}{{Rosas-Guevara}
  et~al.}{2015}]{Rosas15}
{Rosas-Guevara} Y.~M.,  et~al., 2015, \mn@doi [\mnras] {10.1093/mnras/stv2056},
  \href {https://ui.adsabs.harvard.edu/abs/2015MNRAS.454.1038R} {454, 1038}

\bibitem[\protect\citeauthoryear{{Rupke}, {G{\"u}ltekin}  \&
  {Veilleux}}{{Rupke} et~al.}{2017}]{Rupke17}
{Rupke} D. S.~N.,  {G{\"u}ltekin} K.,   {Veilleux} S.,  2017, \mn@doi [\apj]
  {10.3847/1538-4357/aa94d1}, \href
  {https://ui.adsabs.harvard.edu/abs/2017ApJ...850...40R} {850, 40}

\bibitem[\protect\citeauthoryear{{Saglia} et~al.,}{{Saglia}
  et~al.}{2016}]{Saglia16}
{Saglia} R.~P.,  et~al., 2016, \mn@doi [\apj] {10.3847/0004-637X/818/1/47},
  \href {https://ui.adsabs.harvard.edu/abs/2016ApJ...818...47S} {818, 47}

\bibitem[\protect\citeauthoryear{{Saintonge} et~al.,}{{Saintonge}
  et~al.}{2017}]{Saintonge17}
{Saintonge} A.,  et~al., 2017, \mn@doi [\apjs] {10.3847/1538-4365/aa97e0},
  \href {https://ui.adsabs.harvard.edu/abs/2017ApJS..233...22S} {233, 22}

\bibitem[\protect\citeauthoryear{{Saintonge} et~al.,}{{Saintonge}
  et~al.}{2018}]{Saintonge18}
{Saintonge} A.,  et~al., 2018, \mn@doi [\mnras] {10.1093/mnras/sty2499}, \href
  {https://ui.adsabs.harvard.edu/abs/2018MNRAS.481.3497S} {481, 3497}

\bibitem[\protect\citeauthoryear{{Salim} et~al.,}{{Salim}
  et~al.}{2007}]{Salim07}
{Salim} S.,  et~al., 2007, \mn@doi [\apjs] {10.1086/519218}, \href
  {https://ui.adsabs.harvard.edu/abs/2007ApJS..173..267S} {173, 267}

\bibitem[\protect\citeauthoryear{{Santini} et~al.,}{{Santini}
  et~al.}{2009}]{Santini09}
{Santini} P.,  et~al., 2009, \mn@doi [\aap] {10.1051/0004-6361/200811434},
  \href {https://ui.adsabs.harvard.edu/abs/2009A&A...504..751S} {504, 751}

\bibitem[\protect\citeauthoryear{{Scannapieco} \& {Oh}}{{Scannapieco} \&
  {Oh}}{2004}]{Scannapieco04}
{Scannapieco} E.,  {Oh} S.~P.,  2004, \mn@doi [\apj] {10.1086/386542}, \href
  {https://ui.adsabs.harvard.edu/abs/2004ApJ...608...62S} {608, 62}

\bibitem[\protect\citeauthoryear{{Schaye} et~al.,}{{Schaye}
  et~al.}{2015}]{Schaye15}
{Schaye} J.,  et~al., 2015, \mn@doi [\mnras] {10.1093/mnras/stu2058}, \href
  {https://ui.adsabs.harvard.edu/abs/2015MNRAS.446..521S} {446, 521}

\bibitem[\protect\citeauthoryear{{Shakura} \& {Sunyaev}}{{Shakura} \&
  {Sunyaev}}{1973}]{Shakura73}
{Shakura} N.~I.,  {Sunyaev} R.~A.,  1973, \aap, \href
  {https://ui.adsabs.harvard.edu/abs/1973A&A....24..337S} {500, 33}

\bibitem[\protect\citeauthoryear{{Shankar}, {Weinberg}  \&
  {Miralda-Escud{\'e}}}{{Shankar} et~al.}{2009}]{Shankar09}
{Shankar} F.,  {Weinberg} D.~H.,   {Miralda-Escud{\'e}} J.,  2009, \mn@doi
  [\apj] {10.1088/0004-637X/690/1/20}, \href
  {https://ui.adsabs.harvard.edu/abs/2009ApJ...690...20S} {690, 20}

\bibitem[\protect\citeauthoryear{{Sijacki}, {Springel}, {Di Matteo}  \&
  {Hernquist}}{{Sijacki} et~al.}{2007}]{Sijacki07}
{Sijacki} D.,  {Springel} V.,  {Di Matteo} T.,   {Hernquist} L.,  2007, \mn@doi
  [\mnras] {10.1111/j.1365-2966.2007.12153.x}, \href
  {https://ui.adsabs.harvard.edu/abs/2007MNRAS.380..877S} {380, 877}

\bibitem[\protect\citeauthoryear{{Sijacki}, {Vogelsberger}, {Genel},
  {Springel}, {Torrey}, {Snyder}, {Nelson}  \& {Hernquist}}{{Sijacki}
  et~al.}{2015}]{Sijacki15}
{Sijacki} D.,  {Vogelsberger} M.,  {Genel} S.,  {Springel} V.,  {Torrey} P.,
  {Snyder} G.~F.,  {Nelson} D.,   {Hernquist} L.,  2015, \mn@doi [\mnras]
  {10.1093/mnras/stv1340}, \href
  {https://ui.adsabs.harvard.edu/abs/2015MNRAS.452..575S} {452, 575}

\bibitem[\protect\citeauthoryear{{Silk} \& {Rees}}{{Silk} \&
  {Rees}}{1998}]{Silk98}
{Silk} J.,  {Rees} M.~J.,  1998, \aap, \href
  {https://ui.adsabs.harvard.edu/abs/1998A&A...331L...1S} {331, L1}

\bibitem[\protect\citeauthoryear{{Simard}, {Mendel}, {Patton}, {Ellison}  \&
  {McConnachie}}{{Simard} et~al.}{2011}]{Simard11}
{Simard} L.,  {Mendel} J.~T.,  {Patton} D.~R.,  {Ellison} S.~L.,
  {McConnachie} A.~W.,  2011, \mn@doi [\apjs] {10.1088/0067-0049/196/1/11},
  \href {https://ui.adsabs.harvard.edu/abs/2011ApJS..196...11S} {196, 11}

\bibitem[\protect\citeauthoryear{{Snyder} et~al.,}{{Snyder}
  et~al.}{2015}]{Snyder15}
{Snyder} G.~F.,  et~al., 2015, \mn@doi [\mnras] {10.1093/mnras/stv2078}, \href
  {https://ui.adsabs.harvard.edu/abs/2015MNRAS.454.1886S} {454, 1886}

\bibitem[\protect\citeauthoryear{{Somerville} \& {Dav{\'e}}}{{Somerville} \&
  {Dav{\'e}}}{2015}]{Somerville15}
{Somerville} R.~S.,  {Dav{\'e}} R.,  2015, \mn@doi [\araa]
  {10.1146/annurev-astro-082812-140951}, \href
  {https://ui.adsabs.harvard.edu/abs/2015ARA&A..53...51S} {53, 51}

\bibitem[\protect\citeauthoryear{{Sorai} et~al.,}{{Sorai}
  et~al.}{2019}]{Sorai19}
{Sorai} K.,  et~al., 2019, \mn@doi [\pasj] {10.1093/pasj/psz115}, \href
  {https://ui.adsabs.harvard.edu/abs/2019PASJ...71S..14S} {71, S14}

\bibitem[\protect\citeauthoryear{{Sparre} et~al.,}{{Sparre}
  et~al.}{2015}]{Sparre15}
{Sparre} M.,  et~al., 2015, \mn@doi [\mnras] {10.1093/mnras/stu2713}, \href
  {https://ui.adsabs.harvard.edu/abs/2015MNRAS.447.3548S} {447, 3548}

\bibitem[\protect\citeauthoryear{{Springel}}{{Springel}}{2005}]{Springel05}
{Springel} V.,  2005, \mn@doi [\mnras] {10.1111/j.1365-2966.2005.09655.x},
  \href {https://ui.adsabs.harvard.edu/abs/2005MNRAS.364.1105S} {364, 1105}

\bibitem[\protect\citeauthoryear{{Springel}}{{Springel}}{2010}]{Springel10}
{Springel} V.,  2010, \mn@doi [\mnras] {10.1111/j.1365-2966.2009.15715.x},
  \href {https://ui.adsabs.harvard.edu/abs/2010MNRAS.401..791S} {401, 791}

\bibitem[\protect\citeauthoryear{{Springel} \& {Hernquist}}{{Springel} \&
  {Hernquist}}{2003}]{Springel03}
{Springel} V.,  {Hernquist} L.,  2003, \mn@doi [\mnras]
  {10.1046/j.1365-8711.2003.06206.x}, \href
  {https://ui.adsabs.harvard.edu/abs/2003MNRAS.339..289S} {339, 289}

\bibitem[\protect\citeauthoryear{{Springel} et~al.,}{{Springel}
  et~al.}{2018}]{Springel18}
{Springel} V.,  et~al., 2018, \mn@doi [\mnras] {10.1093/mnras/stx3304}, \href
  {https://ui.adsabs.harvard.edu/abs/2018MNRAS.475..676S} {475, 676}

\bibitem[\protect\citeauthoryear{{Sternberg}, {Le Petit}, {Roueff}  \& {Le
  Bourlot}}{{Sternberg} et~al.}{2014}]{Sternberg14}
{Sternberg} A.,  {Le Petit} F.,  {Roueff} E.,   {Le Bourlot} J.,  2014, \mn@doi
  [\apj] {10.1088/0004-637X/790/1/10}, \href
  {https://ui.adsabs.harvard.edu/abs/2014ApJ...790...10S} {790, 10}

\bibitem[\protect\citeauthoryear{{Strateva} et~al.,}{{Strateva}
  et~al.}{2001}]{Strateva01}
{Strateva} I.,  et~al., 2001, \mn@doi [\aj] {10.1086/323301}, \href
  {https://ui.adsabs.harvard.edu/abs/2001AJ....122.1861S} {122, 1861}

\bibitem[\protect\citeauthoryear{{Tacconi} et~al.,}{{Tacconi}
  et~al.}{2018}]{Tacconi18}
{Tacconi} L.~J.,  et~al., 2018, \mn@doi [\apj] {10.3847/1538-4357/aaa4b4},
  \href {https://ui.adsabs.harvard.edu/abs/2018ApJ...853..179T} {853, 179}

\bibitem[\protect\citeauthoryear{{Tal} et~al.,}{{Tal} et~al.}{2014}]{Tal14}
{Tal} T.,  et~al., 2014, \mn@doi [\apj] {10.1088/0004-637X/789/2/164}, \href
  {https://ui.adsabs.harvard.edu/abs/2014ApJ...789..164T} {789, 164}

\bibitem[\protect\citeauthoryear{{Taylor}, {Boylan-Kolchin}, {Torrey},
  {Vogelsberger}  \& {Hernquist}}{{Taylor} et~al.}{2016}]{Taylor16}
{Taylor} C.,  {Boylan-Kolchin} M.,  {Torrey} P.,  {Vogelsberger} M.,
  {Hernquist} L.,  2016, \mn@doi [\mnras] {10.1093/mnras/stw1522}, \href
  {https://ui.adsabs.harvard.edu/abs/2016MNRAS.461.3483T} {461, 3483}

\bibitem[\protect\citeauthoryear{{Teimoorinia}, {Bluck}  \&
  {Ellison}}{{Teimoorinia} et~al.}{2016}]{Teimoorinia16}
{Teimoorinia} H.,  {Bluck} A. F.~L.,   {Ellison} S.~L.,  2016, \mn@doi [\mnras]
  {10.1093/mnras/stw036}, \href
  {https://ui.adsabs.harvard.edu/abs/2016MNRAS.457.2086T} {457, 2086}

\bibitem[\protect\citeauthoryear{{Terrazas}, {Bell}, {Henriques}, {White},
  {Cattaneo}  \& {Woo}}{{Terrazas} et~al.}{2016}]{Terrazas16}
{Terrazas} B.~A.,  {Bell} E.~F.,  {Henriques} B. M.~B.,  {White} S. D.~M.,
  {Cattaneo} A.,   {Woo} J.,  2016, \mn@doi [\apjl]
  {10.3847/2041-8205/830/1/L12}, \href
  {https://ui.adsabs.harvard.edu/abs/2016ApJ...830L..12T} {830, L12}

\bibitem[\protect\citeauthoryear{{Terrazas}, {Bell}, {Woo}  \&
  {Henriques}}{{Terrazas} et~al.}{2017}]{Terrazas17}
{Terrazas} B.~A.,  {Bell} E.~F.,  {Woo} J.,   {Henriques} B. M.~B.,  2017,
  \mn@doi [\apj] {10.3847/1538-4357/aa7d07}, \href
  {https://ui.adsabs.harvard.edu/abs/2017ApJ...844..170T} {844, 170}

\bibitem[\protect\citeauthoryear{{Terrazas} et~al.,}{{Terrazas}
  et~al.}{2020}]{Terrazas20}
{Terrazas} B.~A.,  et~al., 2020, \mn@doi [\mnras] {10.1093/mnras/staa374},
  \href {https://ui.adsabs.harvard.edu/abs/2020MNRAS.493.1888T} {493, 1888}

\bibitem[\protect\citeauthoryear{{Trayford} et~al.,}{{Trayford}
  et~al.}{2015}]{Tayford15}
{Trayford} J.~W.,  et~al., 2015, \mn@doi [\mnras] {10.1093/mnras/stv1461},
  \href {https://ui.adsabs.harvard.edu/abs/2015MNRAS.452.2879T} {452, 2879}

\bibitem[\protect\citeauthoryear{{Trump} et~al.,}{{Trump}
  et~al.}{2015}]{Trump15}
{Trump} J.~R.,  et~al., 2015, \mn@doi [\apj] {10.1088/0004-637X/811/1/26},
  \href {https://ui.adsabs.harvard.edu/abs/2015ApJ...811...26T} {811, 26}

\bibitem[\protect\citeauthoryear{{Veilleux} et~al.,}{{Veilleux}
  et~al.}{2013}]{Veilleux13}
{Veilleux} S.,  et~al., 2013, \mn@doi [\apj] {10.1088/0004-637X/776/1/27},
  \href {https://ui.adsabs.harvard.edu/abs/2013ApJ...776...27V} {776, 27}

\bibitem[\protect\citeauthoryear{{Venturi} et~al.,}{{Venturi}
  et~al.}{2021}]{Venturi21}
{Venturi} G.,  et~al., 2021, \mn@doi [\aap] {10.1051/0004-6361/202039869},
  \href {https://ui.adsabs.harvard.edu/abs/2021A&A...648A..17V} {648, A17}

\bibitem[\protect\citeauthoryear{{Villar-Mart{\'\i}n}, {Humphrey}, {Delgado},
  {Colina}  \& {Arribas}}{{Villar-Mart{\'\i}n} et~al.}{2011}]{Villar-Martin11}
{Villar-Mart{\'\i}n} M.,  {Humphrey} A.,  {Delgado} R.~G.,  {Colina} L.,
  {Arribas} S.,  2011, \mn@doi [\mnras] {10.1111/j.1365-2966.2011.19622.x},
  \href {https://ui.adsabs.harvard.edu/abs/2011MNRAS.418.2032V} {418, 2032}

\bibitem[\protect\citeauthoryear{{Vogelsberger} et~al.,}{{Vogelsberger}
  et~al.}{2014a}]{Vogelsberger14b}
{Vogelsberger} M.,  et~al., 2014a, \mn@doi [\mnras] {10.1093/mnras/stu1536},
  \href {https://ui.adsabs.harvard.edu/abs/2014MNRAS.444.1518V} {444, 1518}

\bibitem[\protect\citeauthoryear{{Vogelsberger} et~al.,}{{Vogelsberger}
  et~al.}{2014b}]{Vogelsberger14a}
{Vogelsberger} M.,  et~al., 2014b, \mn@doi [\nat] {10.1038/nature13316}, \href
  {https://ui.adsabs.harvard.edu/abs/2014Natur.509..177V} {509, 177}

\bibitem[\protect\citeauthoryear{{Wake}, {van Dokkum}  \& {Franx}}{{Wake}
  et~al.}{2012}]{Wake12}
{Wake} D.~A.,  {van Dokkum} P.~G.,   {Franx} M.,  2012, \mn@doi [\apjl]
  {10.1088/2041-8205/751/2/L44}, \href
  {https://ui.adsabs.harvard.edu/abs/2012ApJ...751L..44W} {751, L44}

\bibitem[\protect\citeauthoryear{{Wang} et~al.,}{{Wang} et~al.}{2018}]{Wang18}
{Wang} H.,  et~al., 2018, \mn@doi [\apj] {10.3847/1538-4357/aa9e01}, \href
  {https://ui.adsabs.harvard.edu/abs/2018ApJ...852...31W} {852, 31}

\bibitem[\protect\citeauthoryear{{Wang}, {Cappellari}, {Peng}  \&
  {Graham}}{{Wang} et~al.}{2020}]{Wang20}
{Wang} B.,  {Cappellari} M.,  {Peng} Y.,   {Graham} M.,  2020, \mn@doi [\mnras]
  {10.1093/mnras/staa1325}, \href
  {https://ui.adsabs.harvard.edu/abs/2020MNRAS.495.1958W} {495, 1958}

\bibitem[\protect\citeauthoryear{{Weinberger} et~al.,}{{Weinberger}
  et~al.}{2017}]{Weinberger17}
{Weinberger} R.,  et~al., 2017, \mn@doi [\mnras] {10.1093/mnras/stw2944}, \href
  {https://ui.adsabs.harvard.edu/abs/2017MNRAS.465.3291W} {465, 3291}

\bibitem[\protect\citeauthoryear{{Weinberger} et~al.,}{{Weinberger}
  et~al.}{2018}]{Weinberger18}
{Weinberger} R.,  et~al., 2018, \mn@doi [\mnras] {10.1093/mnras/sty1733}, \href
  {https://ui.adsabs.harvard.edu/abs/2018MNRAS.479.4056W} {479, 4056}

\bibitem[\protect\citeauthoryear{{Werner}, {McNamara}, {Churazov}  \&
  {Scannapieco}}{{Werner} et~al.}{2019}]{Werner19}
{Werner} N.,  {McNamara} B.~R.,  {Churazov} E.,   {Scannapieco} E.,  2019,
  \mn@doi [\ssr] {10.1007/s11214-018-0571-9}, \href
  {https://ui.adsabs.harvard.edu/abs/2019SSRv..215....5W} {215, 5}

\bibitem[\protect\citeauthoryear{{Willmer} et~al.,}{{Willmer}
  et~al.}{2006}]{Willmer06}
{Willmer} C.~N.~A.,  et~al., 2006, \mn@doi [\apj] {10.1086/505455}, \href
  {https://ui.adsabs.harvard.edu/abs/2006ApJ...647..853W} {647, 853}

\bibitem[\protect\citeauthoryear{{Woo} et~al.,}{{Woo} et~al.}{2013}]{Woo13}
{Woo} J.,  et~al., 2013, \mn@doi [\mnras] {10.1093/mnras/sts274}, \href
  {https://ui.adsabs.harvard.edu/abs/2013MNRAS.428.3306W} {428, 3306}

\bibitem[\protect\citeauthoryear{{Woo}, {Dekel}, {Faber}  \& {Koo}}{{Woo}
  et~al.}{2015}]{Woo15}
{Woo} J.,  {Dekel} A.,  {Faber} S.~M.,   {Koo} D.~C.,  2015, \mn@doi [\mnras]
  {10.1093/mnras/stu2755}, \href
  {https://ui.adsabs.harvard.edu/abs/2015MNRAS.448..237W} {448, 237}

\bibitem[\protect\citeauthoryear{{Wu}, {Lu}, {Zhang}  \& {Lu}}{{Wu}
  et~al.}{2013}]{Wu13}
{Wu} S.,  {Lu} Y.,  {Zhang} F.,   {Lu} Y.,  2013, \mn@doi [\mnras]
  {10.1093/mnras/stt1811}, \href
  {https://ui.adsabs.harvard.edu/abs/2013MNRAS.436.3271W} {436, 3271}

\bibitem[\protect\citeauthoryear{{Wyder} et~al.,}{{Wyder}
  et~al.}{2007}]{Wyder07}
{Wyder} T.~K.,  et~al., 2007, \mn@doi [\apjs] {10.1086/521402}, \href
  {https://ui.adsabs.harvard.edu/abs/2007ApJS..173..293W} {173, 293}

\bibitem[\protect\citeauthoryear{{Xue} et~al.,}{{Xue} et~al.}{2010}]{Xue10}
{Xue} Y.~Q.,  et~al., 2010, \mn@doi [\apj] {10.1088/0004-637X/720/1/368}, \href
  {https://ui.adsabs.harvard.edu/abs/2010ApJ...720..368X} {720, 368}

\bibitem[\protect\citeauthoryear{{Yang}, {Mo}, {van den Bosch}, {Pasquali},
  {Li}  \& {Barden}}{{Yang} et~al.}{2007}]{Yang07}
{Yang} X.,  {Mo} H.~J.,  {van den Bosch} F.~C.,  {Pasquali} A.,  {Li} C.,
  {Barden} M.,  2007, \mn@doi [\apj] {10.1086/522027}, \href
  {https://ui.adsabs.harvard.edu/abs/2007ApJ...671..153Y} {671, 153}

\bibitem[\protect\citeauthoryear{{Yang}, {Mo}  \& {van den Bosch}}{{Yang}
  et~al.}{2009}]{Yang09}
{Yang} X.,  {Mo} H.~J.,   {van den Bosch} F.~C.,  2009, \mn@doi [\apj]
  {10.1088/0004-637X/695/2/900}, \href
  {https://ui.adsabs.harvard.edu/abs/2009ApJ...695..900Y} {695, 900}

\bibitem[\protect\citeauthoryear{{York} et~al.,}{{York} et~al.}{2000}]{York00}
{York} D.~G.,  et~al., 2000, \mn@doi [\aj] {10.1086/301513}, \href
  {https://ui.adsabs.harvard.edu/abs/2000AJ....120.1579Y} {120, 1579}

\bibitem[\protect\citeauthoryear{{Zakamska} \& {Greene}}{{Zakamska} \&
  {Greene}}{2014}]{Zakamska14}
{Zakamska} N.~L.,  {Greene} J.~E.,  2014, \mn@doi [\mnras]
  {10.1093/mnras/stu842}, \href
  {https://ui.adsabs.harvard.edu/abs/2014MNRAS.442..784Z} {442, 784}

\bibitem[\protect\citeauthoryear{{Zhu} et~al.,}{{Zhu} et~al.}{2018}]{Zhu18}
{Zhu} Q.,  et~al., 2018, \mn@doi [\mnras] {10.1093/mnrasl/sly111}, \href
  {https://ui.adsabs.harvard.edu/abs/2018MNRAS.480L..18Z} {480, L18}

\bibitem[\protect\citeauthoryear{{Zinger} et~al.,}{{Zinger}
  et~al.}{2020}]{Zinger20}
{Zinger} E.,  et~al., 2020, \mn@doi [\mnras] {10.1093/mnras/staa2607}, \href
  {https://ui.adsabs.harvard.edu/abs/2020MNRAS.499..768Z} {499, 768}

\bibitem[\protect\citeauthoryear{{van den Bosch}, {Aquino}, {Yang}, {Mo},
  {Pasquali}, {McIntosh}, {Weinmann}  \& {Kang}}{{van den Bosch}
  et~al.}{2008}]{Bosch08}
{van den Bosch} F.~C.,  {Aquino} D.,  {Yang} X.,  {Mo} H.~J.,  {Pasquali} A.,
  {McIntosh} D.~H.,  {Weinmann} S.~M.,   {Kang} X.,  2008, \mn@doi [\mnras]
  {10.1111/j.1365-2966.2008.13230.x}, \href
  {https://ui.adsabs.harvard.edu/abs/2008MNRAS.387...79V} {387, 79}

\makeatother
\end{thebibliography}



\appendix

\section{sSFR function}

Due to the lack of \halpha\ detections in the passive sequence
in the SDSS,  one is not able to measure accurate SFRs to 
arbitrarily low specific star formation rate values. This problem 
is dealt with in \cite{Brinchmann04} by imposing an upper limit of
log(sSFR/${\rm yr^{-1}}$) $\sim -12$ for quiescent systems, as indicated
by the lack of emission lines and a large magnitude of the D4000 break.

In the simulations there is no limit on sSFR, since it is possible for 
a given galaxy to have SFR=0 if none of its gas cells/particles reach 
density and temperature thresholds required for star formation. As shown 
in the left panel in Fig.~\ref{fig:appendix-ssfr} the distribution of sSFR 
extracted directly from the simulations coincides with the SDSS quite well 
at high sSFR values, however it extends far beyond the observational limit 
on the low-sSFR end. Because simulations can, in principle, 
have $\log({\rm sSFR})$ distributions extending down to~$- \infty$, showing 
a~comparison between them and the observations proves difficult in 
logarithmic units.

Hence, in order for us to include all simulated galaxies in 
\mbox{Figs.~\ref{fig:ms},~\ref{fig:msfr-fgas} and \ref{fig:msfr-sfe}}, 
we redistribute all objects with log(sSFR/${\rm yr^{-1}}$) $< -12$ 
to values drawn from a distribution of $\log({\rm sSFR})$ in the 
passive population in the SDSS. More precisely, we treat the sSFR 
density function for all galaxies with log(sSFR/${\rm yr^{-1}}$) $< -11$
in the SDSS as a probability distribution and for each qualifying 
galaxy we draw a new `redistributed' sSFR value at random. As we show 
in the right panel of Fig.~\ref{fig:appendix-ssfr}, the amended sSFR 
density function in all simulations matches the observed distribution
quite well at the low-sSFR end. Most importantly, there are no more 
low-sSFR objects extending below the imposed sSFR limit in the SDSS, 
allowing us to show the whole simulated quenched population 
in the \msfr\ plane.

Finally, we would like to stress that this redistribution is performed 
for visualisation purposes only and the newly assigned sSFR values 
do not enter any of our quantitative analysis presented in 
Sec.~\ref{sec:qf}-\ref{sec:gas-line}.

\label{ssfr-function}
\begin{figure}
\includegraphics[width=\columnwidth]{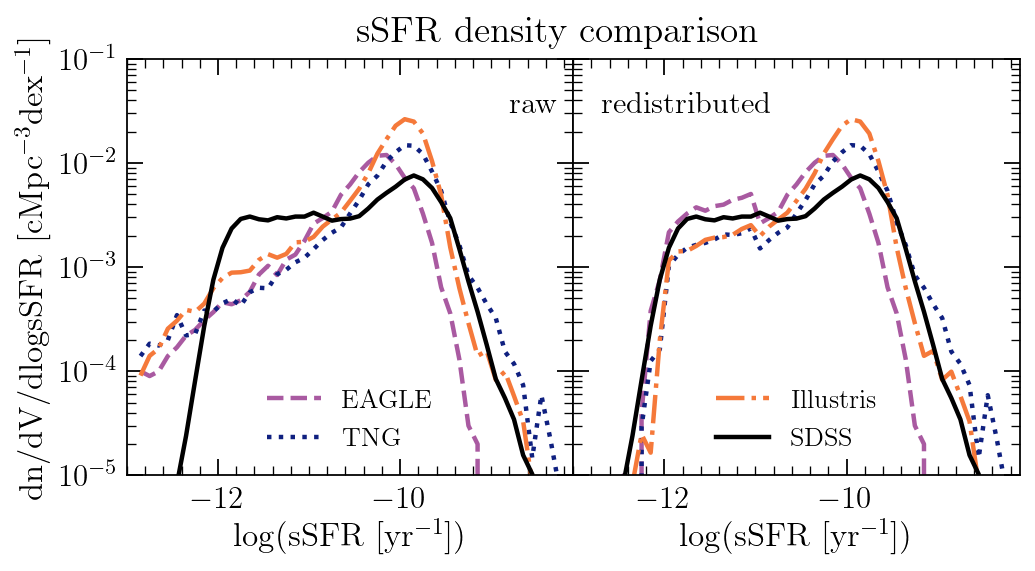} 
    \caption{Comparison of the sSFR comoving density in simulations and observations:
    		left: as simulated, right: with simulation results redistributed to reflect the
    		technical limitations associated with measuring low sSFRs in the observable Universe.
    		The RHS result was constructed by redistributing each object in the simulations
    		with log(sSFR/${\rm yr^{-1}}$) $<-12$, randomly drawing from the SDSS distribution for objects 
    		with log(sSFR/${\rm yr^{-1}}$) $<-11$. The SDSS distribution was 
    		weighted by $1/V_{\rm max}$ to correct
    		for the Malmquist bias.}
    \label{fig:appendix-ssfr}
\end{figure}

\section{Robustness of the Random Forest classifier results}
\label{sec:discussion:rf-robust}

One of the key parts of our analysis relies on Random Forest
classification to determine which of the galactic parameters
is most predictive for deciding whether a~given galaxy has 
ceased its star formation by redshift $z=0$. 
Although the use of machine learning techniques is
gradually becoming more common in astronomy, we appreciate 
how the choice of our tool can raise potential questions about the 
result interpretation, as well as its reproducibility. 
In order to better understand the RF result and how the characteristics of the
data influence the inferred feature importances, we present a~range of tests
of this method in this section.

\subsection{Restricting the number of features available in each split}
\label{sec:appendix:sqrt}

\begin{figure}
		\includegraphics[width=\columnwidth]{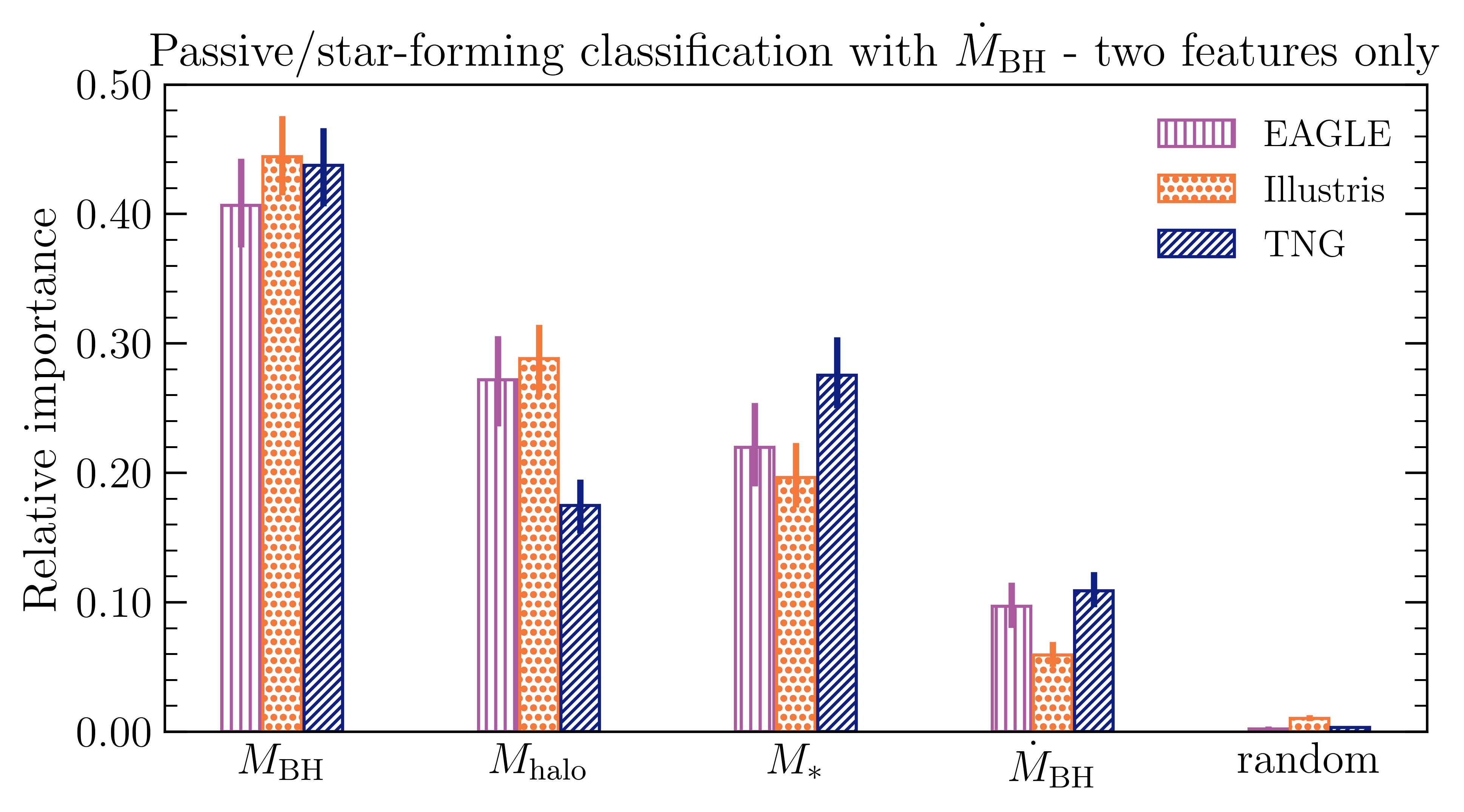}
		\caption{Same as left panel in Fig.~\ref{fig:appendix-rf-combined} with
				 \mdot\ parameter added to the list of features. Among all other
				 investigated parameters \mdot\ carries least predictive power
				 in determining galaxy classification.}
		\label{fig:appendix-rf-mdot}
\end{figure}

\begin{figure*}
	\includegraphics[width=\textwidth]{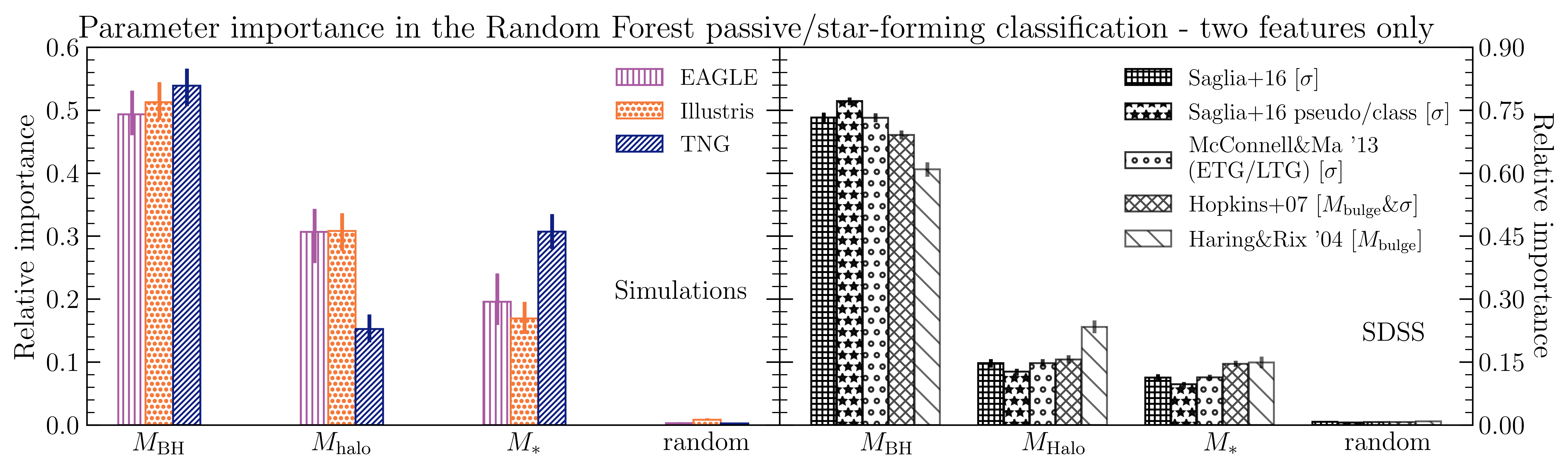}
	\caption{Same as Fig.~\ref{fig:rf-combined} with features restricted to
			 a random draw of two features at each split.
    		 Error bars indicate the 16th and 84th 
    		 percentiles of the distribution in parameter importance
    		 drawn from 500 realisations of the experiment.  
    		 Left panel: in simulations \mstar\ and \mhalo\ pick up 
    		 importance owing to their inter-correlation with \mbh.
      		 Right panel: in SDSS the rise in importance of other parameters
      		 is less pronounced than in the case of simulations.}
      \label{fig:appendix-rf-combined}
\end{figure*}

As we mention earlier in Sec.~\ref{sec:rf-method}, one means of additional
randomisation in an RF architecture is forcing the algorithm to randomly
sample features available for a given split. Empirical tests with
different RF architectures suggest that restricting the number of available 
features improves algorithm performance on previously unseen data
in the presence of noise associated with its measurement
\citep{Breiman01}. This increase in robustness comes at a cost
of a decreased ability of the algorithm to remove inter-correlation within
the data. Because our work does not focus on predicting unknown sSFR
classes and, instead, aims to extract causality in our simulated and observed
data sets, we allow our algorithm to choose from all features at every 
split by default. In order to check whether restricting the number of available
features can influence our main conclusions, we conduct
our experiment again, randomly choosing 2 features at each node. 
We check that our conclusions are robust against this choice in 
the algorithm architecture.

In Figs.~\ref{fig:appendix-rf-mdot}~\&~\ref{fig:appendix-rf-combined}
we explore the change in feature
importances when the decision trees randomly select a subset
of two features to split on at every node. The visible dominance
of \mbh\ over other parameters is well preserved in both the
simulations and the observations, however this time other parameters 
gain on importance in comparison with the RF architecture allowed
to choose from all features at every split in 
Figs.~\ref{fig:rf-combined}~\&~\ref{fig:rf-mdot}. When comparing the
left panel in Fig.~\ref{fig:appendix-rf-mdot} and 
Fig.~\ref{fig:appendix-rf-combined} alone, we see once again that
the \mdot\ bar height is picked up primarily at the expense of the 
\mbh\ score, most likely owing to the connection between the two 
variables in the subgrid black hole accretion model.  

The difference between all features and their random subsampling 
is more pronounced in the simulations, where black hole mass'
relative importance is only about twice the value of the next 
best ranked parameter. The difference between the SDSS and
simulations is primarily dictated by how tight the relationships
between \mbh\ and other parameters are in the data sets.
Given the result in Fig.~\ref{fig:appendix-rf-combined}, we can safely assume
that whenever the algorithm has a choice between black hole 
mass and another parameter (which statistically happens for 
a half of all splits) it is always \mbh\ which results in the highest reduction 
in impurity. For the remaining half of splits the tree has to make
a decision among \mhalo, \mstar\ and a random number, in which
case the feature in the closest numerical association with \mbh\
takes the split. 

As we show in Sec.~\ref{sec:correlations}
the \mbh-\mstar\ and \mbh-\mhalo\ relations are significantly
tighter in the simulations and hence a split on a secondary feature
results in a higher reduction in impurity in the simulations
than in the observations. However, regardless of the change
in relative importance of \mhalo\ and \mstar\,
Fig.~\ref{fig:appendix-rf-combined} demonstrates that black hole mass
is unanimously predicted to hold the most predictive power in 
determining whether central galaxies are star-forming or quenched
in the simulations. This theoretical prediction is then validated
in the observations, regardless of the \mbh\ calibration or 
the algorithm architecture.

\subsection{Feature importances in correlated variables}
\label{sec:discussion:rf-corr}

\begin{figure}
\includegraphics[width=\columnwidth]{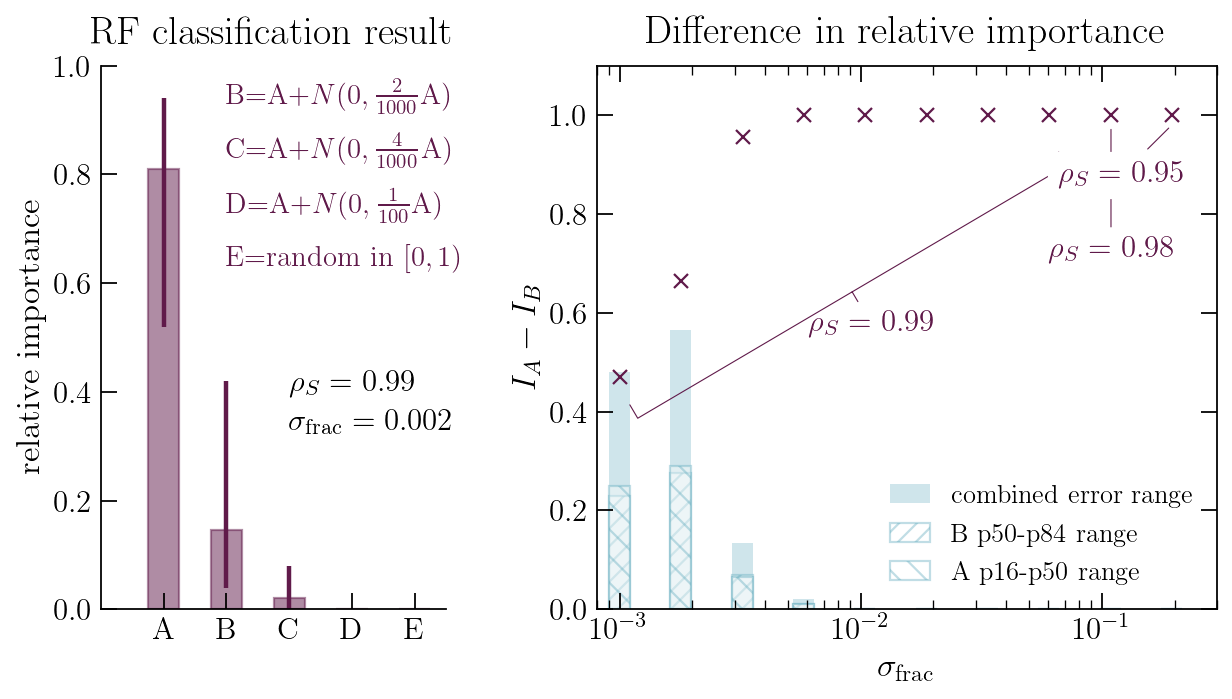} 
    \caption{Left: feature importances of strongly correlated variables in an~RF classification.
    				Spearman's rank correlation coefficient $\rho_S$ between A~and B is 0.99
    				and the result corresponds to the 5th data point from the left in the right plot. 
					Variable A~assigns the class labels and hence performs best, followed by B, C, D 
					in the order of increasing noise.
					Right: same RF experiment repeated for different levels of noise in B, showing the
					difference in feature importance (A-B) along with associated error ranges. Crosses
					in blue shaded regions show where the result is inconclusive.
					The figure demonstrates how successful the RF algorithm is at picking the most
					relevant parameter at~$\sim 1\sigma$ confidence at~$\rho_S=0.99$ between the
					two most successful variables.} 
	\label{fig:rf-corr}
\end{figure}

As we explore in detail in Sec~\ref{sec:correlations}, 
all variables of interest in both simulations and the 
SDSS are relatively strongly inter-correlated with
one another. This close connection among
galactic parameters can raise concerns as to whether 
the Random Forest is able to identify the true feature carrying 
the most importance in the classification, with an appropriate degree 
of confidence. 

We illustrate this potential issue in the EAGLE suite as an example, 
where \mbh\ and  \mhalo\ are the two variables scoring highest out of all 
parameters in Fig.~\ref{fig:appendix-rf-mdot}. When we plot \mbh\ 
against \mhalo\ in Fig.~\ref{fig:corr-mhalo}, we see a~strong correlation between
the two. Without the PCC analysis we cannot a~priori assume which of the
parameters is intrinsically connected to quenching and which of them
earns feature importance owing to the correlation strengths with other
parameters. Hence, the question is how strongly would \mhalo\ and \mbh\
have to be correlated, in order for the black hole mass dominance to be
an ambiguous result in the Random Forest classification.

To check when the above is true, we design a~test of the RF framework,
using synthetic variables which are related to one another via straightforward
functional forms. 
In our synthetic data set we create 10 000 objects to mimic
the sample sizes in this work, each with an entry for variables A, B, C, D 
and E. We first choose the variable A, which is a~set of real numbers $y=x^2$, 
where $x$ is drawn at random from a~flat distribution in $[0.5, 1.5)$.
The exact choice of the functional form of $y(x)$ does not 
affect the results of this test. 
The range of values in A~is chosen such that we can think 
of it as a~measurement of SFR in the real data. 

We then create B, C and D by adding random draws from 
a~Gaussian centred on 0 to A~such that ${\rm B = A} + N(0, \sigma_{\rm B} \times {\rm A})$, 
${\rm C = A} + N(0, \sigma_{\rm C} \times {\rm A})$ and 
${\rm D = A} + N(0, \sigma_{\rm D} \times {\rm A})$.
The amount of scatter we add to a~given measurement in A~changes with A
such that the scatter around $\log({\rm A})=\log({\rm B})$ is constant
across the whole range in A. We choose $\sigma$ values to increase from B to D such that these
parameters are respectively most and least closely correlated with A.
In particular, we choose $\sigma_{\rm C}=2\sigma_{\rm B}$
and $\sigma_{\rm D}=5\sigma_{\rm B}$ to ensure significant differences
among the parameters. Finally, like in the case of our fiducial RF analysis, we also
include a~random draw from a~flat distribution in $[0,1)$ which becomes the 
E variable. 

Once a~given set of synthetic parameters is created, we assign a~class label
for each object based on A, such that all objects with ${\rm A} \leq 1.5$ are
labelled as `passive' and the rest are `star forming'. We then repeat the
Random Forest classification 500 times, calculate the resulting median
feature importances and estimate their uncertainties from the 16th and
84th percentiles of each distribution. The full test consists of 10 different
realisations of the synthetic data in which we increase $\sigma_{\rm B}$
from 0.001 to 0.2 in equally spaced logarithmic steps, decreasing 
the strength of correlation between variables A~and B in each step. 
Finally, in order to quantify their numerical association we calculate their
Spearman's rank correlation coefficient $\rho_S$ in each realisation.

\begin{table} 
\caption{Sperman's rank correlation coefficient $\rho_S$ calculated 
			 between black hole mass and other RF parameters for all
			 data sets. In SDSS $\rho_S$ is calculated separately for
			 each \mbh\ estimator and takes into account the 
			 $\flatfrac{1}{V_{\rm max}}$ and inclination corrections.}
\centering
\begin{tabular}{lccc}
\toprule
 &   \mbh--\mhalo &  \mbh--\mstar &  \mbh--\mdot\  \\
\midrule 
EAGLE & 0.85 & 0.82 & 0.62 \\
Illustris & 0.84 & 0.87  &  0.49 \\  
TNG & 0.90 & 0.95 & 0.19 \\
Saglia+16 $[\sigma]$ & 0.80 & 0.82 & -- \\
\rule{0pt}{3ex}   
Saglia+16 & \multirow{2}{*}{0.79} & \multirow{2}{*}{0.81} & \multirow{2}{*}{--} \\
pseudo/class $[\sigma]$ &  &  &  \\
\rule{0pt}{3ex}   
McConell\&Ma '13 & \multirow{2}{*}{0.79} & \multirow{2}{*}{0.81} & \multirow{2}{*}{--} \\
(ETG/LTG) $[\sigma]$ & & & \\
\rule{0pt}{3ex}   
Hopkins+07 & \multirow{2}{*}{0.87} & \multirow{2}{*}{0.89} & \multirow{2}{*}{--} \\
$[M_{\rm bulge} \& \sigma]$ & & & \\
\rule{0pt}{3ex}   
Haring\&Rix '04 & \multirow{2}{*}{0.91} & \multirow{2}{*}{0.92} & \multirow{2}{*}{--} \\
$[M_{\rm bulge}]$ & & & \\
\bottomrule
\end{tabular}
\label{tab:rhos}
\end{table}

Fig.~\ref{fig:rf-corr} presents the result of our Random Forest test,
showing how strongly two test variables A~and B need to be correlated 
in order for the classifier to produce an ambiguous ranking of feature importances.
We show a representative RF classification result for $\sigma_{\rm B}=0.002$
in the left panel, where bar heights indicate feature importances, while 
the error bars encapsulate the 16th and 84th percentiles of feature
importance distributions. The functional form connecting the variables
is also listed in the left panel. In the right panel we show how the difference
between median importances in A~and B ($I_A - I_B$, bordeaux crosses) 
changes as a~function of $\sigma_{\rm B}$ ($\sigma_{\rm fractional}$ in the figure).
The hatched bars indicate the magnitude of the negative uncertainty on A
and positive uncertainty on B, while the blue shaded bars show the total 
of the two. Whenever a~cross is within the blue bar, the feature importance 
of A~and B agree within their errors. Additionally, the panel lists the
measurement of $\rho_S$ between the two variables for each of the
10 realisations of synthetic data.

The first thing to notice in Fig~\ref{fig:rf-corr} is that the value 
of \mbox{$I_A-I_B$} is always positive, hence the most important parameter in 
the classification is always identified, 
regardless of how closely the next best variable is associated with it. 
The significance of this result, however, differs as indicated by the location
of bordeaux crosses with respect to the blue shaded bars. 
Most importantly, we can see how the significance of $I_A-I_B$ varies
as a~function of $\rho_S$. It is clear that even for the extreme 
case of $\rho_S=0.99$ the most relevant parameter is selected by the
RF algorithm at a~confidence level of $\sim 1 \sigma$. 

The Random Forest robustness test in Fig.~\ref{fig:rf-corr} demonstrates
how incredibly successful the method is at selecting the most important
parameter among strongly inter-correlated variables. In order to see how this
test supports our RF results in Sec~\ref{sec:rf} and 
Appendix~\ref{sec:appendix:sqrt}, we calculate $\rho_S$
between \mbh\ and each of \mhalo, \mstar\ and \mdot\ in turn, 
summarising the values in Table~\ref{tab:rhos}. It is apparent that
galactic parameters in both the observed and simulated data in our
analysis never reach the threshold required for the Random Forest
result to be ambiguous. The highest correlation coefficient we measure
is between \mbh\ and \mstar\ in TNG, reaching $\rho=0.95$ -- 
a correlation strength at which the most important parameter is
identified with full confidence. Therefore we conclude
that the unanimous victory of black hole mass over other galactic
parameters in the Random Forest experiment 
is a~robust result, unaffected by the strong inter-correlations present
among the variables in the investigated data.

\begin{figure}
	\includegraphics[width=\columnwidth]{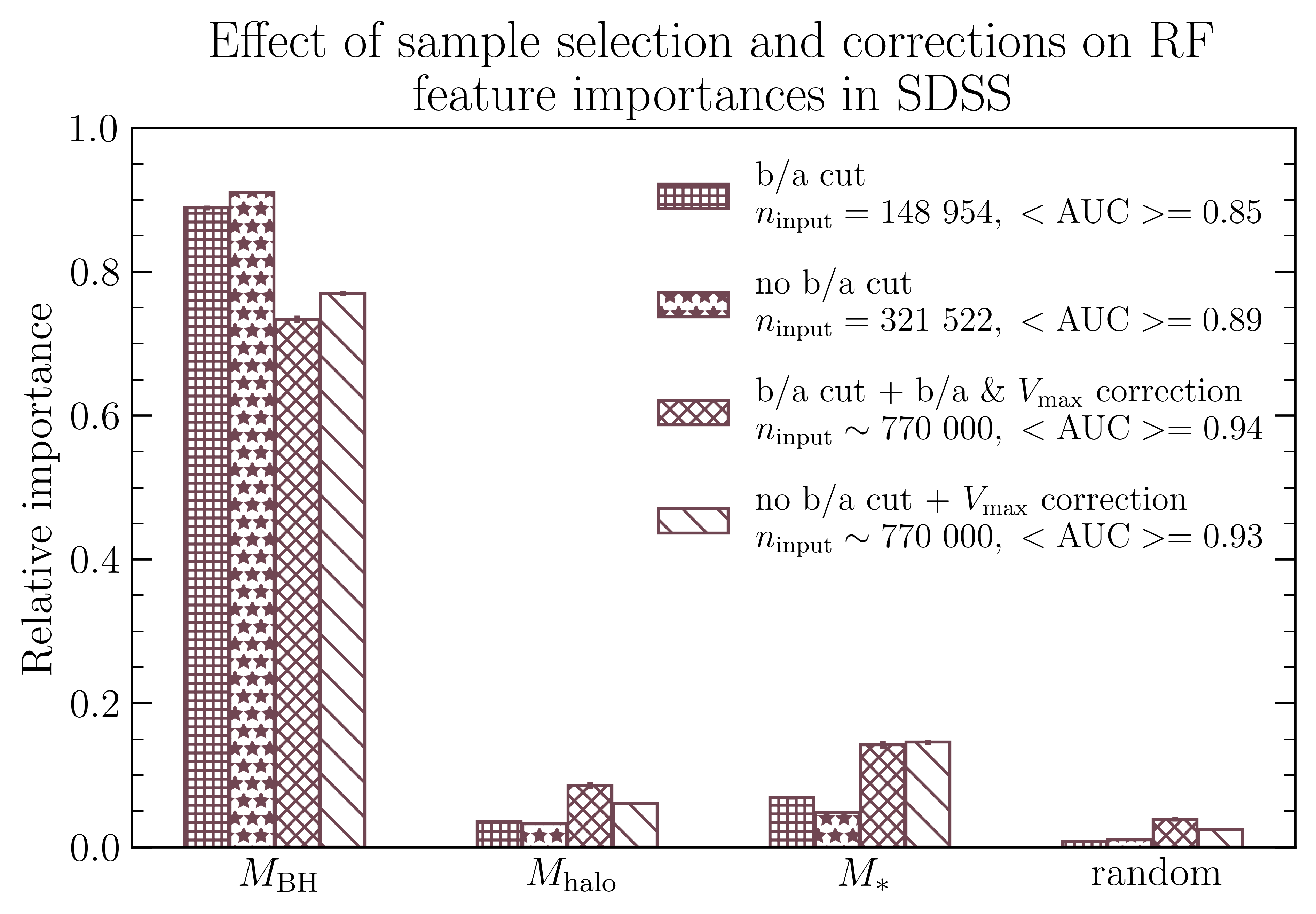}
	\caption{Testing the effect of adding corrections and/or removing an inclination cut on
				RF feature importances in the SDSS. It is apparent that \mbh\ remains clearly the most
				important parameter, regardless of the tested input data set.}
	\label{fig:corrections-rf}
\end{figure}

\subsection{Including sample corrections in the RF}
\label{sec:discussion:sample-cuts}

Another issue to consider when approaching our classification result
for the SDSS galaxies,
is accounting for the volume weighting and inclination correction
within the Random Forest experiment. We can achieve such corrections 
by drawing a~large sample of objects with replacement from the following 
probability distribution:
\begin{equation}
p_i = \frac{\omega_i}{\sum_{i} \omega_i},
\end{equation}
where $\omega_i$ is the combined weight defined earlier in Eq.~\ref{eq:sdss-weight}.
In this way objects with higher weights (predominantly low-mass and star-forming galaxies)
are more likely to be drawn and hence enter the input sample several times. 
The astronomical instinct to correct for the unobservable parts of the Universe, 
however, is in conflict with the machine learning algorithm, which assumes that
the training and testing sets are disjoint and hence the testing exclusively 
involves previously unseen data. Upon random splitting into the test and training
sets the same repeated object will undoubtedly be found in both of them, violating 
this assumption. In order to convince ourselves, however, that ignoring individual 
object weights does not affect our main conclusions, we implement the RF algorithm
in four different scenarios, testing how adding corrections and/or removing the 
inclination cut on disks affects the resulting feature importances.

Fig.~\ref{fig:corrections-rf} shows the effect of adding corrections and/or 
removing an inclination cut on RF feature importances in the SDSS. 
It demonstrates that \mbh\ is by far the most important parameter 
among all three, regardless of the applied sample 
cuts and corrections in the Random Forest algorithm.
Hence we conclude that our conclusions in Sec.~\ref{sec:rf} for the SDSS 
which do  not include the $1/V_{\rm max}$ and inclination corrections are both: 
a) consistent with the ones reached with
these corrections implemented; and, 
b) what is crucial, consistent with the main assumptions
of the Random Forest algorithm. 

Finally, we also learn that introducing corrections inflates the AUC score, 
as expected from the repeated exposure to the same objects in both the training
and testing samples. This repetition is also responsible for the increased 
importance of a~random number in classification. 
Removing the inclination cut with no correction
effectively adds uncertainty to the \mbh\ values, since the \vdisp\ measurement 
in highly inclined disks becomes a~noisy tracer of disk rotation. 
This is apparent in the decreased importance of \mbh\ in comparison 
with other parameters. One can also notice that accounting for
the Malmquist bias and the inclination correction swaps the order of \mhalo\ and 
\mstar, potentially owing to a~higher precision with which the latter is measured.

\begin{figure}	
\includegraphics[width=\columnwidth]{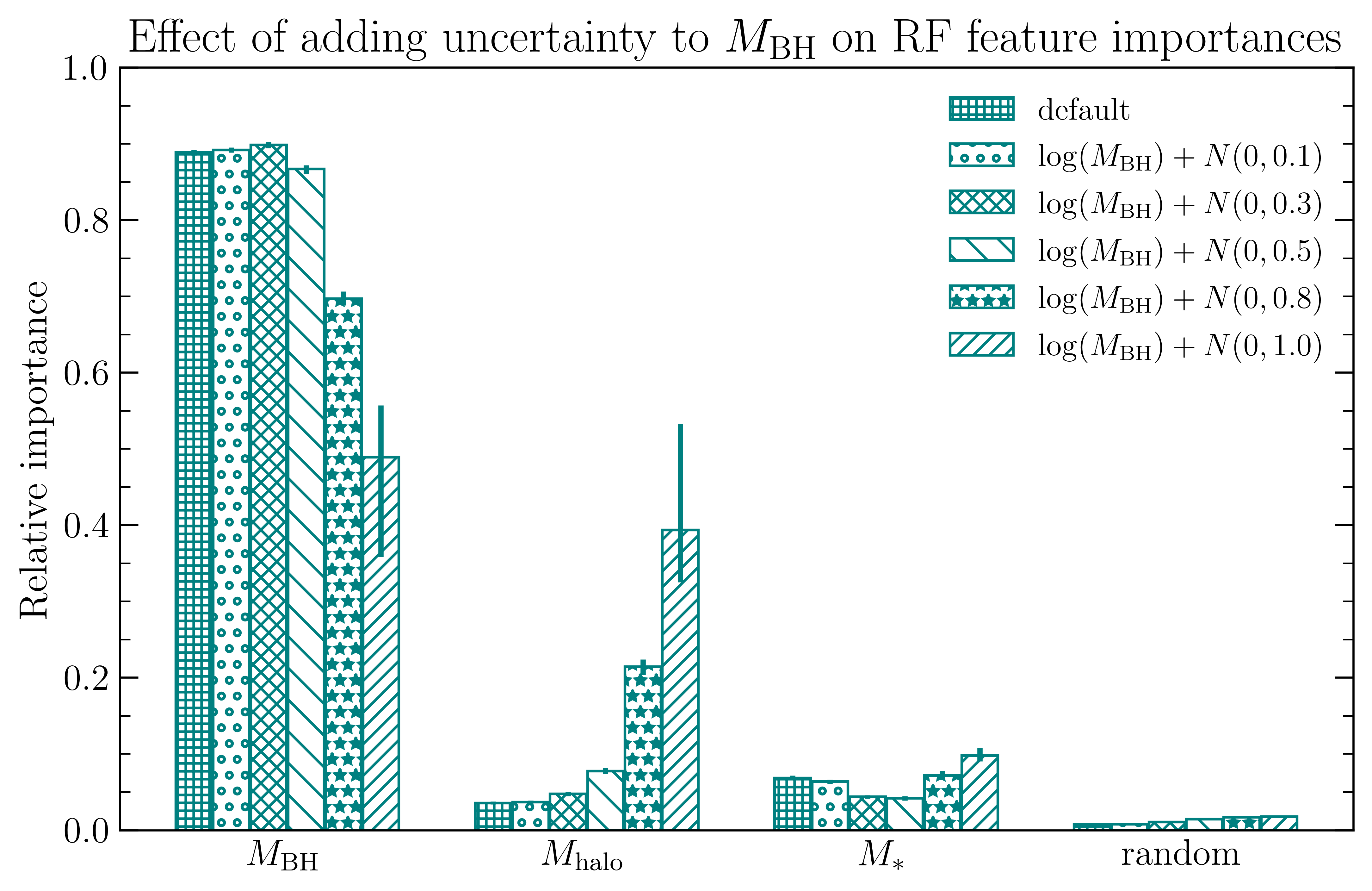} 
    \caption{The effect of adding scatter to \mbh\ estimates in the SDSS on RF feature
    		 importances. This test addresses the concern of the RF result being impacted
    		 by a~hypothetical better precision in the \vdisp\ measurement as compared to \mhalo\ and
    		 \mstar. The figure shows that adding even as much as 0.8~dex of Gaussian random 
    		 noise to \mbh\ does not allow other parameters to take the lead. This behaviour
    		 breaks down at around $\sim$1~dex added uncertainty, when \mhalo\ and \mbh\ 
    		 carry the same importance within $1\sigma$ uncertainty. 
    		 This is significantly more than the estimated difference in precision with which 
    		 we measure both parameters, equal to $\sim$0.1~dex (i.e. an order of magnitude
    		 less than required).}
    	\label{fig:rf-err}
\end{figure}

\subsection{The influence of measurement uncertainty on feature importances}
\label{sec:discussion:err}

The last item in our Random Forest robustness discussion is the potential effect that 
parameter uncertainty can have on the order of feature importances output by the
algorithm. This issue, of course, is only relevant for the SDSS, where each variable
is associated with either a~measurement error or an uncertainty introduced by
the use of a~calibration. Given that all parameters of interest are estimated with
a range of unique methods, it is naturally expected for all parameters to be 
measured or inferred with different degrees of precision. The immediate worry
in this scenario is for the algorithm to select the feature measured with the 
smallest error as the most informative one in the classification. 

The reason for this is the fact that 
parameters with increased levels of noise (poorer measurement quality) are more
likely to bury their connection to the class labels in their increased scatter. 
Similarly, more precise measurements are less likely to result 
in a~misclassification because the observed and true values are closer together. 
Hence, in the hypothetical case of two parameters being equally predictive 
of quenching, the one with smaller measurement uncertainty would be 
identified as a stronger predictor in the pair.

In the context of our Random Forest experiment, we want to check whether the strong
dominance of \mbh\ over \mhalo\ and \mstar\ in the SDSS could be driven by the
differences in precision with which all three parameters are measured. According to
the published uncertainty estimates in our sample, stellar mass is the most precisely 
measured variable, with a~median error of $\sim 0.15\ {\rm dex}$ on \mstar. 
For black hole masses, the main source of uncertainty comes from the scatter
about the $M_{\rm BH}-\sigma_\ast$ relation, which amounts to $\sim 0.45\ {\rm dex}$,
around 3 times more than in the case of \mstar. Finally, the halo masses are
the least precise estimates in the SDSS and their standard error is assumed 
to be $\sim 0.5\ {\rm dex}$. 

Given that the importance of stellar mass for the
passive galaxy classification is lower than that of black hole mass, it is certain
that \mbh\ is the more relevant parameter between the two. Increasing the quality
of \mbh\ estimates in the SDSS would only lead to a~bigger gap between these two
feature importances, further highlighting the black hole dominance. In the case of halo
mass the situation is different. If \mhalo\ and \mbh\ are 
hypothetically equally predictive of quenching classification, then a slightly 
worse precision in \mhalo\ measurement would cause its feature importance to 
be less than that of \mbh. 

In order to check whether the order of feature importances in the SDSS could
be driven by the difference in measurement quality, we repeat the RF experiment
for different levels of scatter added to \mbh. 
In each trial we create an augmented dataset by adding random 
Gaussian noise centred on 0 to the \mbh\ measurement,
such that $\log(M_{\rm BH,\ new})=\log(M_{\rm BH}) + N(0,\sigma)$. 
Throughout this test we increase the value of $\sigma$ until the 
relative importance of \mbh\ and \mhalo\ are comparable, to estimate 
how much more precisely black hole mass 
would have to be measured in order to drive our main conclusion.
What we check for in this test is not how the individual uncertainty 
affects the order between \mbh\ and \mhalo\ but rather what is the impact
of a `differential measurement error', i.e. the difference in precision 
between the two parameters.  

In Fig.~\ref{fig:rf-err} we present the result of our test, which clearly shows
how the difference between \mbh\ and \mhalo\ decreases with increasing 
level of differential noise added to \mbh\ alone. 
It takes as much as an additional 1~dex scatter for the importance 
of \mhalo\ to be comparable with \mbh. This is a factor of 10 higher
than the estimated difference in measurement quality between the two variables 
of $\sim 0.1$ dex. Therefore, we conclude that our Random Forest 
result is not driven by the differences in parameter uncertainty and 
that the overwhelming \mbh\ dominance is an intrinsic quality of 
AGN feedback quenching in massive central galaxies.

\section{Comparison with dynamical measurements of \mbh}
\label{sec:appendix:terrazas}

\begin{figure*}
\includegraphics[width=\textwidth]{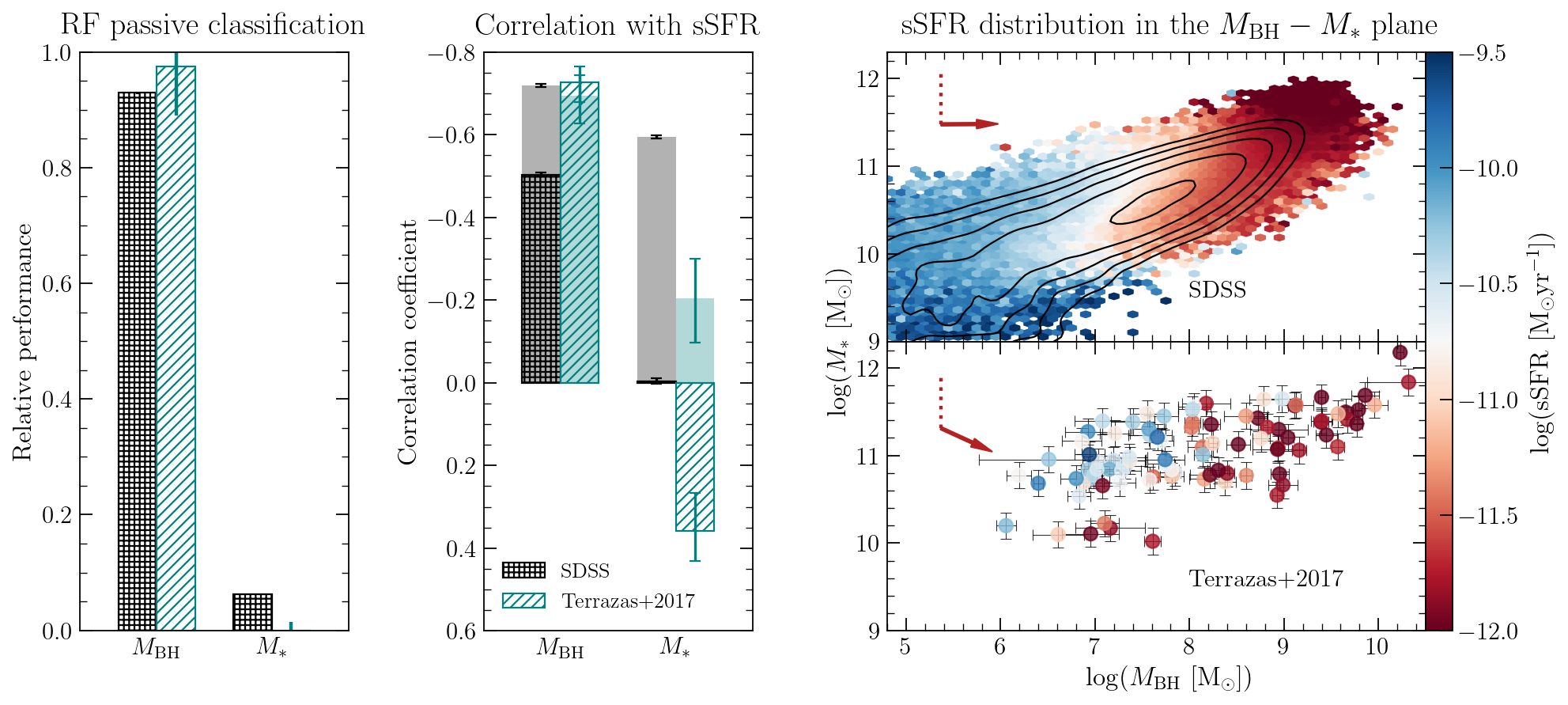} 
    \caption{RF quenching classification analysis repeated with Terrazas+2017 data, 
    		 compared to SDSS, showing that the conclusions we draw from our indirectly 
    		 inferred \mbh\ are consisted with dynamically measured black hole masses. 
    		 It is important to recognise, however, 
    		 that the dynamically measured sample of black hole masses is not representative 
    		 of the local Universe.}
    	\label{fig:appendix-terrazas}
\end{figure*}

As a final consistency check of our method we investigate the reliability 
of calibrations to estimate \mbh\ in the SDSS galaxies. To this end 
we repeat our Random Forest and correlation analyses in a sample of 90 dynamically
measured black hole masses in \cite{Terrazas17}. Because we only have access to two 
parameters: \mbh\ and \mstar, we repeat the RF classification in the SDSS using 
these two parameters only, in order to compare both sets of observations side-by-side. 
We use the same algorithm architectures as before, requiring a minimum of 5 and 200 
objects per leaf node in the \cite{Terrazas17} sample and the SDSS respectively. 
In the SDSS we show the results  using the calibration in Eq.~\ref{eq:S16} to estimate 
\mbh\ from \vdisp\ \citep{Saglia16}, however our results are consistent for all other 
calibrations explored in this work.

The leftmost panel in Fig.~\ref{fig:appendix-terrazas} shows the relative importance
of \mbh\ and \mstar\ for determining whether a galaxy is star forming or quenched,
using \mbh, \mstar\ and a random number as input features.
Both observations unanimously agree that black hole mass holds the most predictive
power in determining the star formation state of a~galaxy and that it dwarfs \mstar\
in comparison. 

The middle panel in Fig.~\ref{fig:appendix-terrazas} 
shows PCC values, following the same structure as
Fig.~\ref{fig:corr-mstar}, where shaded bars correspond to the Spearman's
rank correlation coefficient between a given parameter and sSFR where hatched bars
indicate the PCC. In the sample of dynamical measurements of \mbh, stellar mass
shows a positive correlation with sSFR once its connection to black hole mass
is accounted for, which is not seen in our SDSS analysis. This discrepancy
could be driven by both the differences in sample selection as well as 
the precision with which \mbh\ is measured in the data sets.
We also note that a~mild inversion of the PCC with 
stellar mass was previously shown in \cite{Bluck20a}, where the authors 
used a measurement of $\Delta {\rm SFR}$ instead of sSFR to 
calculate the correlations. In consequence, the quenching direction of 
$\theta= 116.2^{+ 4.6}_{- 6.7}$ in the \cite{Terrazas17} sample is mildly 
oriented towards decreasing \mstar, 
unlike $\theta=89.5^{+ 0.8}_{-0.8}$ in the SDSS, which is almost exactly 
horizontal. Both quenching directions, however, show that increasing 
black hole mass is the more efficient way to quench a galaxy, between
\mstar\ and \mbh. 

Results presented in Fig.~\ref{fig:appendix-terrazas} show an excellent qualitative
agreement between our SDSS analysis and a sample of dynamically measured
black hole masses. Therefore we are confident that our use of calibrations to 
estimate \mbh\ from other galactic parameters like \vdisp\ and \mbulge\
does not drive artificial trends between quenching and the AGN activity
integrated over time.

\section{Observational realism in the simulations}
\label{sec:sim-realism}

In order to extract theoretical predictions which we can reliably compare against
the observations in Secs.~\ref{sec:gas} and \ref{sec:gas-line}, we apply three layers
of observational realism to the raw output from Illustris and TNG. We restrict the
simulated galaxies to central regions covered by the SDSS fibres, imitate the 
quality cuts on emission line fluxes and account for measurement uncertainty by
adding random scatter to the fibre-restricted, S/N-ratio selected galaxy population.

In order to mimic the field of view of SDSS fibres we first compute the distributions of 
fibre radii ($r_{\rm fib}$) in units of kpc, splitting the SDSS sample into three bins in stellar mass.
In Fig.~\ref{fig:appendix-rfib} we show the distributions in $r_{\rm fib}$ along with
Gaussian fits in each \mstar\ bin separately. As expected, the fibres in the least massive
objects cover the smallest central regions due to their proximity, while the most massive 
objects are found at furthest distances away, hitting our imposed sample selection limit
of $z=0.2$ (fibre physical radii of just under 5 kpc). For each galaxy in a given stellar
mass bin in Illustris and TNG we take a random draw in $r_{\rm fib}$ from its corresponding
Gaussian distribution. Then, we integrate the galaxy profile in $\rho_{{\rm MH2}}$,
 $\rho_{M_\ast}$ and $\rho_{\rm SFR}$ up to this radius to obtain `in-fibre' estimates
in galaxy molecular gas mass, stellar mass and star formation rate. 

Whenever this procedure yields ${\rm SFR_{fib}}=0$, we set 
\mbox{${\rm sSFR_{fib}}=10^{-12} {\rm yr^{-1}}$} and
${\rm SFE_{fib}}=10^{-11}{\rm yr^{-1}} $ and mark regions in 
Figs.~\ref{fig:line-plots-sdss},~\ref{fig:appendix:line-plots}~and~\ref{fig:appendix:h2-calibrations}
in grey hatching to indicate where these objects dominate median trends.
Whenever the integrated in-fibre $M_{\rm H2,\, fib}=0$, we set 
$\log(f_{\rm gas})= -3.5$ and mark regions where these objects dominate with
maroon hatching. Objects which have both 
${\rm SFR_{fib}}=0$ and $M_{\rm H2,\, fib}=0$ are removed from the SFE analysis
due to their undefined value of ${\rm SFE_{fib}}$.
Because the published profiles
are discretised into radial bins we always choose the radial extent closest to the
random draw. In the case of very extended galaxies we use a single, most central radial bin, 
regardless of its extent in kpc, which can be higher than the \mbox{$\sim 5$ kpc} cut-off we 
see in the SDSS.

\begin{figure}
\includegraphics[width=\columnwidth]{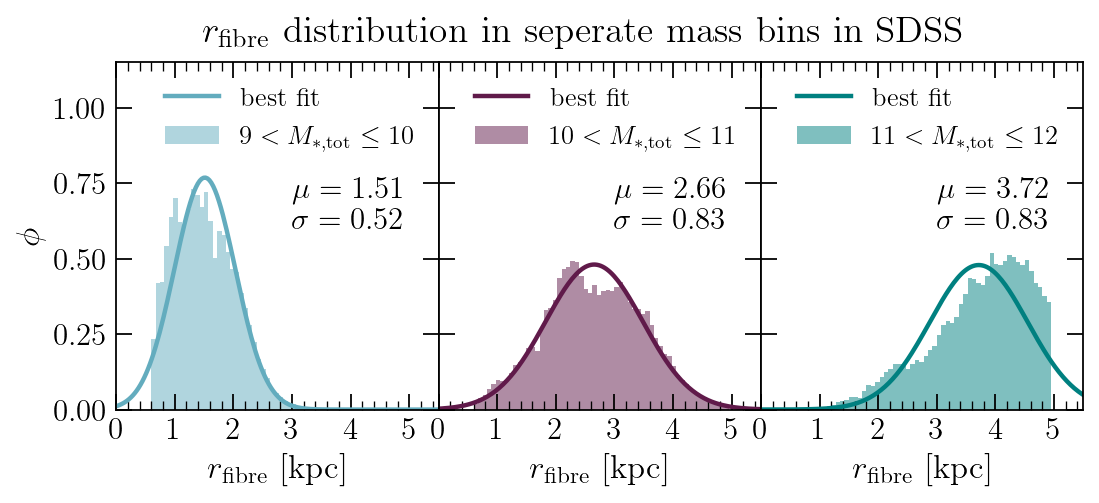} 
    \caption{Gaussian fits to fibre radii distributions in the SDSS in bins of stellar mass.
    		Each galaxy in the simulations is assigned a~random draw from the fits and relevant
    		quantities are integrated up to this radial extent to check how restricting the simulated
    		results to galactic centres influences raw trends for entire galaxies.}
    \label{fig:appendix-rfib}
\end{figure}

The second step in our observational realism imitates object selection based on 
emission line fluxes in the SDSS. This sample cut effectively removes objects with low
SFR from our sample of observed galaxies, substantially decreasing the size of its
quiescent population. In order to apply a similar selection in the simulations
we first calculate the sample completeness as a function of $\log({\rm sSFR})$ in the 
SDSS, once we select the objects based on cuts in S/N ratios on emission 
line fluxes. In Fig.~\ref{fig:appendix-sfr-completeness} we present the SDSS completeness
as a black solid line, overplotted on the raw SFR distribution in Illustris and TNG.
For each simulation we then split the objects into 0.2~dex wide bins in $\log({\rm sSFR})$
and randomly select a fraction of galaxies for further analysis, 
corresponding to the SDSS completeness in a~given bin. As we show in 
Fig.~\ref{fig:appendix-sfr-completeness} the drop in SDSS completeness coincides
with the low-sSFR tail of the distributions in both simulations, hence this
realism step preferentially rids the simulations of their quiescent objects, as
we originally expected. Most importantly, we remove all objects with global 
SFR=0 in the simulations, which form a non-negligible population in TNG.

\begin{figure}
	\includegraphics[width=\columnwidth]{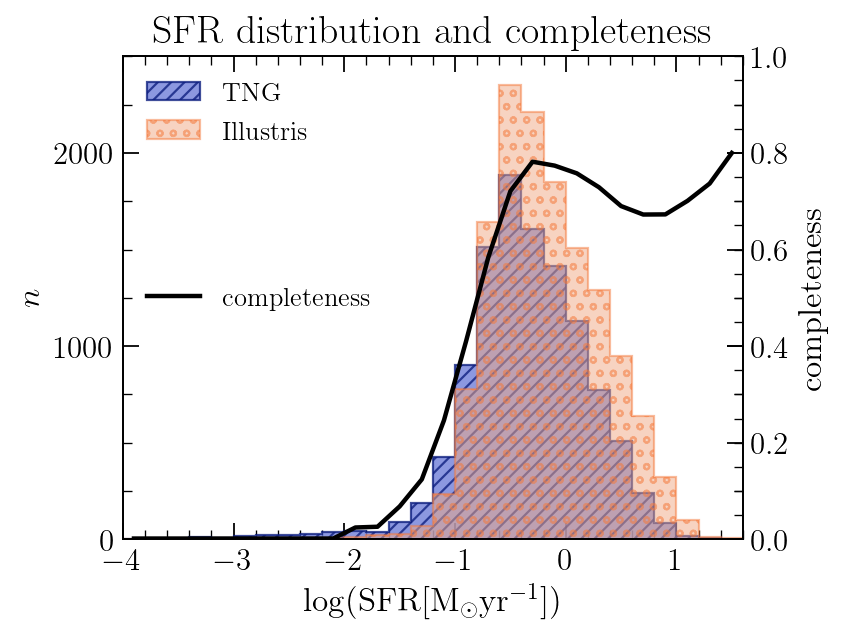} 
    \caption{SDSS sample completeness as function of SFR calculated after imposing an emission line
    		S/N ratio cut, which is overlaid on the distribution of SFR values in the simulations. For each
    		SFR bin, a~fraction of galaxies corresponding to the SDSS completeness is drawn at random
    		and only this selected subset of objects is used in further analysis. This step
    		mimics the effect of emission line quality cut in the observations.}
    \label{fig:appendix-sfr-completeness}	
\end{figure}

As a final step in the simulation post-processing we add random scatter to the
data, mimicking measurement uncertainty associated with quantities estimated 
in the SDSS. As we explain in the main text, for each simulated galaxy we add
a random draw from a Gaussian centred on 0, such that a given parameter $X$
is assigned a new value of $\log(X_{\rm new}) = \log(X) + N(0, \sigma_X)$ 
and $\sigma_X$ is the median measurement error on this quantity in the SDSS.
In Figs.~\ref{fig:appendix-fgas-realism}-\ref{fig:appendix:line-plots} we treat
\mbh, \mstar, $M_{\rm \ast, \, fib}$, SFR, ${\rm SFR_{fib}}$ and $M_{\rm H2, \, fib}$
in this fashion, using $\sigma_X$ values listed in Table~\ref{tab:sigmax}. 
In the case of SFR (both the in-fibre and global) we only add scatter to star forming
objects (i.e. classified according to log(sSFR/${\rm yr^{-1}}$) $> -11$) 
since we treat SFR values in the quenched SDSS
galaxies as upper limits estimated from the strength of D4000.

The second columns in 
Figs.~\ref{fig:appendix-fgas-realism},~\ref{fig:appendix-sfe-realism}~\&~\ref{fig:appendix:line-plots}
show the effect of restricting galaxies to SDSS fibre 
apertures in the simulations. Overall, this primarily acts 
to highlight and strengthen trends previously seen in entire galaxies.
When we look at Fig.~\ref{fig:appendix-fgas-realism}, we see that raw \fgas\ 
(labelled as `default' in the figure) in both simulations
show decreasing trends between the star forming and passive populations.
These trends are also perfectly recovered in the galactic centres, albeit with significant
quantitative differences. In Illustris both the SF and PA galaxies have lower \fgas\ in
their cores, however the difference is bigger for the Main Sequence objects (a decrease
of 0.58~dex in the median value, in contrast with 0.13~dex in the passive population). In
TNG, on the other hand, aperture correction has a dramatic effect on \fgas, revealing
passive galaxies which are overwhelmingly devoid of gas in their centres (the median
value of  $\log(f_{\rm gas})$ in the PA population is equal to -4.60~dex, in contrast
with -1.92~dex in the case of whole galaxies). We see very similar effects of 
aperture correction on median \fgas\ as a function of \mbh\ in 
Fig.~\ref{fig:appendix:line-plots} as well, where the pronounced decline in raw \fgas\ in Illustris 
is only marginally steepened, while the central regions of galaxies in TNG record
a dramatic drop in \fgas\ around $\log(M_{\rm BH}/M_\ast) \sim 8$.

Aperture correction in the simulations also acts to strengthen the trends in SFE
seen in the raw data. As demonstrated in Fig.~\ref{fig:appendix-sfe-realism}, 
galactic centres of quenched objects in Illustris 
are rather highly efficient at forming new stars, in contrast with their MS counterparts.
With the exception of a few high-\mstar\ outliers, galaxies overall classified as
passive in the suite are making the most of the limited gas available in their
centres. When we then look at the trends in SFE as a function of \mbh\ in 
Fig.~\ref{fig:appendix:line-plots} we can see that the formerly present turn-off 
at high-\mbh\ is no longer present in the galactic centres and that,
in fact, the most massive black holes in the sample are associated
with the highest SFE values in Illustris. In TNG the trend in SFE goes in the
exact opposite direction to Illustris in Fig.~\ref{fig:appendix-sfe-realism} 
and aperture correction steepens it by increasing the median difference between SF and PA populations
by 0.2~dex. We also see that the galactic centres in the passive sequence
are almost exclusively not star forming at all, reaching the SFE floor value.
Similarly, when we look at SFE as function of \mbh\ in Fig.~\ref{fig:appendix:line-plots},
galaxies above $\log(M_{\rm BH}/{\rm M_\odot}) \sim 8$ have zero star formation
efficiency and only recover their star-forming ability at the highest black hole masses.

Mimicking S/N ratio cuts in the simulations has straightforward consequences
for the results in Figs.~\ref{fig:appendix-fgas-realism} and \ref{fig:appendix-sfe-realism}.
The passive sequence in both Illustris and TNG is significantly less numerous
and the majority of outliers in \fgas\ and SFE are removed from the \msfr\ plane.
The median values in PA and SF populations are only slightly affected by this realism
step and hence the trends remain virtually unchanged. In Fig.~\ref{fig:appendix:line-plots}
the S/N cut affects primarily galaxies with the highest \mbh, decreasing the scatter
around median trends in \fgas\ and SFE at the high black hole mass end.

\begin{figure*}
	\includegraphics[width=\textwidth]{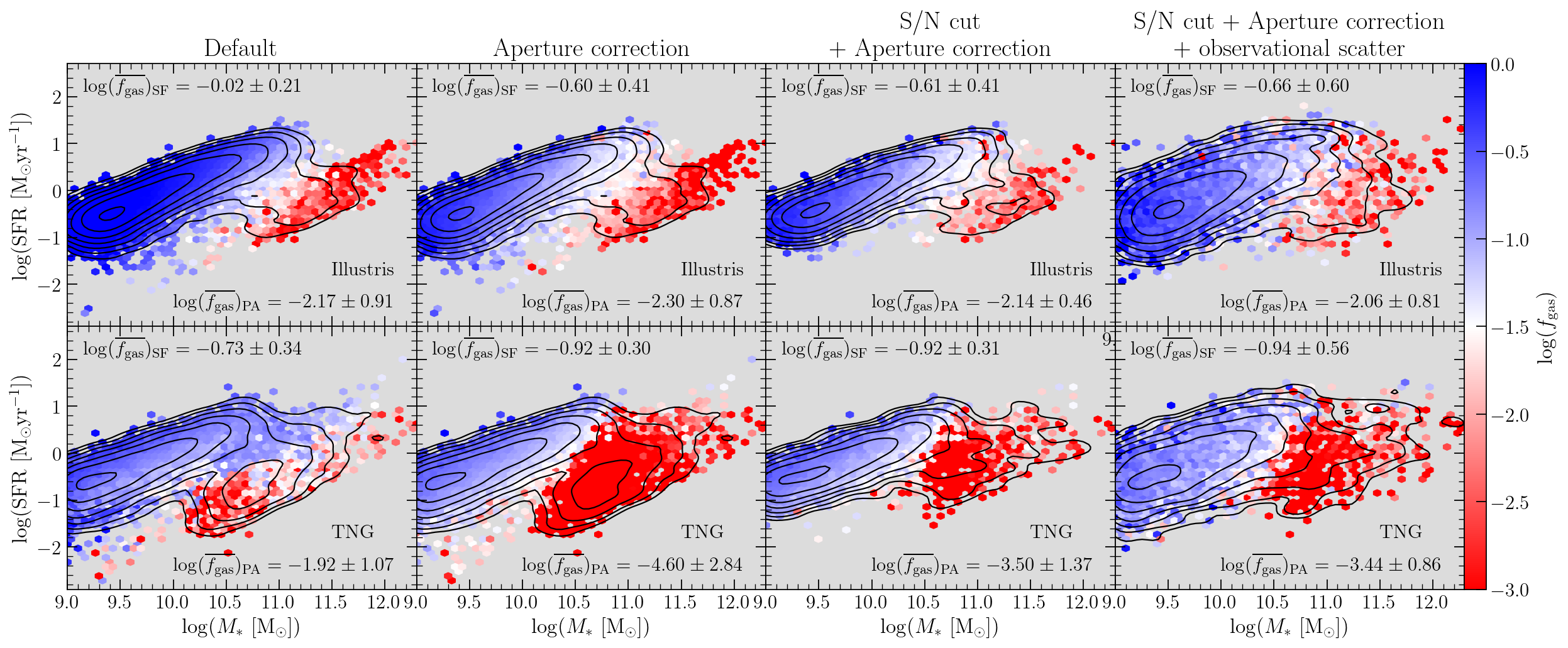}
	\caption{The influence of all observational realism steps on the raw \fgas\ values extracted
					from Illustris (top row) and TNG (bottom row) in Fig.~\ref{fig:msfr-fgas-sdss}. 
					Colour indicates median \fgas\ in
					each hexagonal bin, while black contours trace object density in the plane.
					In both simulations the aperture correction (second column) results in a decrease 
					in \fgas, however the effect is more pronounced in TNG. 
					Imitating the S/N ratio cut on emission lines (third column) visibly decreases the 
					quenched population, dramatically decreasing its size. Mimicking measurement
					error by adding scatter to the data (rightmost column) causes the Main Sequence
					to swell, however the decreasing trend in \fgas\ is preserved well.}
	\label{fig:appendix-fgas-realism}	
\end{figure*}

\begin{figure*}
	\includegraphics[width=\textwidth]{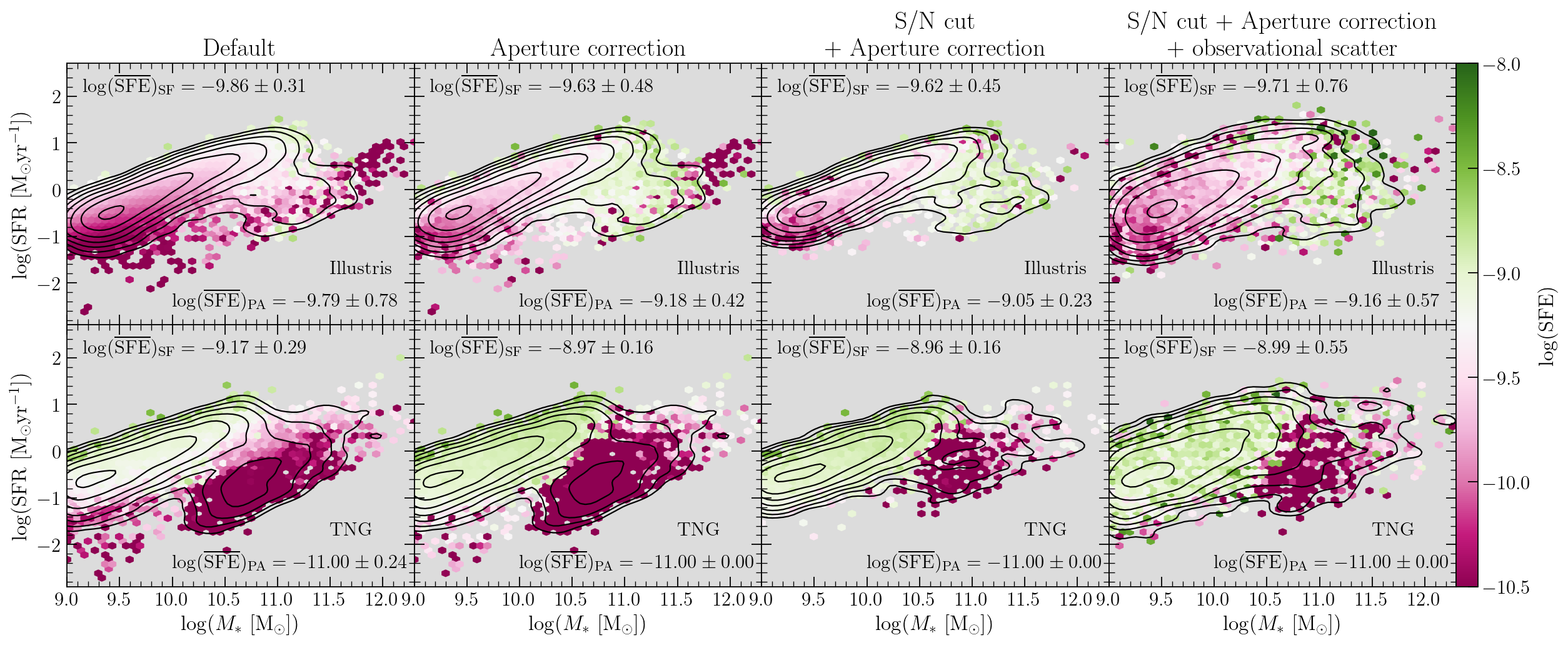}
	\caption{Same as Fig.~\ref{fig:appendix-fgas-realism} but with SFE instead of \fgas.
				 The aperture correction (second column) strongly enhances raw trends
				 in SFE, showing no star formation in galactic centres in TNG and 
				 very efficient formation of new stars in Illustris. Adding observation-like 
				 scatter and mimicking the S/N ratio cut in the data only further highlights
				 the contrast between Illustris and TNG. }
	\label{fig:appendix-sfe-realism}			
\end{figure*}

Accounting for measurement error in the simulations preserves all trends seen
previously in the \msfr\ plane, however it has profound consequences for 
the distribution of galaxies on the star forming Main Sequence.
Added scatter significantly spreads out the MS relations in the simulations,
causing them to swell and reach down towards the region in the plane
occupied by quenched galaxies in the observations. Albeit generous,
the random scatter added to raw data is motivated by the observed measurement
errors and hence this behaviour of simulated MS might suggest the existence
of potential Main Sequence regularisation mechanisms which are not
currently captured in cosmological models of structure formation.
In Fig.~\ref{fig:appendix:line-plots} measurement error primarily
acts to smoothen median trends in sSFR, \fgas\ and SFE with \mbh, 
in particular concealing abrupt changes in these quantities in TNG.
The difference between the third and fourth columns in the figure
suggests an interesting possibility, where even the most dramatic
underlying trends in galactic properties could be significantly smeared
out by the uncertainty associated with the observations to produce
smooth relationships we see in the SDSS.

\begin{figure*}
\includegraphics[width=\textwidth]{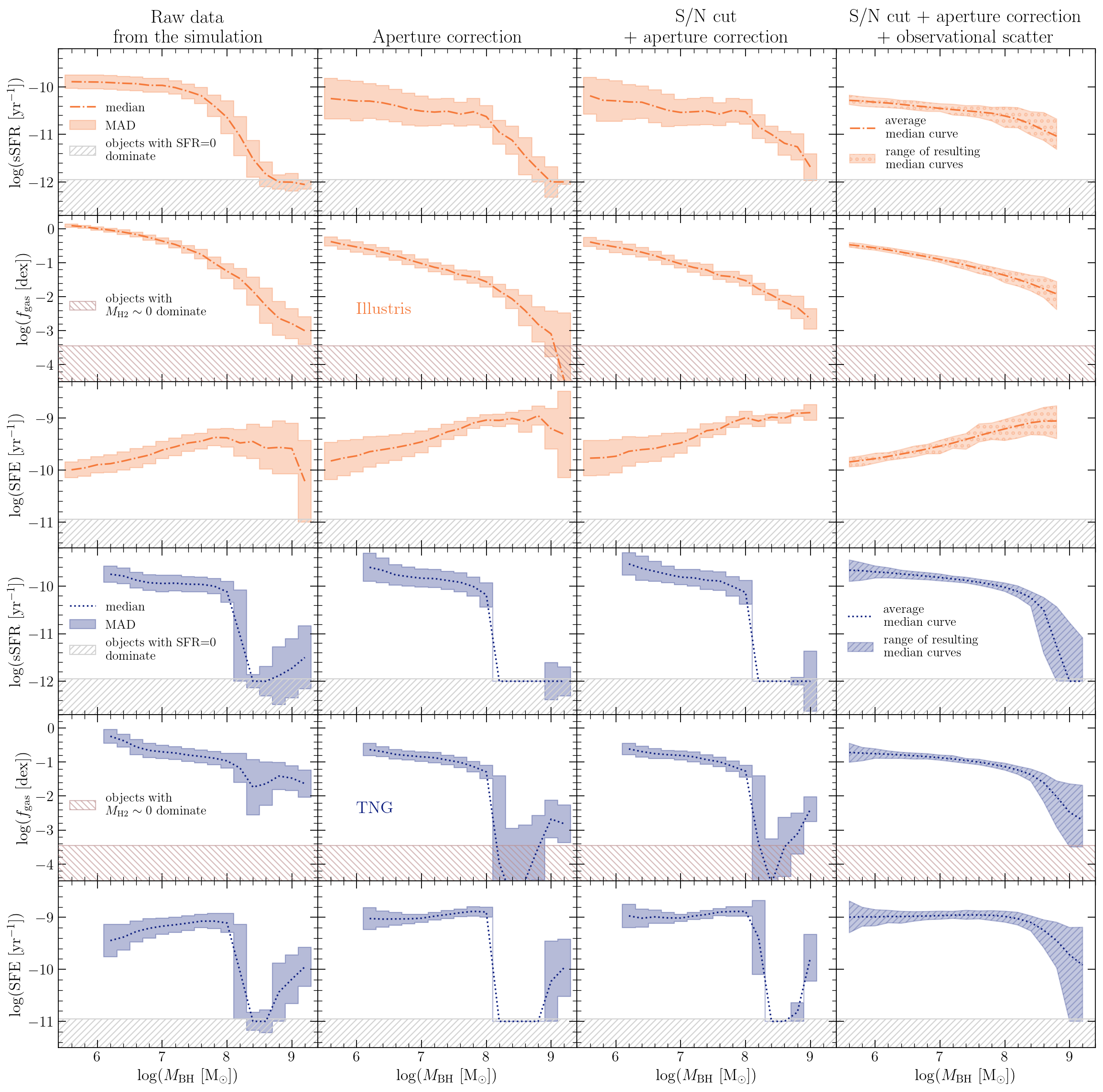} 
    \caption{All steps in simulation realism necessary to move between Figs.~\ref{fig:line-plots-raw}
    					and \ref{fig:line-plots-sdss}. As seen previously in Figs.~\ref{fig:appendix-fgas-realism}
    					and \ref{fig:appendix-sfe-realism}, aperture correction reveals more gas-poor centres
    					in both simulation suites, which are efficiently forming stars in Illustris and are completely
    					deprived of star formation in TNG. Imitating the S/N ratio cut on emission lines in the SDSS
    					affects primarily the quenched population associated with the highest black hole masses in
    					the sample. Adding observation-like measurement uncertainty to the raw data does 
    					not have a very strong impact on the median trends in Illustris, while in TNG it acts to 
    					smoothen out abrupt drops in sSFR, \fgas\ and SFE around 
    					$\log(M_{\rm BH}/{\rm M_\odot}) \sim 8$.}
    \label{fig:appendix:line-plots}
\end{figure*}

\section{Testing different HI/H\textsubscript{2} transition models}
\label{sec:appendix:h2-models}

\begin{figure*}
\includegraphics[width=\textwidth]{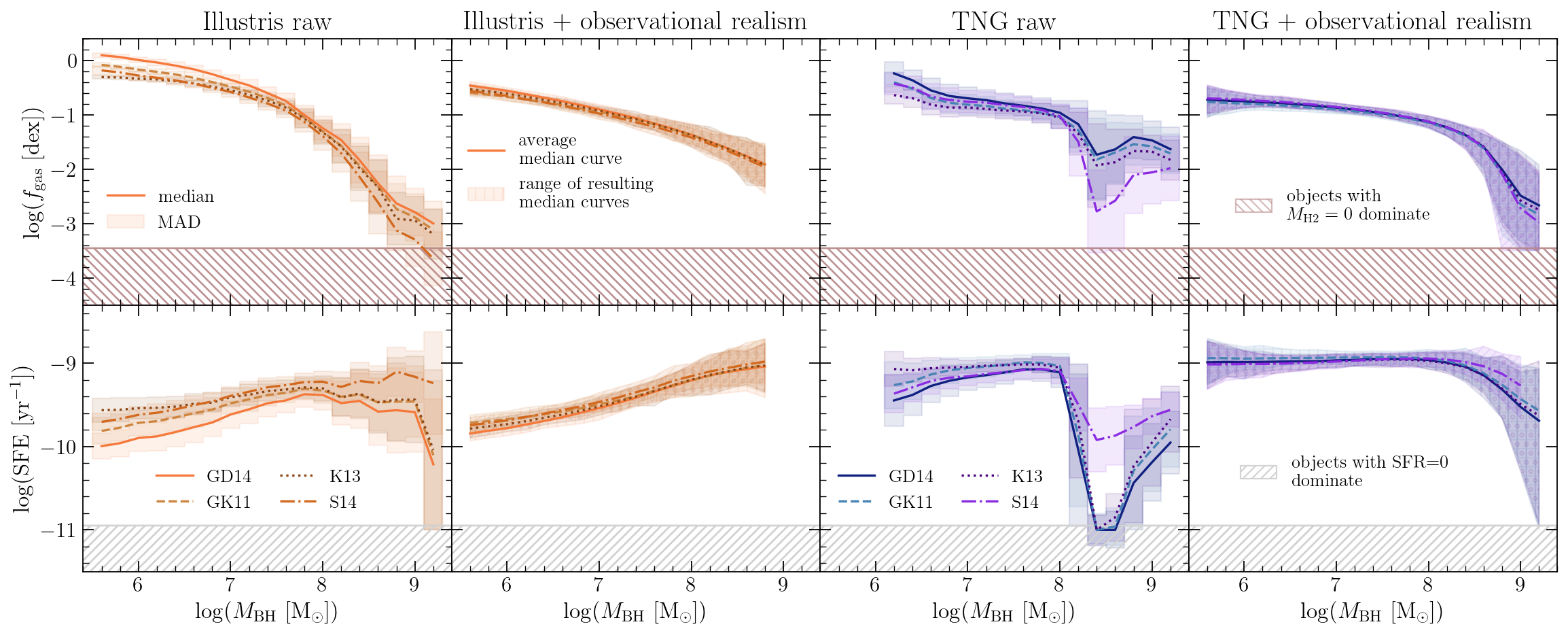}
    \caption{Comparison of\fgas\ and SFE trends among all HI/${\rm H2}$ transition models used 
    				to calculate molecular gas fractions in the publicly available HI/H2 galaxy content 
    				catalogues. When we focus on the raw output from the simulations we see 
    				minimal differences among the models. Once observational realism is included, 
    				these differences vanish completely, leaving behind perfectly consistent trends
    				in median quantities.}
    \label{fig:appendix:h2-calibrations}
\end{figure*}

In order to successfully capture large cosmological volumes 
within a finite computation tim}, simulations need to 
settle for a~relatively low resolution of about a~kiloparsec in 
spatial extent and a~million \msol\ in mass. Moreover,
the computational cost associated with including chemical networks 
and radiative transfer calculations on-the-fly
prevents cosmological simulations from tracing the detailed 
evolution of elements in different phases present within the 
modelled gas. As a consequence of this limitation, a range of 
physical properties of gas cannot be directly predicted in 
a~simulation run and requires additional post-processing to extract 
quantities of interest. An example of such property is the multiphase 
structure of hydrogen, in particular the abundance of 
HI and \hmol. In order to estimate those in a given 
simulation snapshot one uses a HI/\hmol\ transition model 
- a prescription for calculating the fraction of neutral hydrogen
in its molecular form given gas state variables, the presence of dust 
and local ionising radiation. Due to the resolution limit, which
is significantly larger than the size of individual molecular clouds 
in cosmological simulations, these prescriptions depend
on locally averaged properties estimated for individual gas cells
like e.g. gas metallicity, density or the UV background.
 
The publicly available HI/\hmol\ catalogues for Illustris and IllustrisTNG \citep{Diemer18} 
contain results obtained using four different models for the HI/\hmol\ transition.
The \cite{Gnedin14} (GD14) and \cite{Gnedin11} (GK11) models are based on high-resolution
simulations of isolated disk galaxies, while \cite{Krumholz13} (K13) and \cite{Sternberg14} (S14)
use analytic prescriptions to compute molecular gas fractions. In this work we present all 
gas-oriented results using GD14, however we do not have a preference for the
choice of model among the available set. For each step in our analysis we also check
that the choice of HI/\hmol transition model does not strongly influence our results 
and demonstrate this explicitly using median trends in \fgas\ and SFE 
as an example in Fig.~\ref{fig:appendix:h2-calibrations}.

Fig.~\ref{fig:appendix:h2-calibrations} presents both the raw results (columns 1 and 3)
and simulations post-processed with observational realism (columns 2 and 4) for 
all four HI/\hmol transitions models in Illustris and TNG. The top row shows
the behaviour of gas fractions, while the bottom one compares the trends in SFE,
both as a function of \mbh. It is apparent that the choice of model leads to only very subtle
differences in trends inferred from the direct output of the simulations. TS14 
deviates most from the other three models and these differences are more striking
in SFE in both simulations. They are mainly caused by the differences in galaxy
samples among the four curves, because S14 estimates many more objects with
\mgas=0 which also have SFR=0. These need to be removed from the SFE
panels because their SFE value is undefined. When we add observational realism to all quantities,
however, all minor differences among the models virtually disappear, yielding
perfectly consistent median trends in \fgas\ and SFE.


\bsp	
\label{lastpage}
\end{document}